\documentclass{aa}

\usepackage{graphicx}
\usepackage{txfonts}
\usepackage{siunitx}
\usepackage{xcolor}
\definecolor{alizarin}{rgb}{0.82, 0.1, 0.26}

\newcommand{\oiii}{[O\,\textsc{iii}]}
\newcommand{\nii}{[N\,\textsc{ii}]}
\newcommand{\sii}{[S\,\textsc{ii}]}

\newcommand{\siii}{[S\,\textsc{iii}]}
\newcommand{\oii}{[O\,\textsc{ii}]}
\newcommand{\hii}{H\,\textsc{ii}}
\newcommand{\ha}{H$\alpha$}
\newcommand{\hb}{H$\beta$}
\newcommand{\re}{R$_\mathrm{e}$}
\begin{document}

   \title{SDSS IV MaNGA - Metallicity and ionisation parameter in local star-forming galaxies from Bayesian fitting to photoionisation models}

 \author{M. Mingozzi\inst{1}\fnmsep\inst{2,3}
        \and
F. Belfiore\inst{2, 4}
\and
G. Cresci\inst{2}
\and
K. Bundy\inst{5}
\and
M. Bershady\inst{6, 7}
\and
D. Bizyaev\inst{8,9}
\and
G. Blanc\inst{10,11}
\and
M. Boquien\inst{12}
\and
N. Drory\inst{13}
\and
H. Fu\inst{14}
\and
R. Maiolino\inst{15,16}
\and
R. Riffel\inst{17,18}
\and
A. Schaefer\inst{6}
\and
T. Storchi-Bergmann\inst{17,18}
\and
E. Telles\inst{19}
\and
C. Tremonti\inst{6}
\and
N. Zakamska\inst{20}
\and
K. Zhang\inst{21}
}

   \institute{Dipartimento di Fisica e Astronomia, Universit\`{a} degli Studi di Bologna, Via Piero Gobetti 93/2, I-40129, Bologna, Italy\\
              \email{matilde.mingozzi@inaf.it}
            \and
            INAF -- Osservatorio Astrofisico di Arcetri, Largo E. Fermi 5, I-50157, Firenze, Italy
            \and
            INAF -- Osservatorio Astrofisico di Padova, Vicolo dell'Osservatorio 5, I-35122, Padova, Italy
            \and
            European Southern Observatory, Karl-Schwarzschild-Str. 2, Garching bei M{\"u}nchen, D-85748, Germany.
            \and
            University of California Observatories, University of California Santa Cruz, 1156 High St., Santa Cruz, CA 95064, USA
            \and
            Department of Astronomy, University of Wisconsin-Madison, 475N. Charter St., Madison, WI 53703, USA
            \and
            South African Astronomical Observatory, P.O. Box 9, Observatory 7935, Cape Town, South Africa
            \and
            Apache Point Observatory and New Mexico State University, Sunspot, NM, 88349, USA
            \and
            Sternberg Astronomical Institute, Moscow State University, Moscow, Russia
            \and
            Observatories of the Carnegie Institution for Science, 813 Santa Barbara St., Pasadena, CA, USA
            \and
            Departamento de Astronomía, Universidad de Chile, Camino del Observatorio 1515, Las Condes, Santiago, Chile
            \and
            Centro de Astronomía (CITEVA), Universidad de Antofagasta, Avenida Angamos 601, Antofagasta, Chile
            \and
            McDonald Observatory, The University of Texas at Austin, 1 University Station, Austin, TX 78712, USA
            \and 
            Department of Physics \& Astronomy, The University of Iowa, 203 Van Allen Hall, Iowa City, IA 52242, USA	
            \and
            Cavendish Laboratory, University of Cambridge, 19 J. J. Thomson Avenue, Cambridge CB3 0HE, UK
            \and
            Kavli Institute for Cosmology, University of Cambridge, Madingley Road, Cambridge CB3 0HA, UK
            \and
            Laboratório Interinstitucional de e-Astronomia - LIneA, Rua General José Cristino 77, Rio de Janeiro, RJ 20921-400, Brazil
            \and
            Universidade Federal do Rio Grande do Sul, IF, CP 15051, Porto Alegre 91501-970, RS, Brazil
            \and
            Observatório Nacional, Rua José Cristino, 77, Rio de Janeiro, RJ, 20921-400, Brasil
            \and
            Center for Astrophysical Sciences, Department of Physics and Astronomy, Johns Hopkins University, 3400 North Charles Street, Baltimore, MD 21218, USA
            \and
            Lawrence Berkeley National Laboratory, 1 Cyclotron Road, Berkeley, CA 94720, USA
}

   \date{Received XXX; accepted XXX}

 
  \abstract
{
We measured gas-phase metallicity, ionisation parameter and dust extinction for a representative sample of 1795 local star-forming galaxies using integral field spectroscopy from the SDSS-IV MaNGA survey. We self-consistently derive these quantities by comparing observed line fluxes with photoionisation models using a Bayesian framework. We also present the first comprehensive study of the \siii$\lambda\lambda$9069,9532 nebular lines, which have long been predicted to be ideal tracers of the ionisation parameter.
Unfortunately, we find that current photoionisation model predictions substantially over-predict the intensity of the \siii\,lines, while broadly reproducing other observed optical line ratios. We discuss how to nonetheless make use of the information provided by the \siii\, lines by setting a prior on the ionisation parameter. Following this approach, we derive spatially-resolved maps and radial profiles of metallicity and ionisation parameter. 
The metallicity radial profiles derived are comparable with previous works, with metallicity declining toward the outer parts and a flattening in the central regions, in agreement with infall models of galaxy formation, that predict that spiral discs build up through accretion of material, which leads to an inside-out growth.
On the other hand, ionisation parameter radial profiles are flat for low-mass galaxies, while their slope becomes positive as galaxy mass increases. However, the ionisation parameter maps we obtain are clumpy, especially for low-mass galaxies. The ionisation parameter is tightly correlated with the equivalent width of \ha\, [$EW$(\ha)], following a nearly universal relation, which we attribute to the change of the spectral shape of ionising sources due to ageing of H\textsc{ii} regions. We derive a positive correlation between ionisation parameter and metallicity at fixed $EW$(\ha), in disagreement with previous theoretical work expecting an anti-correlation. 
}

\titlerunning{SDSS IV MaNGA - metallicity and ionisation parameter}
\authorrunning{M. Mingozzi et al.}
\maketitle
%

\section{Introduction}\label{sec:intro}
The metal content in galaxies is mainly governed by the interplay of inflows of gas and metals from the circum- and intergalactic medium,  metal production via stellar nucleosynthesis and outflows via galactic winds. Because of these connections, the abundance and distribution of metals in the interstellar medium (ISM) of local galaxies provides strong constraints on current models of galaxy evolution (e.g. \citealt{lilly2013, maiolino2019}). 

Emission lines, originating from nebulae photoionised by massive stars, represent one of the most powerful tracers of the metal abundance in both nearby galaxies and the high-redshift Universe.
Gas-phase metallicity is commonly traced by the abundance of oxygen, because of its high relative abundance and the availability of lines of the dominant ionic species in the optical wavelength range. Moreover, the production of oxygen is less complex compared to other elements such as nitrogen and carbon. Therefore, in this paper we use metallicity and oxygen abundance (12+log(O/H)) interchangeably, as commonly done in studies of chemical abundances based on \hii\, regions.
In particular, measurement of the \oiii$\lambda$4363 auroral line together with the \oiii$\lambda\lambda$4959,5007 nebular lines allows the determination of the electron temperature of the \hii\, region. When combined with other temperature-sensitive lines the ionisation structure of the nebula and metal content can be derived with a set of minimal assumptions (e.g. \citealt{stasinska2004}). 
Auroral lines, however, are often too weak to be detected in extragalactic sources at even modest redshifts and in large galaxy surveys (e.g. \citealt{stasinska2005}), and thus this method is still mostly limited to a low number of galaxies and often biased towards luminous and metal-poor regions (e.g. \citealt{andrews2013, jones2015, sanders2016, yates2019}). 

In light of this, \citet{pagel1979} and \citet{alloin1979} suggested methods to estimate the metallicity of \hii\, regions using strong emission lines (SEL) only.
Traditionally, this approach relies on the choice of a metallicity \textit{diagnostic} (i.e. one or several line ratios) combined with an abundance calibration prescription.
Among the most widely used metallicity diagnostics there are $\rm R23=$ (\oiii$\lambda\lambda$4959,5007 + \oii$\lambda\lambda$3726,29)/\hb \, \citep{pagel1979,pilyugin2005}, O3N2 = (\oiii$\lambda\lambda$4959,5007/\hb)/(\nii$\lambda\lambda$6548,84/\ha) \citep{pettini2004,marino2013,curti2017} and $ \rm N2= $ \nii$\lambda\lambda$6548,84/\ha \, \citep{storchi-bergmann1994, denicolo2002}, together with other combinations of multiple line ratios (e.g. \citealt{pilyugin2011,pilyugin2012}).
Each diagnostic has advantages and drawbacks, making it suitable in different situations. Nevertheless, different metallicity calibrations, even when based on the same diagnostic, are generally not mutually consistent leading to e.g. offsets of 0.2-0.6~dex in the absolute abundance scale \citep{kewley2008, blanc2015}. The origins of this discrepancy are still largely debated in the literature.

An important limitation of the ``strong-line approach'' to determining metallicity is the secondary dependence of metallicity diagnostics on other parameters of the ISM, such as the ionisation parameter (q), as well as relative chemical abundances and the shape of the ionising continuum.
The ionisation parameter is defined as the ratio of the number flux of ionising photons to the density of hydrogen atoms. In a spherical geometry it can be defined as
\begin{equation}\label{eq:ionu}
    q=\frac{Q_{ion}}{4 \pi r^2 n_{\rm e}},
\end{equation} 
where $Q_{ion}$ is the number of ionising photons emitted per unit time, $r$ is the distance between the source and the emitting cloud, and $n_e$ is the electron density. 

In this work we move past the traditional SEL diagnostics of metallicity and instead make use of a Bayesian approach, based on \citet{blanc2015}, and inspired by the previous work of \citet{charlot2001}, \citet{brinchmann2004} and \citet{tremonti2004}. 
In this framework, observed SEL fluxes are directly compared with a grid of photoionisation models. 
This approach has the advantage of modelling all the available SEL ratios self-consistently, but is also subject to specific limitations. 
First, the current generation of photoionisation models are not capable of correctly reproducing all the observable SEL ratios \citep{dagostino2019}. 
Second, photoionisation models are based on a large number of simplifying assumptions and it is not guaranteed that the obtained solution is unique and that the real uncertainties are correctly estimated.
Moreover, any physical correlation between metallicity and ionisation parameter, which is hard-wired into empirical models, cannot be easily ported to Bayesian methods. 
When using a Bayesian approach it is therefore necessary to use a wide enough set of SELs to break the well-known degeneracy between metallicity and q.

Empirically the ionisation parameter is best determined by ratios of emission lines of different ionisation stages of the same element, with $\rm O3O2 = $ \oiii$\lambda\lambda$4959,5007/\oii$\lambda\lambda$3726,29 being the most widely used proxy in the optical wavelength range \citep{diaz2000,kewley2002,kobulnicky2004}. 
O3O2, however, is strongly dependent on metallicity, as shown for example in \citet{kewley2002}, since the \oii\, lines can only be excited by relatively high temperatures, and thus disappear in cooler, high-metallicity \hii\, regions.
\citet{kewley2002} demonstrated that the ratio between S$^{++}$ and S$^+$, traced by \siii$\lambda\lambda9069,9532$/\sii$\lambda6717,32$ (S3S2), provides a more accurate measure of the ionisation parameter, as already suggested by \citet{mathis1982,mathis1985,dopita1986}.
Given the redder wavelengths of the sulphur emission lines, and consequently their lower excitation energies with respect to the oxygen lines, S3S2 is roughly independent of metallicity (e.g. variation of 0.3~dex in the range of $\rm 7.63 < log(O/H) + 12 < 8.93$, \citealt{kewley2019}). Furthermore, \citet{kewley2019} argued that S3S2 is also insensitive to ISM pressure P for $ 4<{\rm log(P}/k{\rm)}<7$.

Until recently, however, the near-infrared \siii$\lambda\lambda9069,9532$ lines (from now on referred to as the \siii\, lines), have only rarely been used to investigate the ionisation parameter (see work by \citealt{diaz2000,kehrig2006}) because they are weaker than their oxygen counterparts and no large spectroscopic survey of nearby galaxies covered the required wavelength range. 
This situation is now changing, with more facilities covering the full wavelength range between 6000 and 10000 \AA, including the X-shooter and MUSE instruments on the VLT \citep{cresci2017,mingozzi2019, kreckel2019}, MODS on LBT, or the SDSS-IV MaNGA survey and the first detection of the \siii\, lines at high redshift with near-IR instruments (e.g. \citealt{curti2019b, sanders2019}).

In this paper, we make use of data from the MaNGA survey \citep{bundy2015, yan2016}, part of the last generation Sloan Digital Sky Survey IV (SDSS IV, July 2014--2020), to study metallicity and ionisation parameter of local galaxies in a Bayesian framework.  Specifically, we present the first study of the \siii\, lines, and therefore a detailed characterisation of the ionisation parameter, in a large sample of local galaxies.
Based on our improved measurements of the ionisation parameter, we aim to characterise how $q$ varies within galaxies, across different stellar masses, and investigate what relation (if any) exists between ionisation parameter and other key physical parameters of the H\textsc{ii} regions, such as metallicity, star formation rate surface density and age. 

The paper is structured as follows. In Sec.~\ref{sec:data_and_sample} we introduce the galaxy sample and the data analysis, while in Sec.~\ref{sec:method} we discuss the Bayesian method we employ to derive metallicity, ionisation parameter and gas extinction, highlighting the fundamental role played by \siii\, lines. In Sec.~\ref{sec:results} we present and discuss our results, while in Sec.~\ref{sec:conclusion} we conclude, highlighting benefits and drawbacks of our approach.

Throughout this work, we use a $\Lambda$ cold dark matter cosmology with $\Omega_{\rm M} = 0.3$, $\Omega_{\Lambda} = 0.7$ and $H_0 = 70$~km$^{-1}$~s$^{-1}$~Mpc$^{-1}$.

\section{Data and sample}
\label{sec:data_and_sample}

\subsection{The galaxy sample}
\label{sec:galaxy_sample}

The MaNGA survey is one of the three components of SDSS-IV \citep{blanton2017}. Observations are carried out at the 2.5~m telescope at Apache Point Observatory \citep{gunn2006}. The survey aims to map 10000 galaxies in the redshift range $0.01 < z < 0.15$ by 2020. MaNGA galaxies are selected to be representative of the overall galaxy population for $\rm log(M_{\star}/M_\odot)>9$.
The sample is drawn from an extended version of the NASA-Sloan catalogue (NSA \texttt{v1\_0\_1}\footnote{\url{https://www.sdss.org/dr15/manga/manga-target-selection/nsa/}}, \citealt{blanton2011}). 
MaNGA observations are carried out with 17 hexagonal fiber-bundle integral field units (IFUs) that vary in diameter from 12" (19 fibers) to 32" (127 fibers). Each fiber has a diameter of 2" \citep{drory2015}. The IFUs feed light into the two dual-channel BOSS spectrographs, that provide simultaneous wavelength coverage in the 3600--10300~\AA\ wavelength range, with a resulting spectral resolution of $R\sim1400$ at $\lambda\sim4000$~\AA \, and $R\sim2600$ near $\lambda\sim9000$~\AA \,($R\sim2000$ corresponds to a velocity dispersion of $\sigma \sim70$~km/s, \citealt{smee2013}).
A uniform radial coverage to radii of 1.5~\re\, and 2.5~\re\, is achieved for 2/3 (primary sample) and 1/3 (secondary sample) of the final sample, respectively. In order to compensate for light loss during observations, a three-point dithering pattern is used, allowing also to obtain a uniform point spread function (PSF, \citealt{Law2015}).

The data analysed in this paper are part of the fifteenth SDSS Data Release (DR15, \citealt{aguado2019}), reduced according to the algorithms described in \citet{Law2016} and \citet{yan2016} and subsequent updates. The data release includes the output of the MaNGA data analysis pipeline (DAP, \citealt{westfall2019, belfiore2019}) for a sample of 4688 spatially-resolved galaxies observed until July 2017.
Integrated galaxy global properties such as redshift, total stellar mass ($M_*$), elliptical effective radius (\re) and inclinations are drawn from the extended version of NSA v1\_0\_1, i.e. the parent targeting catalog described in Section 2 of \citet{wake2017}, which includes the calculation of elliptical Petrosian aperture photometry. Elliptical Petrosian effective radii and inclinations are used throughout this work to construct de-projected radial gradients.

From this initial sample of galaxies, we selected 4099 objects with $z<0.08$, in order to have access to the \siii $\lambda$9532 line within the observed wavelength range. 
We select star-forming galaxies according to the `excitation morphology', as described in \citet{belfiore2016}. 
This scheme makes use of the classical Baldwin-Phillips-Terlevich (BPT) diagnostic diagram \citep{baldwin1981, veilleux1987, kauffmann2003b, kewley2006} to map the excitation source of the ISM throughout galaxies. In particular, we exclude galaxies that show central or extended low-ionisation emission-line regions (cLIERs and eLIERs), line-less galaxies (no line emission detected), and galaxies with Seyfert-like central regions. We retain only objects classified as star-forming, defined as objects dominated by star formation (SF) at all radii. This selection cut limits our sample to mostly main-sequence galaxies, while eliminating the majority of green valley objects \citep{belfiore2018} and all passive galaxies.
Finally, we exclude highly inclined systems, namely galaxies with minor to major axis ratio (b/a) less than than 0.4, leading to a final sample of 1795 galaxies. 

In the rest of this work we further exclude spaxels which are not classified as star-forming according to either the \nii- or the \sii-BPT diagrams (i.e. all the spaxels above the demarcation lines defined by \citealt{kauffmann2003b} and by \citealt{kewley2001}, respectively), for which we assume that the \ha \, flux is contaminated by other physical processes aside from star formation. 

\subsection{Spectral fitting}\label{sec:spectralfitting}
We make use of the flux measurements obtained by Gaussian fitting for the \oii$\lambda\lambda$3726,29, \hb, \oiii$\lambda\lambda$4959,5007, \nii$\lambda\lambda$6548,84, \ha, \sii$\lambda$6717 and \sii$\lambda$6731 emission lines in stellar-continuum subtracted spectra from the MaNGA DAP v2.2.1. 
In the fit performed by the DAP the flux ratios of doublets are fixed when such ratios are determined by atomic physics, with transition probabilities taken from \citet{osterbrock2006} (e.g. intrinsic ratios of 0.340 and 0.327 for \oiii\, and \nii\, doublets, respectively). 
Furthermore, in DR15 the velocities of all the fitted emission lines are tied, while the velocity dispersions are free to vary, with the exception of the doublets with fixed flux ratios and the \oii$\lambda\lambda$3726,29 lines. 
\citet{belfiore2019} (see Sec.~2.2.3) found that the impact of these algorithmic choices on the line fluxes is negligible.

In DR15, the stellar continuum fitting is limited by the spectral range of the adopted MILES stellar library \citep{sanchez-blazquez2006,falcon-barroso2011}. 
Because of the absence of a continuum model redder than $\sim 7400$~\AA\, the DAP does not fit the \siii$\lambda\lambda$9069,9532 lines. We therefore performed a bespoke fitting process.

Since the red part of the spectrum around the lines of interest does not feature prominent stellar absorption features, we use a linear baseline to account for the local continuum $\pm30$~\AA\ around the \siii\, emission lines. Each line is then fitted with a single Gaussian, making use of MPFIT \citep{markwardt2009} and masking the remaining part of the spectrum. We fixed the flux ratio of the two \siii\, lines to the theoretical ratio of 2.47 determined from the transition probabilities used in \textsc{pyneb} \citep{pyneb} and their velocities to the velocity obtained by the DAP for the previously fitted emission lines. 
Lower ($\sim40$~km/s) and upper ($\sim400$~km/s) limits are imposed on the velocity dispersion in order to reduce the probability of fitting sky lines, or noise features in the spectra. All line fluxes are corrected for Galactic foreground extinction using the maps of \citet{schlegel1998} and the reddening law of \citet{odonnell1994}. 

\begin{figure*}
    \begin{minipage}{0.6\columnwidth}
                \includegraphics[width=1\columnwidth]{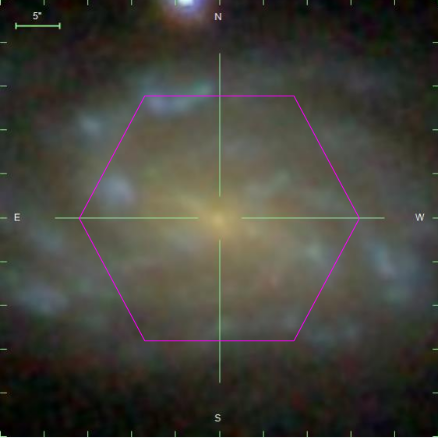}
                \vspace{+3mm}
    \end{minipage} 
    \begin{minipage}{1.4\columnwidth}
                \includegraphics[width=1\columnwidth]{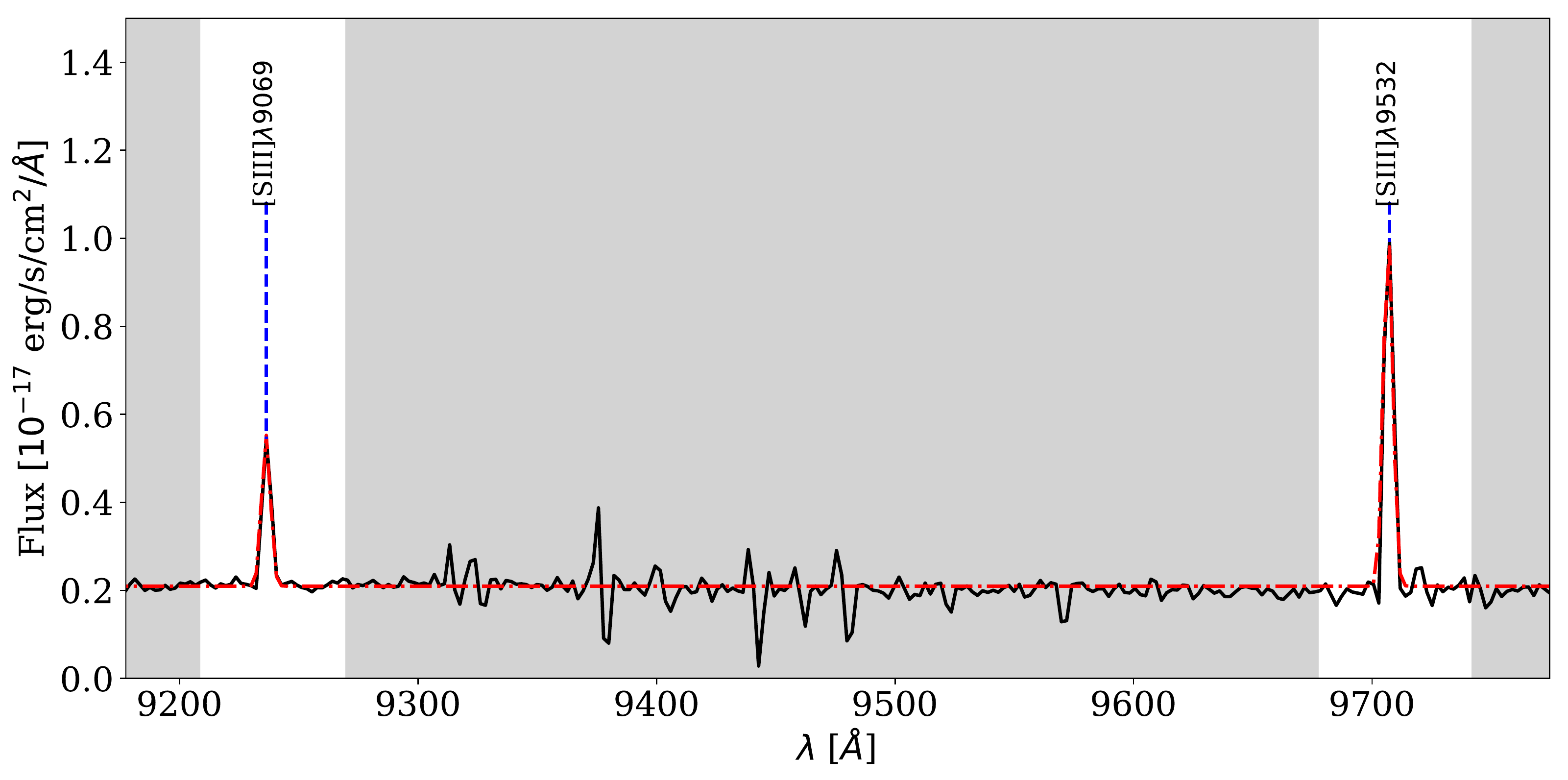}
    \end{minipage} 
    
    \begin{minipage}{2\columnwidth}
                \includegraphics[width=1.\columnwidth]{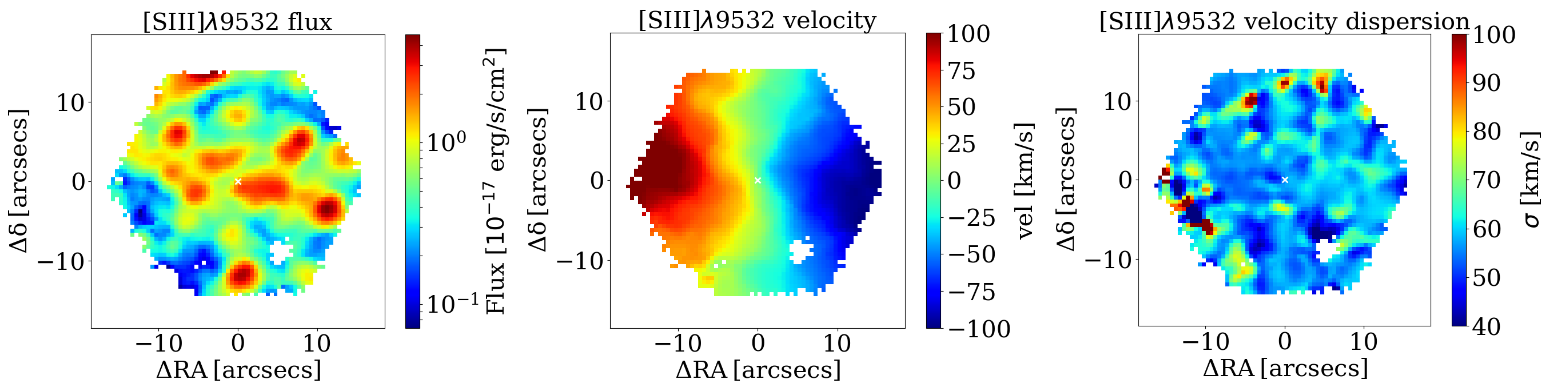}
    \end{minipage} 
    \caption{Top left: $g-r-i$ image composite from SDSS for MaNGA galaxy 8150-6103, with the MaNGA IFU hexagonal FoV overlaid. East is to the left. Top right: An example of our fitting procedure for the \siii\, lines for one representative spaxel in this example galaxy. The black solid line is the observed spectrum, while the dashed-dotted red line is the best-fit. The regions in grey were masked before fitting. 
    Bottom panels: \siii$\lambda$9532 flux, velocity and velocity dispersion maps of MaNGA galaxy 8150-6103.}
        	\label{fig:fit_siii}
   \end{figure*}

The top panel of Fig.~\ref{fig:fit_siii} shows the g-r-i image composite from SDSS of MaNGA galaxy 8150-6103 with the MaNGA hexagonal field of view (FoV) overlaid (on the left), and an example of our fitting procedure for the \siii\, lines for one spaxel (on the right). The observed spectrum is reported in black, while the best-fit is highlighted by the dashed-dotted red line. The regions in grey were masked before fitting. 
The bottom panel of Fig.~\ref{fig:fit_siii} presents maps of the \siii$\lambda$9532 flux, velocity and velocity dispersion for the same galaxy.

In Appendix \ref{app:pipe3d}, we present a comparison between the emission line fluxes considered in this work and those computed by the Pipe3D code \citep{sanchez2016, sanchez2018}, which performs independent continuum subtraction and emission-line measurements of the MaNGA data. There is a good overall agreement between the two (see also the Appendix of \citealt{belfiore2019}), up to the \siii$\lambda9069$ emission line. However, we find a larger discrepancy with respect to the other emission lines for the \siii$\lambda9532$, likely associated with a failure in the extrapolation of the continuum performed by Pipe3D (S. F. S\'anchez, private communication).

\subsection{Signal-to-noise selection}\label{sec:snr}
In this work, we define the signal-to-noise of each emission line $S/N_{\mathrm{line}} = \mathrm{Flux/Err_{flux}}$, as in \citet{belfiore2019}. This quantity is equivalent to the classical amplitude-over-noise ratio often used for line detection.
In order to obtain reliable measurements for the ISM parameters of interest we exclude from further analysis spaxel with $S/N$(\ha)~$ < 15$. We follow this approach, based on the work of \citet{mannucci2010}, because a high $S/N$(\ha) ensures that the main optical lines are generally detected with $S/N>1.5$ without introducing metallicity biases. According to \citet{belfiore2019}, the fluxes computed by the DAP with $S/N \leq 1.5$ are unreliable. We therefore consider them as upper limits.
This cut on $S/N$(\ha) selects 85\% of the star-forming spaxels of the galaxy sample. 
In Appendix \ref{sec:snr_app}, we show the $S/N$ radial profiles for MaNGA galaxies as a function of stellar mass. Overall, all strong lines of interest have $S/N > 5$ at $R  < 2$~\re, except the \siii\, lines, which are particularly weak ($S/N \sim 2-4$) at large radii and for massive galaxies.

Aside from emission from \hii\ regions, galaxies are observed to contain low surface brightness diffuse ionised gas (DIG). The origin of this ionised gas emission is likely to be a combination of different processes, such as radiation leaking from classical \hii\, regions, massive stars in the field and radiation from hot evolved stars \citep{hoopes2003,oey2007,zhang2017,sanders2017,wylezalek2018}. In this work we only consider \hii\, region models. It is therefore necessary to minimize the contribution of DIG to our observed line fluxes.

Previous works have demonstrated the existence of a tight relation between diagnostic line ratios and \ha\, equivalent width ($EW$) in emission (e.g., \citealt{sanchez2014, belfiore2016, zhang2017, lacerda2018,vale-asari2019}). Specifically, \citet{lacerda2018} argue that $EW$(\ha)~$<3$~\AA \, traces emission from hot low-mass evolved stars \citep{stasinska2008}, while gas with $EW$(\ha)~$=3-14$~\AA\, may include a mixture of DIG and emission from star forming regions. Because of our selection in BPT classification and signal-to-noise described above, virtually all the spaxels taken into account in this work have $EW$(\ha)~$>3$ \AA, and are therefore dominated by flux from H\textsc{ii} regions.

\section{Methods}
\label{sec:method}
\subsection{Bayesian approach to determining ISM properties}
\label{sec:izi}
Our method to derive the oxygen abundance and the ionisation parameter of ionised nebulae is based on the software tool IZI presented by \citet{blanc2015}. In brief, IZI compares an arbitrary set of observed SELs (and their associated errors) with a grid of photoionisation models, calculating the joint and marginalised posterior probability density functions (PDFs) for 12 + log(O/H) and log($q$). This method allows us to include flux upper limits (see \citealt{blanc2015} for details), and it provides a self-consistent way of inferring the physical conditions of ionised nebulae using all available information. In particular, this approach makes it straightforward to test the effect of either a particular choice of SELs (as commonly done in traditional metallicity diagnostics) or a particular set of photoionisation models.

For this work we have re-written the original IDL IZI code in Python and added several modifications to the original code. The main innovation of our revisited version of IZI is the simultaneous estimate of a third parameter, the gas extinction $E(B-V)$. 
The gas extinction along the line-of-sight towards star-forming regions can be directly probed with Balmer recombination line flux ratios, comparing the observed and intrinsic Balmer decrements (e.g. \citealt{calzetti1997}). $E$(B-V) is usually derived assuming a fixed un-attenuated case B recombination, and generally an intrinsic Balmer decrement of \ha/\hb~$ = 2.86$, appropriate for an electron density $n_e = 100$~cm$^{-3}$ and electron temperature $T_e = 10^4 $~K \citep{osterbrock1989}.
The intrinsic ratio depends weakly on density, with \ha/\hb~$ = 2.86-2.81$ over four orders of
magnitude in electron density ($n_e=10^2-10^6$~cm$^{-3}$, \citealt{osterbrock1989}). However, the dependence on temperature is relatively larger, leading to \ha/\hb\, intrinsic ratio value between
$3.04-2.75$ for $5000-20000$~K \citep{dopita2003}.
Consequently, adopting a single intrinsic Balmer ratio neglects the direct effect of metallicity and ionisation parameter on the temperature of the nebula.
For instance, \citet{brinchmann2004} pointed out that ignoring the metallicity-dependence of the case B \ha/\hb\, ratio leads to an overestimate of the dust attenuation by up to $\sim0.5$~mag for the most metal rich galaxies (see e.g. Fig.~6 \citealt{brinchmann2004}).
Therefore, in this work we estimate dust extinction together with metallicity and ionisation parameter, assuming a foreground screen attenuation, a Calzetti attenuation law with $R_V=4.05$ (see \citealt{calzetti2000}) and adopting the intrinsic Balmer decrements self-consistently calculated within the photoionisation model grids. Another big advantage of inferring gas extinction, metallicity and ionisation parameter simultaneously is that we are self-consistently taking into account the covariance in dust-corrected line fluxes after applying a dust correction.

We make use of the Markov chain Monte Carlo (MCMC) Python package \textsf{emcee} to evaluate the PDF of the three parameters \citep{foreman-mackey2013}.
Since photoionisation models cannot fully reproduce all observed line ratios, following \citet{blanc2015}, \citet{kewley2008} and \citet{dopita2013} we adopt a systematic fractional error in the flux predicted by the photoionisation model of 0.1~dex, except for the \ha\, line, for which we took into account 0.01~dex, in order to weight it more when constraining $E$(B-V). This intrinsic uncertainty is the dominant source of the error term in IZI, being generally larger than the measurement error associated to observed fluxes. Consequently, signal-to-noise levels of the strong lines have a minor effect on the IZI output.
Despite the differences from the publicly-available IDL code, for the rest of this work we refer to our rewritten and modified code as IZI unless otherwise noted.

As discussed by \citet{blanc2015}, it is not trivial to define a `best-fit' solution for a specific parameter if its PDF is not Gaussian (e.g. strong asymmetries and/or multiple peaks).
In these cases the mean is not a reliable estimator because it can be significantly different from the most likely solution, given by the mode.
We choose as `best-fit' value the marginalized median (i.e. $50^{\mathrm{th}}$ percentile) of the PDF for each parameter.
The upper and lower errors of the best-fit solution are obtained by computing the $16^{\mathrm{th}}$ and $84^{\mathrm{th}}$ percentiles of the distribution, respectively, and are defined as $\Delta_{up}=84^{\mathrm{th}}-50^{\mathrm{th}}$ and $\Delta_{down}=50^{\mathrm{th}}-16^{\mathrm{th}}$. If the PDF is Gaussian, these two values are equal and correspond to the standard deviation.

As our default photoionisation models set, we use the photoionisation model grids presented in \citet{dopita2013} (D13 models hereafter).
These models are computed with the \textsc{mappings}-IV code, which with respect to the previous version includes new atomic data, an increased number of ionic species treated as full non-local-thermodynamic-equilibrium multilevel atoms, and the ability to use electron energy distributions that differ from a simple Maxwell-Boltzmann distribution \citep{nicholls2013,dopita2013}. The grids that we use here are, however, calculated assuming a Maxwell-Boltzmann distribution, in light of the ongoing controversy on whether the use of alternative distributions is justified for \hii\ regions \citet{draine2018}.

In D13 models, the input ionising spectrum is computed through the population synthesis code
STARBURST99 \citep{leitherer1999}, taking into account the Lejeune/Schmutz extended stellar atmosphere models \citep{schmutz1992,lejeune1997}, a constant star formation history (SFH), a Salpeter initial mass function (IMF, lower and upper mass cut-off at 0.1~M$_{\odot}$ and 120~M$_{\odot}$, respectively), and an age of 4~Myr. D13 models are isobaric with an electron density $n_e \sim 10$~cm$^{-3}$ and are computed for a spherical geometry. The assumed gas phase abundances are taken from \citet{asplund2009}, except for nitrogen and carbon. For these elements empirical fitting functions as a function of O/H are adopted (see Table~3 in \citealt{dopita2013}).
Furthermore, D13 models include dust physics and radiative transfer, assuming a population of silicate grains following \citet{mathis1977} size distribution and a population of small carbonaceous grains within the ionised regions, as described in detail in \citet{dopita2005}.
The metallicities vary in the range 12+log(O/H)~$= [7.39-9.39]$ (the solar oxygen abundance is 8.69), while the ionisation parameter varies in the range log($q$)~$ = [6.5-8.5]$. Intrinsic Balmer decrement values vary in the range ~$ = [2.85-3.48]$, and are systematically higher than the standard Case B recombination value, as reported by \citet{dopita2013}.

\begin{figure}
    \begin{minipage}{1\columnwidth}
                \includegraphics[width=1\columnwidth]{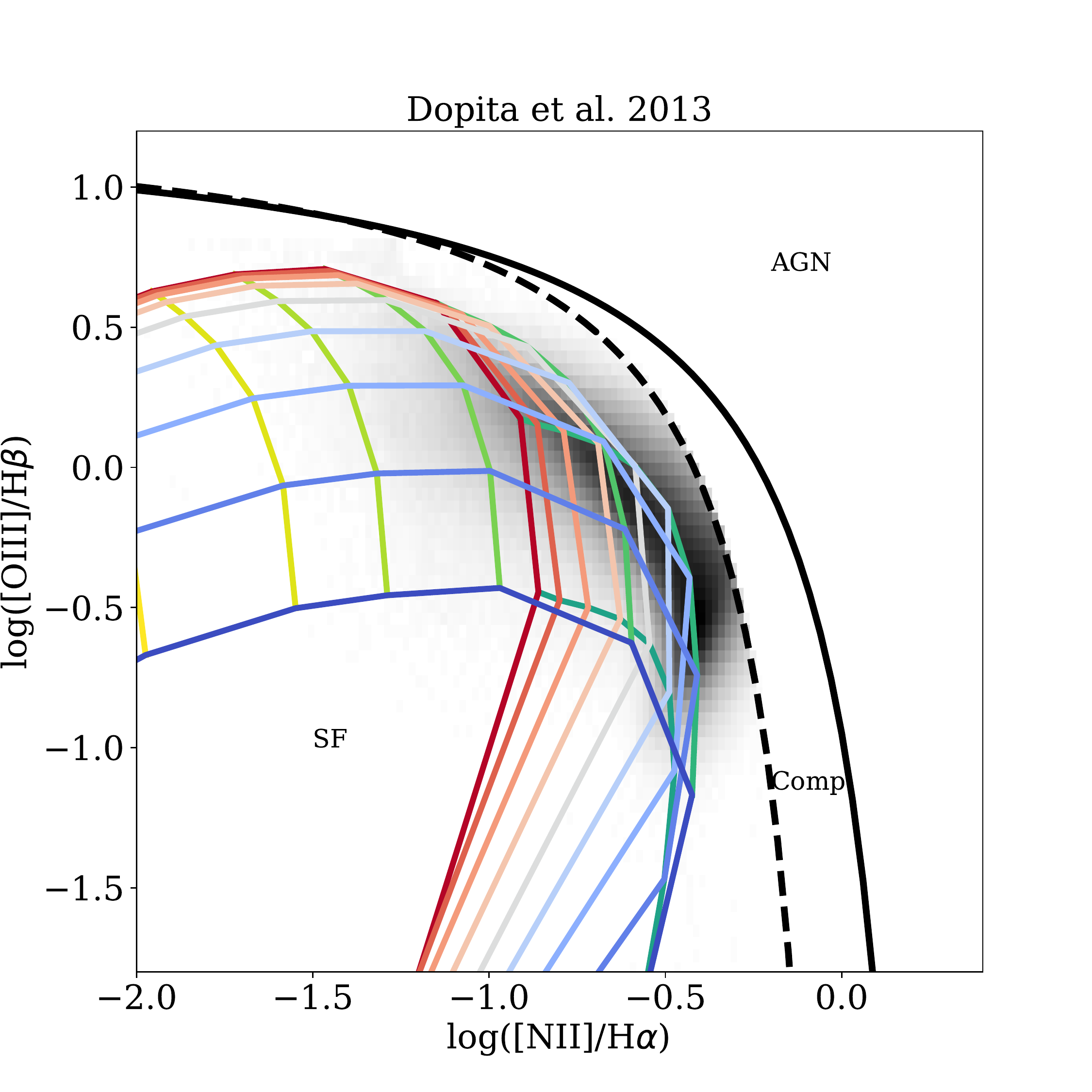}
    \end{minipage} 
    \begin{minipage}{1\columnwidth}
                \includegraphics[width=1\columnwidth]{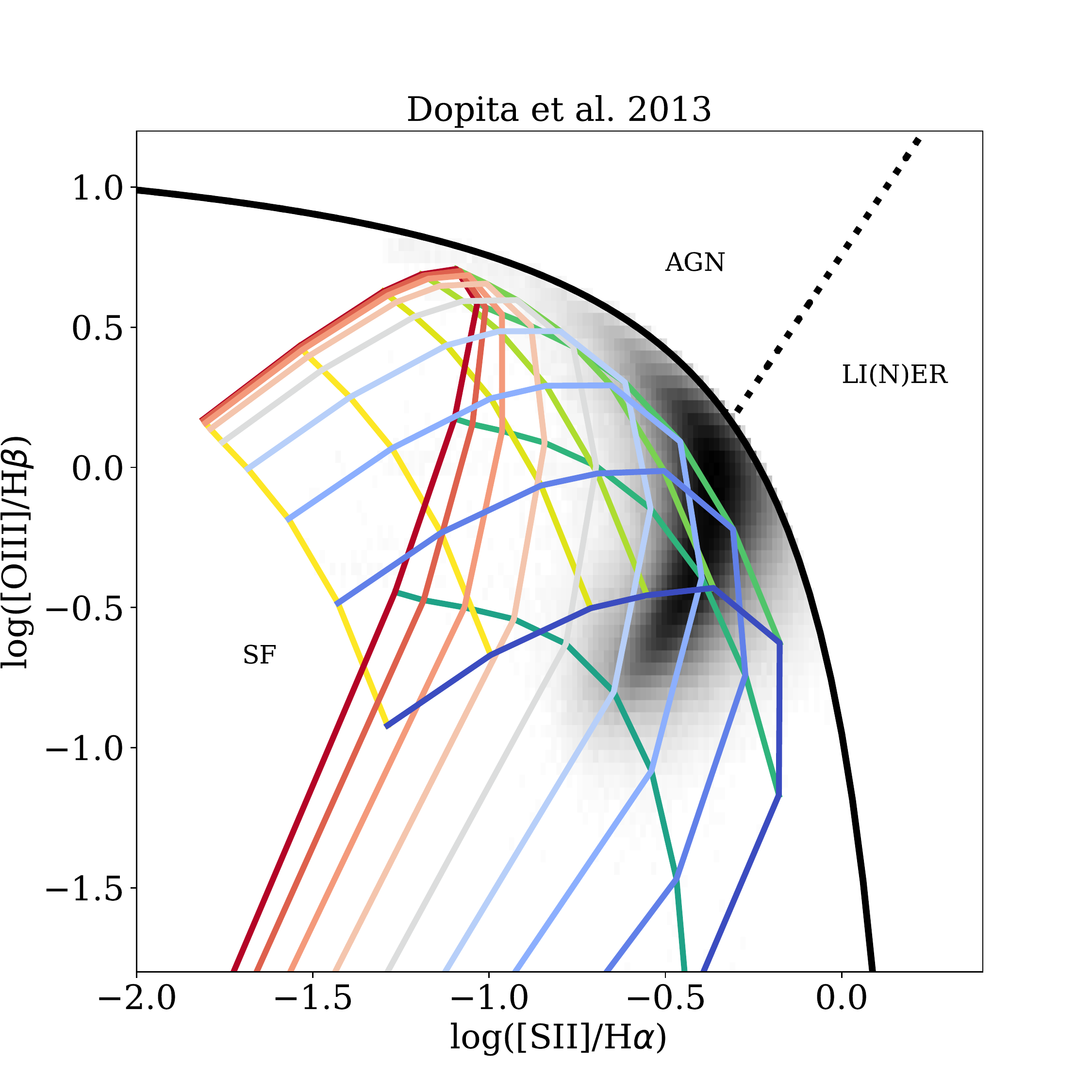}
    \end{minipage} 
        \caption{\nii- and \sii-BPT diagrams for our sample of MaNGA spaxels compared with D13 models, for metallicities in the range [7.39--9.39] (increasing from yellow to green) and ionisation parameters in the range [6.5--8.5] (redder means higher log($q$) and bluer lower log($q$)). D13 models generally well-reproduce the observed line ratios in MaNGA spaxels as ensembles.}
        \label{fig:obsdiagnostics}
  \end{figure}

In Fig.~\ref{fig:obsdiagnostics} we show as an example the comparison between the \nii- and \sii-BPT diagrams for our sample of MaNGA star-forming spaxels with the grid of D13 models described above.
The dashed curve represents the demarcation line defined by \citet{kauffmann2003b} (Ka03), while the solid curve is the theoretical upper limit on SF line ratios proposed by \citet{kewley2001} (Ke01). The dotted line, instead, is the boundary between Seyferts and low-ionisation (nuclear) emission-line regions (LI(N)ERs) introduced by \citet{kewley2006}. 
We noticed that \sii\, lines are predicted to be weaker by $\sim0.1$~dex (see Fig.~\ref{fig:obsdiagnostics}) than observed, as already reported in \citet{dopita2013} (a lower discrepancy is observed also between observed and modelled \nii/\ha\, line ratios). 
This is due to the fact that \sii\, ions are much more sensitive than \nii\, ions to the diffuse radiation field in \hii\, regions \citep{dopita2006b,kewley2019}.
This issue is also discussed in \citet{levesque2010}, where the authors suggest that their models probably do not produce sufficient flux in the far-ultraviolet ionising spectrum. They argue that \sii\, requires a larger partially ionised zone generated by a harder radiation field than the one present in the models.
  
Considering the above caveats, the optical line ratios in MaNGA spaxels are generally well-reproduced by the D13 models as ensembles. We have also reproduced all the diagnostics plots based on optical emission lines presented in \citet{dopita2013} and find the MaNGa data global distribution to be well-covered by the model grids.
As a caveat, we stress that this test does not demonstrate whether single galaxies are also well-reproduced by the models in all the diagnostic plots.

\subsection{The failure of photoionisation models in reproducing the observed \siii\, lines}
\label{sec:failuresiii}
\citet{stasinska2006} suggested that $\rm S3O3= $ \siii$\lambda\lambda9069,9532$/\oiii$\lambda\lambda4959,5007$ versus S3S2 represents a useful diagnostic plot, where the horizontal axis is mostly tracing metallicity while the vertical axis traces the ionisation parameter. Indeed, 
S3O3 is found to vary monotonically with metallicity, similarly to the more commonly employed O3N2 diagnostic \citep{dopita2013}. 

\begin{figure*}
    \begin{minipage}{1\columnwidth}
	\includegraphics[width=.85\columnwidth]{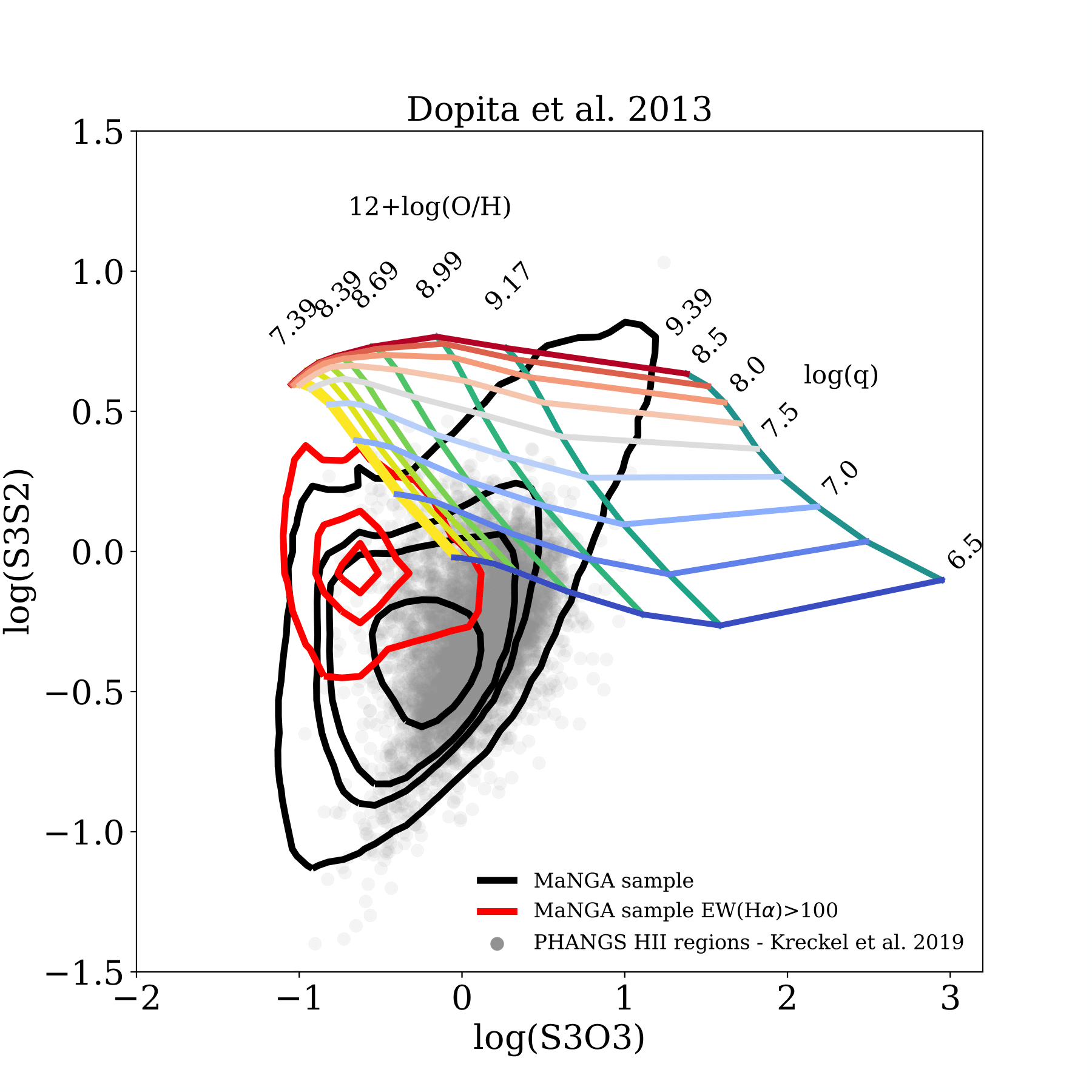}
    \end{minipage}
    \begin{minipage}{1\columnwidth}
	\includegraphics[width=.85\columnwidth]{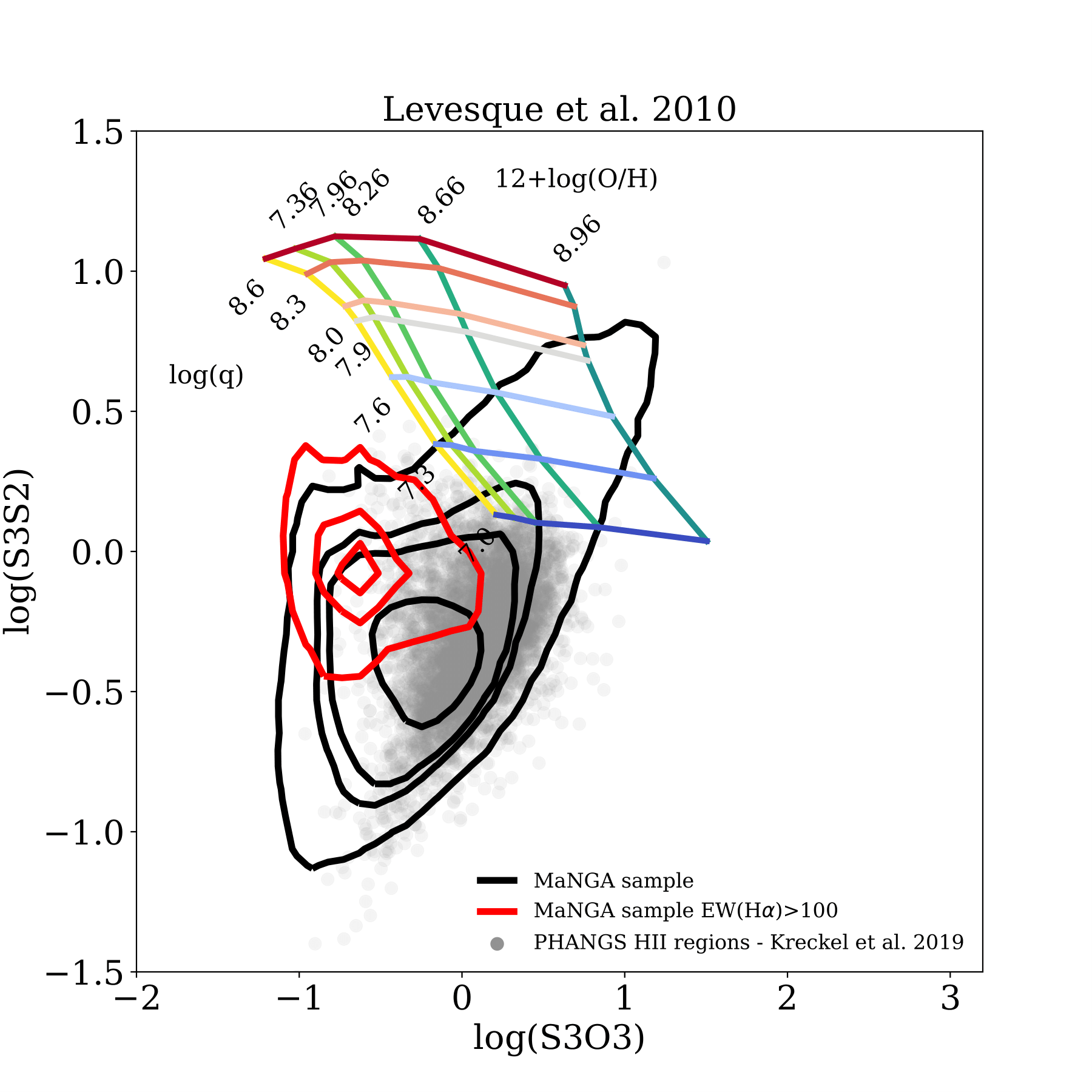}
    \end{minipage}
    \begin{minipage}{1\columnwidth}
	\includegraphics[width=.85\columnwidth]{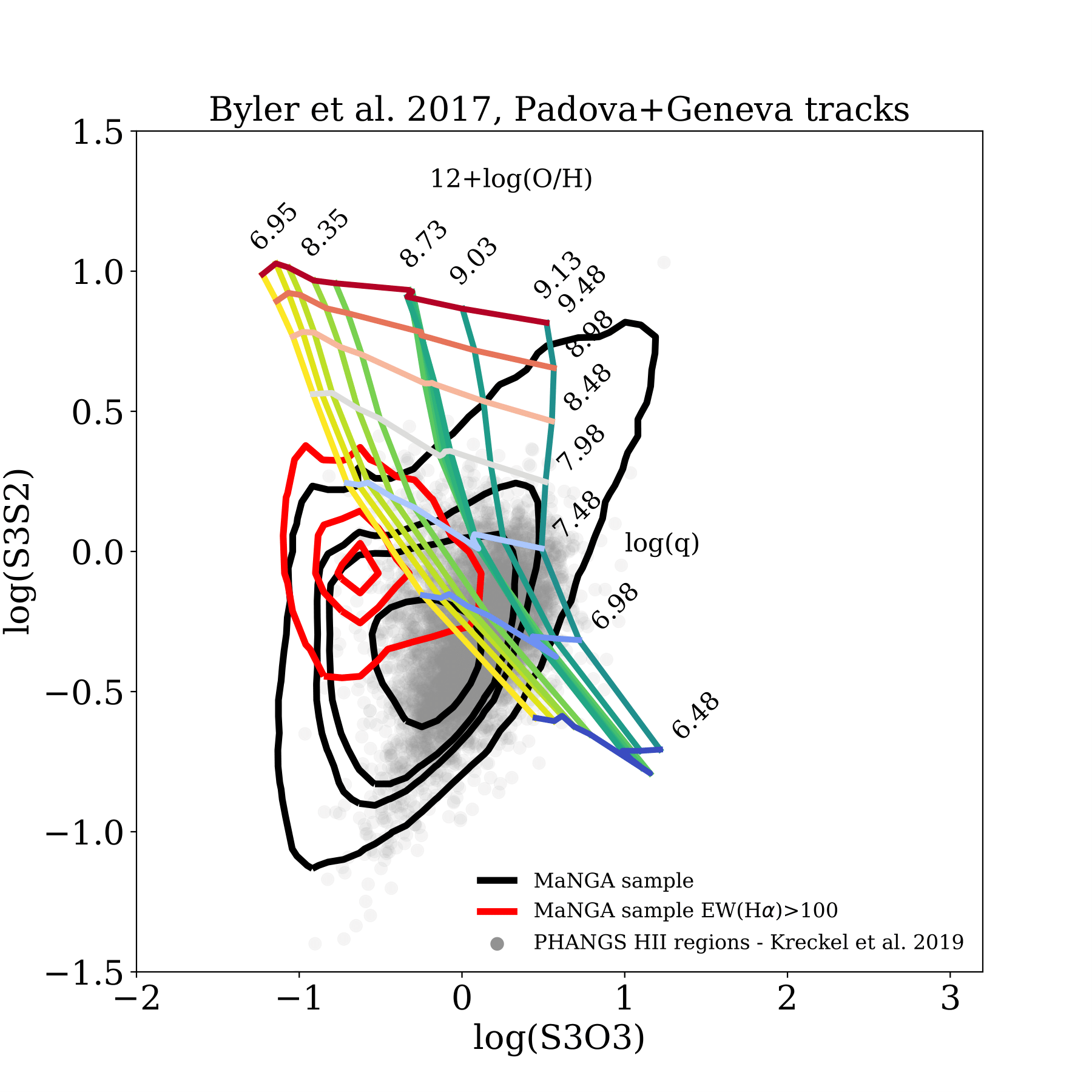}
    \end{minipage}
    \begin{minipage}{1\columnwidth}
	\includegraphics[width=.85\columnwidth]{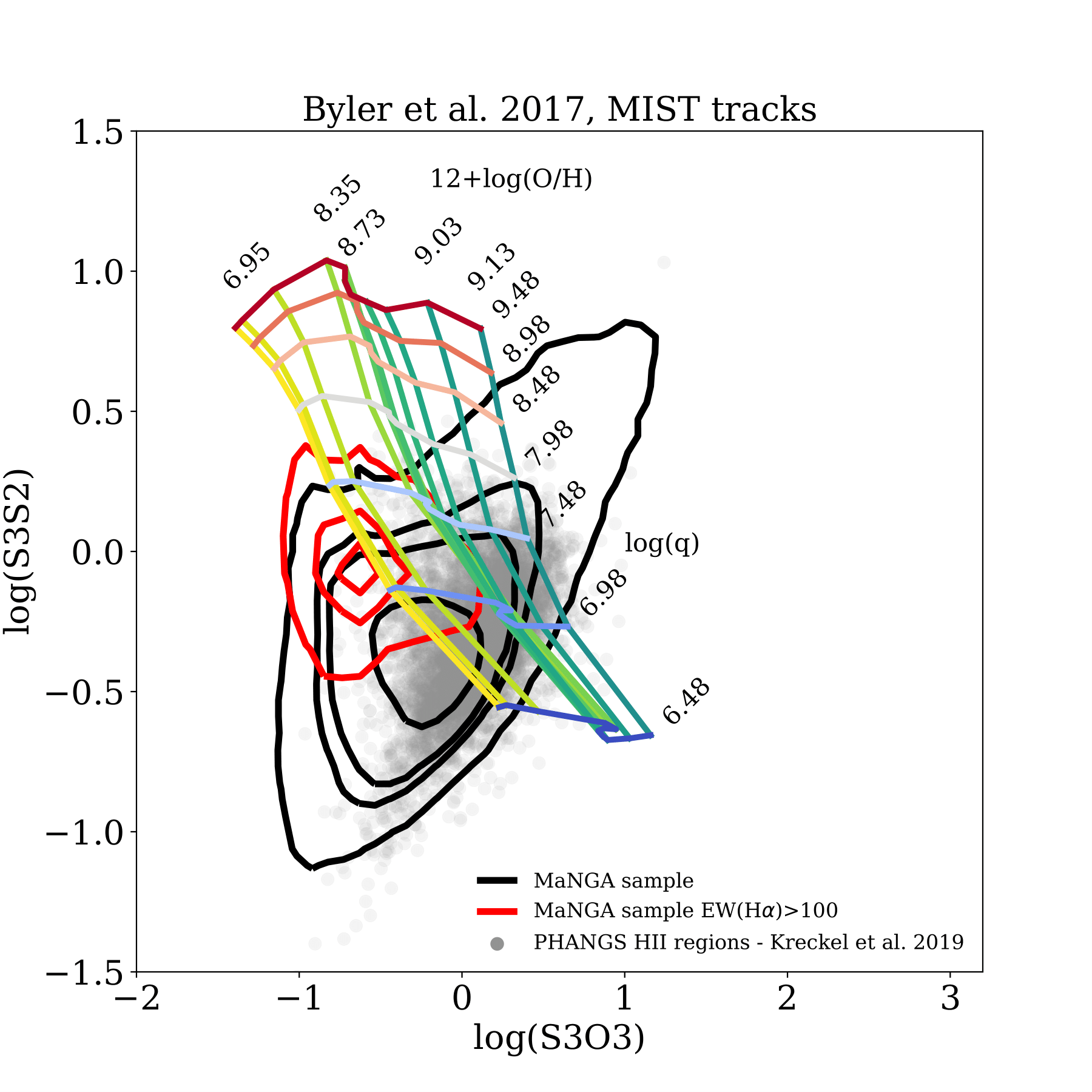}
    \end{minipage}
    
    \centering
    \begin{minipage}{1\columnwidth}
	\includegraphics[width=.85\columnwidth]{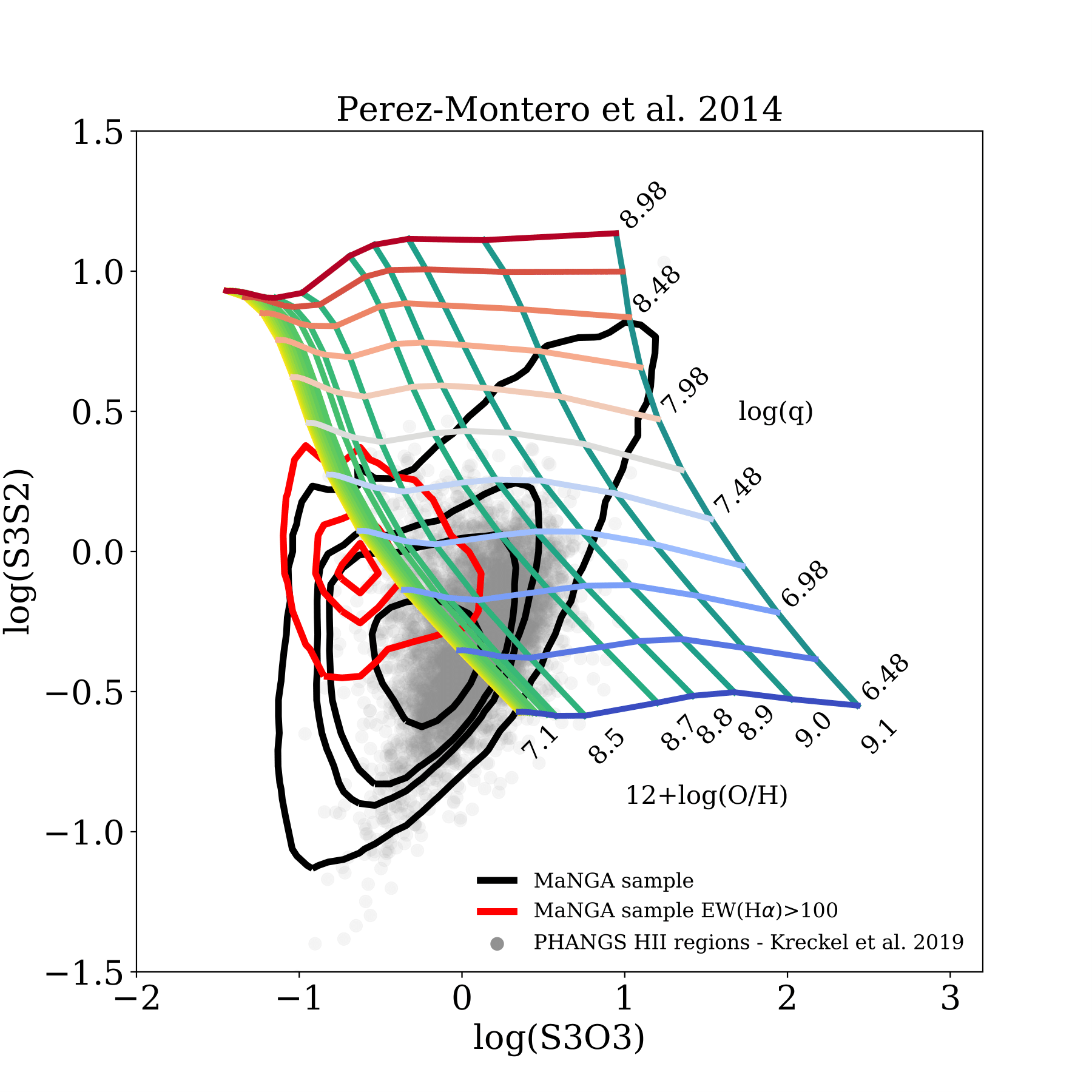}
    \end{minipage}
    \caption{\siii/\oiii\, versus \siii/\sii\, diagnostic diagrams introduced by \citet{stasinska2006} showing the density contours for all the star-forming spaxels of our sample (in black), with superimposed \citet{dopita2013}, \citet{levesque2010}, \citet{byler2017} pdva+geneva, \citet{byler2017} MIST, and \citet{perez-montero2014} models respectively. Metallicity increases horizontally from yellow to green, while ionisation parameter increases vertically [redder means higher log($q$) and bluer lower log($q$)]. The red contours correspond to the brighest MaNGA star-forming regions [$EW$(\ha)~$>100$], while the grey dots are the \hii\, regions analysed by \citet{kreckel2019} with the PHANGS survey. Current photoionisation models fail to reproduce the observed \siii\, lines.} 
    \label{fig:sulfurproblems}
\end{figure*}

Fig.~\ref{fig:sulfurproblems} shows S3O3 versus S3S2 of all the selected MaNGA spaxels (black contours), the brightest MaNGA star-forming regions [$EW$(\ha)~$>100$~\AA)] (red contours) and single \hii\, regions analysed in \citet{kreckel2019}\footnote{Private communication.} within eight galaxies from the Physics at High Angular resolution in Nearby GalaxieS\footnote{\url{https://www.phangs.org}} (PHANGS) survey (grey dots), in comparison with different sets of photoionisation models (described below).
The observed line ratios are corrected for reddening assuming a Calzetti attenuation law with $R_V=4.05$ (see \citealt{calzetti2000}) and adopting the intrinsic Balmer decrement of 2.86.
As shown in the top left panel of Fig.~\ref{fig:sulfurproblems}, the D13 models largely overestimate the \siii \, fluxes (black contours), being consistently shifted both horizontally and vertically with respect to our data.
In order to test whether this discrepancy is caused by any specific feature of the D13 models, we consider three other sets of photoionisation models, based on different ingredients and assumptions: the \citet{levesque2010} models (Fig.~\ref{fig:sulfurproblems}, upper right panel), two different versions of the \citet{byler2017} models (Fig.~\ref{fig:sulfurproblems}, lower left and right panels), and the \citet{perez-montero2014} models.

\citet{levesque2010} (L10) photoionisation models are computed with \textsc{mappings}-III, adopting a version of the STARBURST99 population synthesis code \citep{vazquez2005} that uses the Pauldrach/Hillier stellar atmosphere models \citep{pauldrach2001,hillier1998}, and evolutionary tracks produced by the Geneva group \citep{schaller1992}. Therefore, these models are characterised by a detailed non local thermal equilibrium modelling of metal line blanketing, which significantly affects the shape of the ionising spectrum, unlike the Lejeune/Schmutz models used in D13.

\citet{byler2017} (B17) photoionisation models, instead, are obtained with version 13.03 of \textsc{cloudy} \citep{ferland2013}, using the Flexible Stellar Population Synthesis code (FSPS; \citealt{conroy2009}), based on Padova+Geneva and MESA isochrones and stellar tracks (MIST; \citealt{choi2016, dotter2016}) evolutionary tracks, where the latter include stellar rotation. Stellar rotation affects stellar lifetimes, luminosities, and effective temperatures through rotational mixing and mass loss, implying harder ionising spectra and higher luminosities (see \citealt{choi2016,byler2017} for further details). B17 models are characterised by a spherical shell cloud geometry and a constant gas density of $n_{\rm e}=100$~cm$^{-3}$. The assumed gas phase abundances are taken from \citet{dopita2000}, which are based on the solar abundances from \citet{anders1989}.

Finally, \citet{perez-montero2014} (PM14) photoionisation models are obtained with version 13.03 of \textsc{cloudy} \citep{ferland2013} as well, assuming a plane-parallel geometry with a constant electron density of $n_{\rm e}=100$~cm$^{-3}$.
The input ionising spectrum is derived from PopStar \citep{molla2009} evolutionary synthesis models. 
The assumed gas phase abundances are taken from \citet{asplund2009}, except for nitrogen, that is considered a free parameter, varying in the range [-2,0]. The grid shown in Fig.~\ref{fig:sulfurproblems} corresponds to log(N/O)=-0.875, close to the solar value log(N/O)=-0.86 (\citealt{asplund2009}; note that this choice does not take into account the detailed shape of the N/O versus O/H sequence, but this is of little importance to the \siii\ line fluxes). 

Interestingly, taking into account these different sets of photoionisation models does not resolve the discrepancy with our observations. The \cite{levesque2010} grids are even more further removed from the data than the \cite{dopita2013} grids. The two \citet{byler2017} and \cite{perez-montero2014} grids appear to provide a better fit, since they predict  log(\siii/\sii), line ratios as low as  $\sim$ -0.75, in better agreement with the data, but still overestimate log(\siii/\oiii). As a consequence of the latter fact, only low-metallicity models can be superimposed with the data.

We then test whether this observed discrepancy persists in the high-S/N regime. This regime is also worth exploring because low surface brightness and low $EW$(\ha) regions in MaNGA galaxies tend to be increasingly contaminated by DIG, which is known to display different line ratios than classical \hii\ regions. We therefore select only the brightest star-forming regions by looking for spaxels with $EW$(\ha)~$>100$~\AA\, (red contours), and still find a significant difference between observed and predicted fluxes.

Since MaNGA data have a kpc scale resolution, \hii\, regions, characterised by sizes between $10-100$~pc \citep{azimlu2011,gutierrez2011,whitmore2011} cannot be resolved, and spatial averaging could lead to loose significant information. In order to show that this issue is not the cause of the discrepancy between models (designed for \hii\,regions) and the data with regards to the \siii\ lines, we also take into account single \hii\, regions analysed by \citet{kreckel2019} with the PHANGS survey (grey dots in Fig.~\ref{fig:sulfurproblems}). For these galaxies, we estimated the \siii$\lambda$9069,9532 flux from the extinction-corrected \siii$\lambda$9069 reported in \citet{kreckel2019}, assuming an intrinsic value between the two \siii\, lines of 2.47 \citep{pyneb}. The fact that the discrepancy holds also in this case allows also to exclude that it is related to the presence of DIG.

The discrepancy between observed and modelled \siii\, fluxes has been already reported in the literature (e.g. \citealt{garnett1989,dinerstein1986,ali1991}). Specifically, \citet{garnett1989} showed that the \siii/\sii\, ratio was over-predicted by their photoionisation models. Their models produced \oii/\oiii\, ratios comparable to observations, suggesting the discrepancy may be due to limitations in modelling stellar atmospheres and/or in the atomic data for sulfur. Very recently, \citet{kewley2019} discuss the use of \siii/\sii\ as an ideal ionisation parameter diagnostic, but come to the same conclusions with regards to the limitations in the modelling effort so far. \citet{kewley2019} however, only discuss modelling issues with regards to the \sii\ lines, since no large sample of galaxies with \siii\ observations was available to date.
In this work we finally provide such a dataset and confirm that the discrepancy between models and observations of the \siii\ lines is a lingering problem, which persists in observations of a large sample of local galaxies and latest-generation photoionisation models based on independent state-of-the-art codes.

\subsection{Applying IZI to the MaNGA data}
\label{sec:applyizi}
Given that the \siii\, lines are not well-reproduced by photoionisation models, we first run IZI excluding the \siii\ line fluxes. In particular, we consider the fluxes of \oii$\lambda\lambda$3726,29, \hb, \oiii$\lambda\lambda$4959,5007, \nii$\lambda\lambda$6548,84, \ha, \sii$\lambda$6717 and \sii$\lambda$6731 and make use of the D13 models. We stress that, even though E(B-V) is derived assuming a screen attenuation, the IZI output takes into account also the internal dust taken into account in D13 models (see Sec.~\ref{sec:izi}). Indeed, the internal dust modifies the thermal structure of the \hii\, region, and thus the emission-line spectrum, particularly in the high-excitation regions, affecting the abundances derived by the model (e.g. \citealt{dopita2000}).
We assume uniform priors for metallicity and the log($q$) within the range provided by the selected photoionisation models. The extinction $E$(B-V) is also assumed to follow a flat prior in the range [0, 1] mag. We have checked that adding the \siii\ lines to the list of strong lines ratios does not change the IZI output. This is because the \siii\ lines are so badly reproduced by the models that they are unable to provide useful information on the best-fit parameter values. If the \siii\ lines are introduced in the fit they are characterised by a large $\chi^2$ ($10-25$) compared to all other lines, highlighting how the D13 models are generally capable of reproducing all strong line ratios except those including \siii.

\begin{figure}
	\includegraphics[width=\columnwidth]{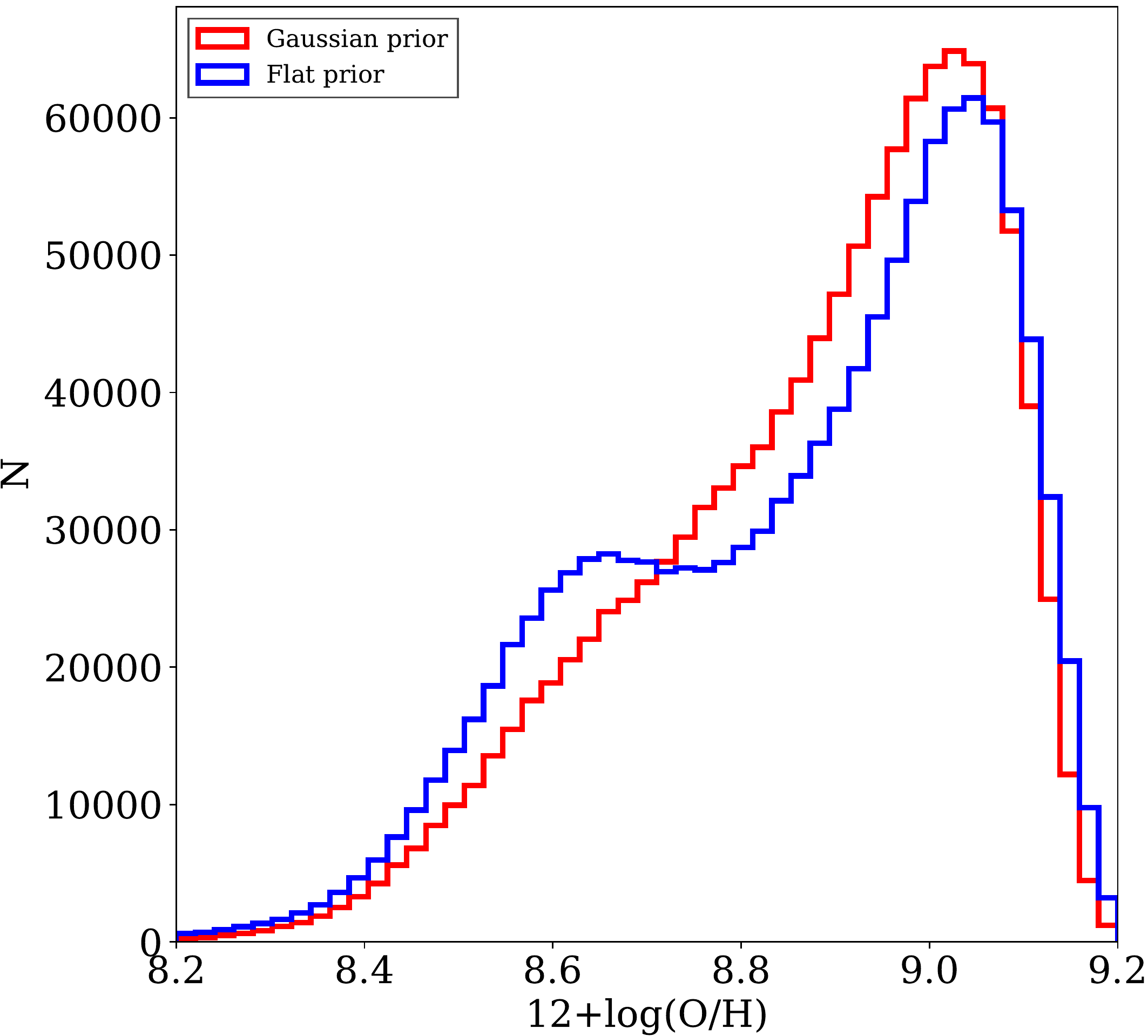}
    \caption{Distribution of the best-fit values of 12+log(O/H) for all MaNGA spaxels considered in this work. Metallicities are derived by IZI with a Gaussian (red) and a flat (blue) prior on the ionisation parameter, respectively. Specifically, the mean of the Gaussian prior is given by the \citet{diaz1991} calibration, that links \siii/\sii\ line and ionisation parameter.}
    \label{fig:priorVSnopriormet}
\end{figure}

Fig.~\ref{fig:priorVSnopriormet} shows in blue the histogram of the best-fit values of the metallicity obtained by IZI without taking into account the \siii\, lines. Interestingly, the metallicity distribution tends to be bimodal, peaking at 12+log(O/H)~$\sim9$ and at 12+log(O/H)~$\sim8.6$, with a gap around 12+log(O/H)~$\sim8.8$. As we discuss below, we argue that the peak at 12+log(O/H)~$\sim8.6$ could be due to an underestimation of metallicity, because of the degeneracy between 12+log(O/H) and log($q$). The results of taking into account the \siii/\sii\,line ratio to determine the ionisation parameter are described in the next section and are shown with the red line.

\begin{figure}
      \begin{minipage}{1\columnwidth}
        	\includegraphics[width=1.\columnwidth]{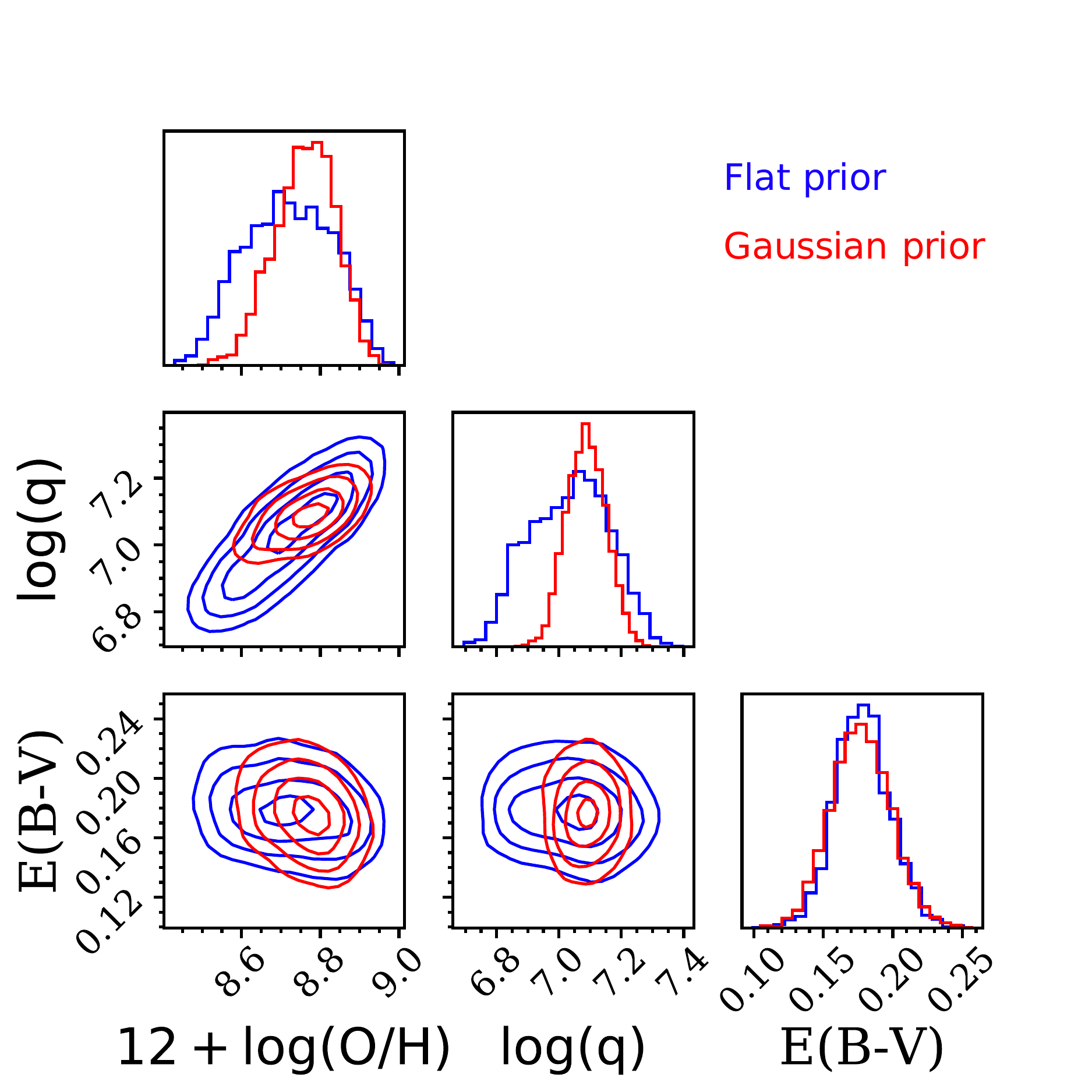}
    \end{minipage} 
        \begin{minipage}{1\columnwidth}
        	\includegraphics[width=1.\columnwidth]{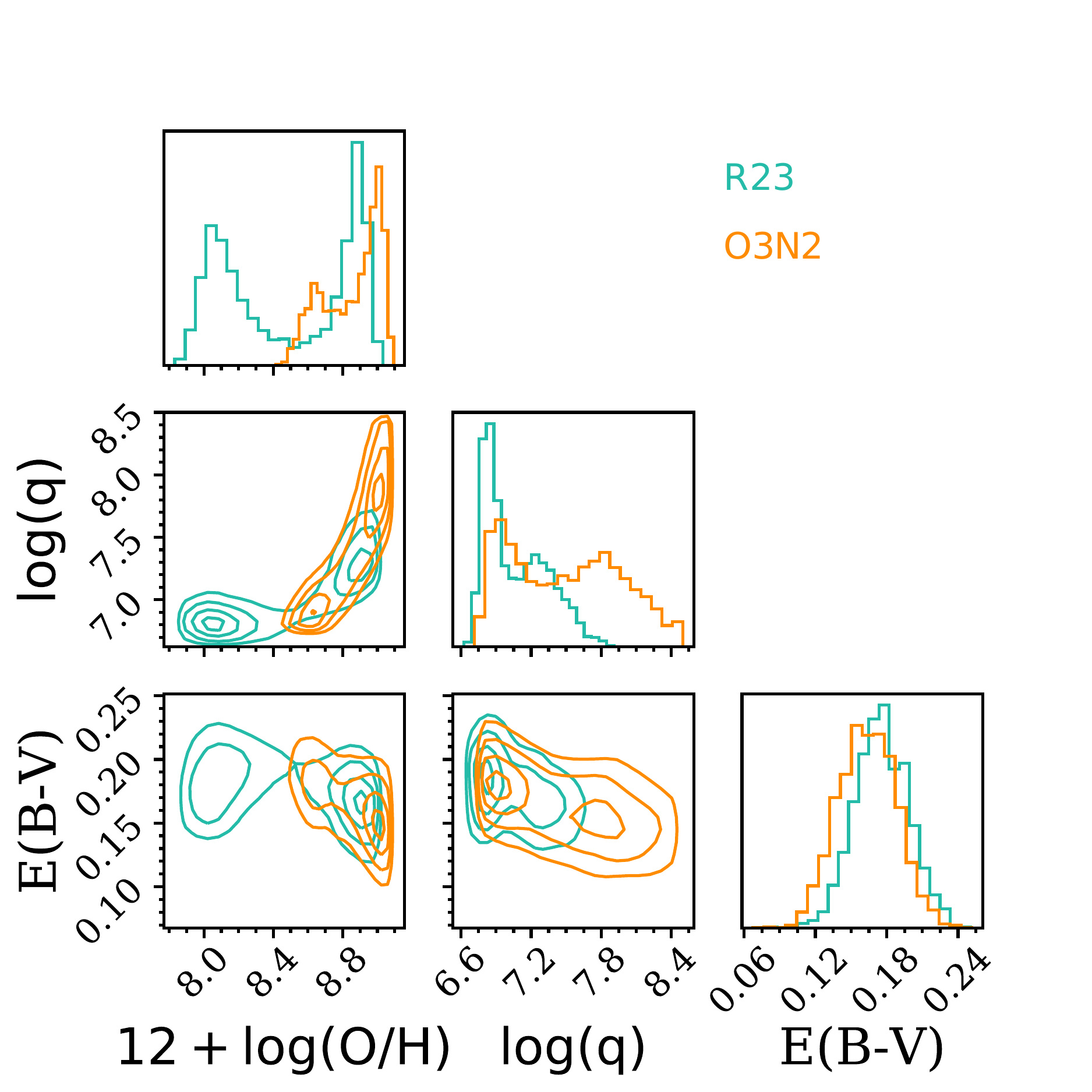}
    \end{minipage} 
         \caption{Top panel: Posterior PDFs relative to 12+log(O/H), log($q$) and $E$(B-V), inferred with a Gaussian (red) and a flat (blue) prior on the ionisation parameter, respectively, with D13 models, for a single spaxel of the galaxy 7990-12703. 
         Bottom panel: Posterior PDFs relative to the three parameters, inferred only taking into account \oii$\lambda\lambda$3726,29, \hb \, and  \oiii$\lambda\lambda$4959,5007 (i.e., R23, in green), and \hb, \oiii$\lambda\lambda$4959,5007, \nii$\lambda\lambda$6548,84 and \ha, (i.e. O3N2, in orange), respectively.}
        \label{fig:tests}
  \end{figure}

In order to better visualise the origin of the bimodality in the metallicity distribution of spaxels and its possible relation to intrinsic degeneracies in the fitting process, we show as an example different tests performed on one spaxel of the galaxy 7990-12703, chosen as a random spaxel with best-fit metallicity close to the lower-metallicity peak in the bimodal metallicity distribution of all MaNGA spaxels (Fig.~\ref{fig:priorVSnopriormet}, 12+log(O/H)~$=8.7$).
In the upper panel of Fig.~\ref{fig:tests} we show in blue the posterior PDFs of 12+log(O/H), log($q$) and $E$(B-V), inferred by IZI without taking into account the \siii\ lines. Strong degeneracy between metallicity and ionisation parameter is evident from the figure. The 12+log(O/H) PDF is found to be very wide, characterised by likely values in the range [8.6,8.9] (best-fit median value of 12+log(O/H)~$=8.7$). We believe that the inability of constraining the metallicity in this example is due to the lack of a strong ionisation parameter diagnostic capable of tightly constraining log($q$), as suggested by the wide PDF of this parameter, covering the range [6.8,7.4].
The lower panel of Fig.~\ref{fig:tests} displays in green and orange the PDFs calculated using only \oii$\lambda\lambda$3726,29, \hb \, and  \oiii$\lambda\lambda$4959,5007 (i.e., R23), and \hb, \oiii$\lambda\lambda$4959,5007, \nii$\lambda\lambda$6548,84 and \ha, (i.e. O3N2), respectively. 
In both cases, there is a clear degeneracy between a low-metallicty low-ionisation parameter and a high-metallicity, high-ionisation parameter solution.
Indeed, R23 is known to be double-valued (e.g., \citealt{pagel1979,dopita2006b}). O3N2 is commonly used as metallicity diagnostic, but it carries a strong dependence on the ionisation parameter, because of the very different ionisation potentials of N$^+$ (14.5 eV) and O$^{++}$ (35.1 eV) \citep{alloin1979}. This leads to a systematic underestimation of 12+log(O/H) for higher ionisation parameters and overestimation of 12 + log(O/H) in regions characterised by lower ionisation parameter values (e.g. \citealt{cresci2017,kruhler2017}). 

\subsection{Imposing a prior on the ionisation parameter based on the \siii/\sii\ ratio}
\label{sec:priorvsnoprior}

The discrepancy between model predictions and \siii\, emission line fluxes prevents us from introducing them directy in our method together with all the other emission lines.
Hence, in order to understand if and to what extent the \siii$\lambda\lambda$9069,9532 lines can help in the determination of the ionisation parameter and in breaking the degeneracy with metallicity illustrated in Fig.~\ref{fig:priorVSnopriormet} and Fig.~\ref{fig:tests}, we compute a second run of IZI using the \siii/\sii\ ratio to set a Gaussian prior on log($q$). 
Specifically, we choose to set the mean of this Gaussian prior by linking \siii/\sii\ and ionisation parameters via a commonly-used calibration obtained by using a large grid of single-star photoionisation models by \citet{diaz1991} (D91 hereafter),
\begin{equation}
\label{eq:diaz}
    \mathrm{log}(q)= - 1.68 \times \mathrm{log}(\sii/\siii) + 7.49,
\end{equation}
taking into account a standard deviation of 0.2~dex.
Recently, \citet{morisset2016} recalibrated this relation using an updated version of \textsc{cloudy} \citep{ferland2013} and atomic data, finding reasonable agreement with Eq.~\ref{eq:diaz}.
This approach allows the \siii\, lines to add extra information to the fit, while at the same time not fully constraining log($q$) to the value obtained via the D91 calibration. We stress that since we do not fix the ionisation parameter to the value inferred by \citet{diaz1991}, the use of \citet{morisset2016} calibration would not change our results, given that \citet{diaz1991} and \citet{morisset2016} are reasonably consistent within $\pm0.2$~dex.
The Gaussian prior can be applied only in those spaxels in which \siii$\lambda$9532, \sii $\lambda$6717, \sii$\lambda$6731, \ha \, and \hb \, are observed with $S/N>1.5$, which are $\sim80$\% of the total. 

\begin{figure}
 	\includegraphics[width=1\columnwidth]{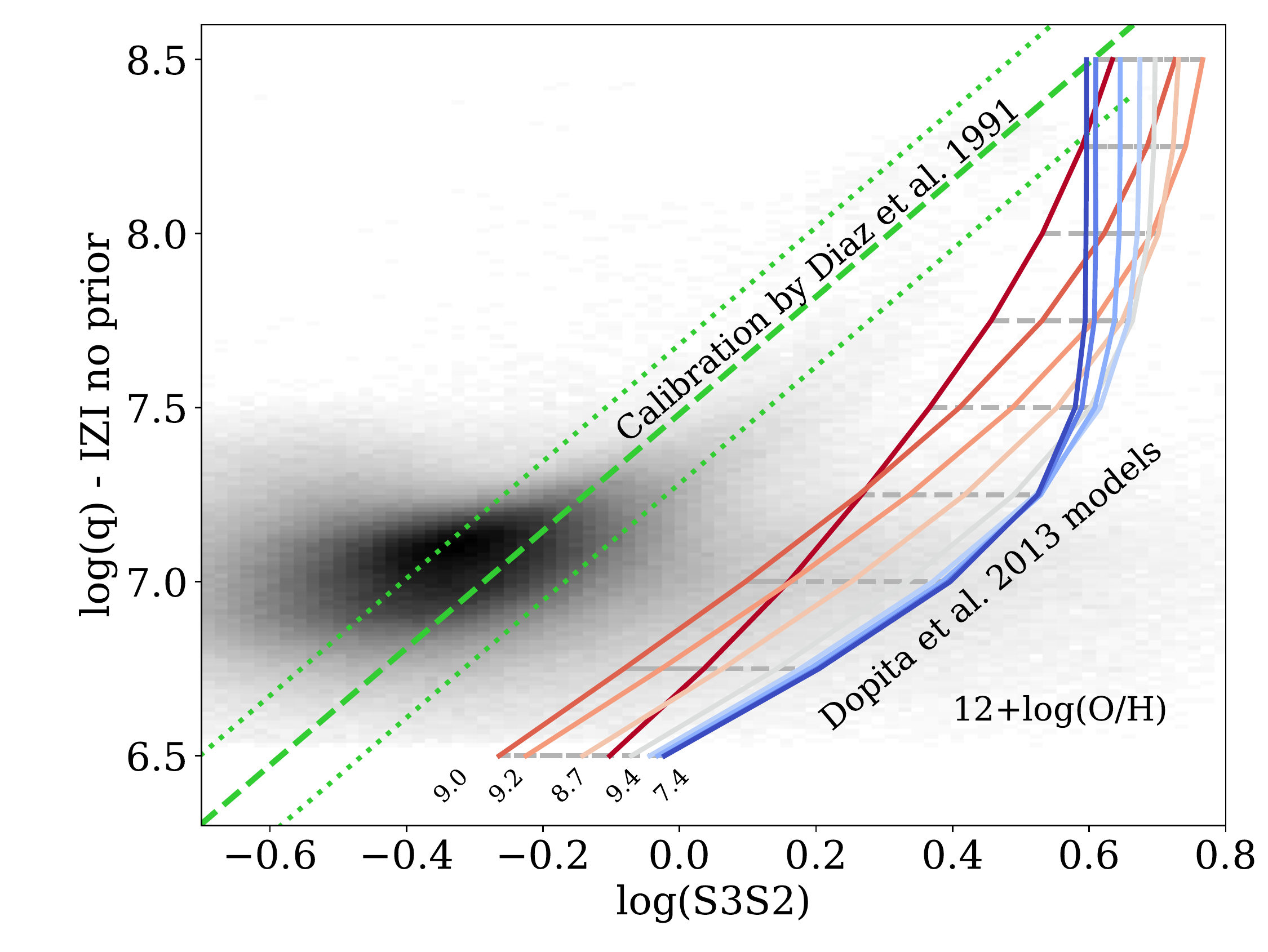}
    \caption{The distribution of log($q$) as a function of log(S3S2) inferred with IZI with a flat prior on the ionisation parameter is shown in shades of grey. The green dashed line shows the log($q$) values obtained with Eq.~\ref{eq:diaz} (\protect\citet{diaz1991}), while the green dotted lines represent the $\pm0.2$~dex scatter around this relation. D13 models are shown in colour (red-blue, going from higher to lower metallicity).} 
    \label{fig:whydiaz}
\end{figure}

Fig.~\ref{fig:whydiaz} shows log($q$) inferred with IZI with the flat prior on the ionisation parameter 
as a function of the observed log(S3S2). The green dashed line illustrates the log($q$) values obtained with Eq.~\ref{eq:diaz}, while the green dotted lines represent the $\pm0.2$~dex scatter around this relation. 
D13 models are shown for comparison (from red to blue going from high to low metallicity).
Given the range of log(S3S2) observed in our data, the ionisation parameters obtained via the D91 calibration are broadly in the same range as the ionisation parameter inferred by IZI using all other strong lines except \siii\, within $\pm0.2$~dex. 

The reason why the D91 calibration, which is itself based on photoionisation models, is in better agreement with the data than all the other models considered in the previous section is still an unresolved problem. 
The main discrepancy between D13, D91 and \citealt{morisset2016} photoionisation models, however, could be related to the input ionising spectrum (see also Sec.~4.1 and Fig.~5 in \citealt{morisset2016}), but also to the underlying assumptions of the models (e.g. changes in atomic data, inclusion/exclusion of dust physics). 

\begin{figure*}
    \begin{minipage}{0.666\columnwidth}
	\includegraphics[width=1.\columnwidth]{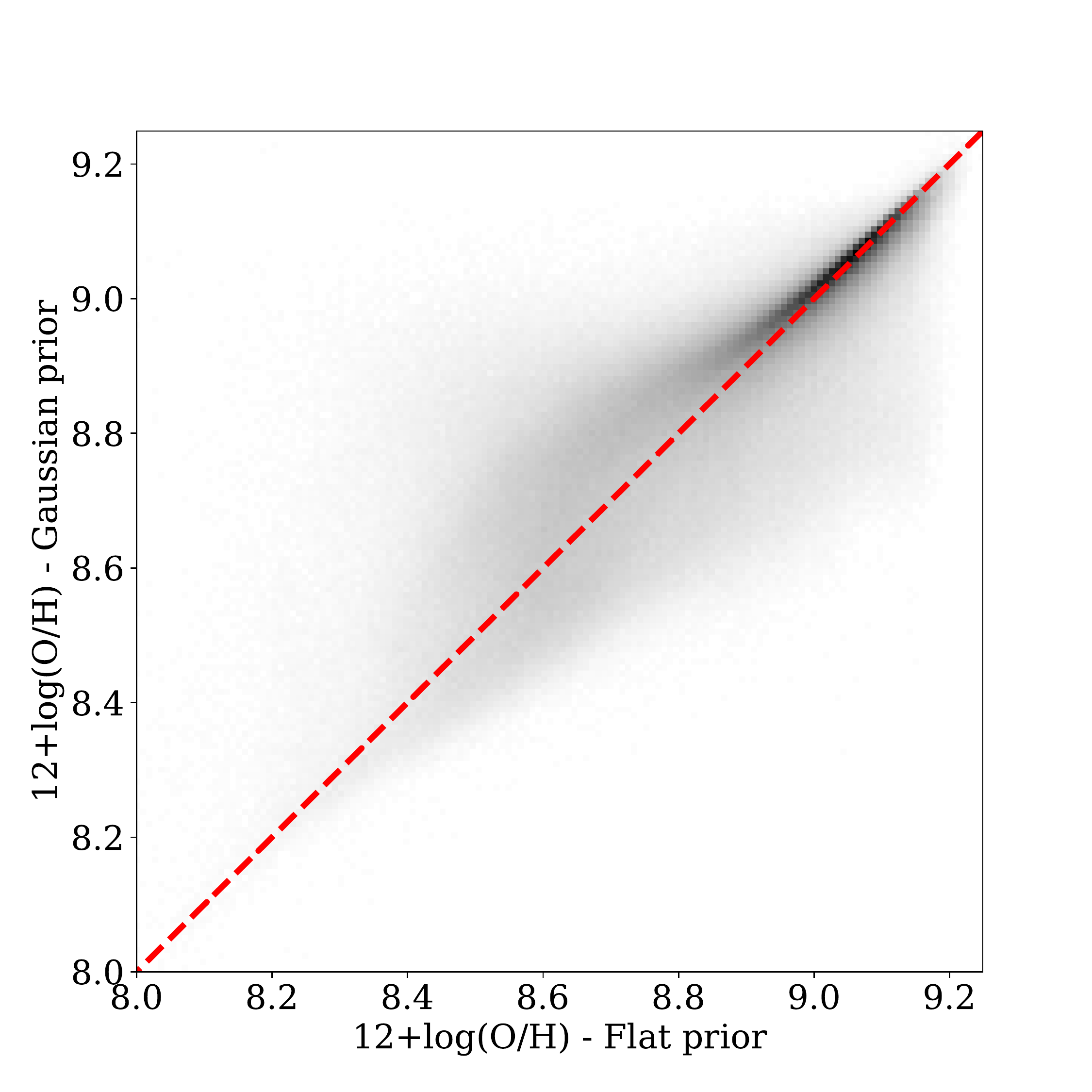}
    \end{minipage}
    \begin{minipage}{0.666\columnwidth}
	\includegraphics[width=1\columnwidth]{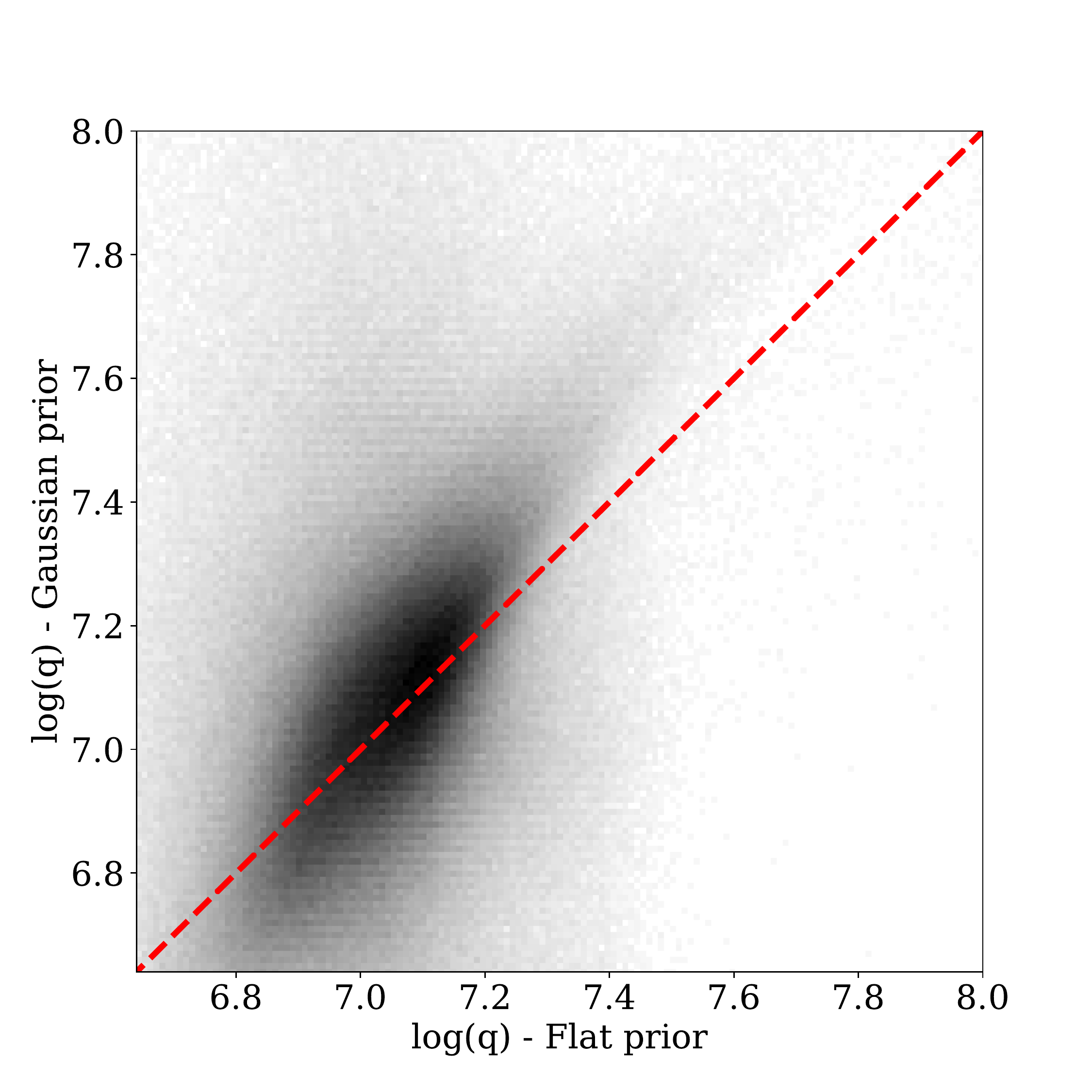}
    \end{minipage}
    \begin{minipage}{0.666\columnwidth}
	\includegraphics[width=1\columnwidth]{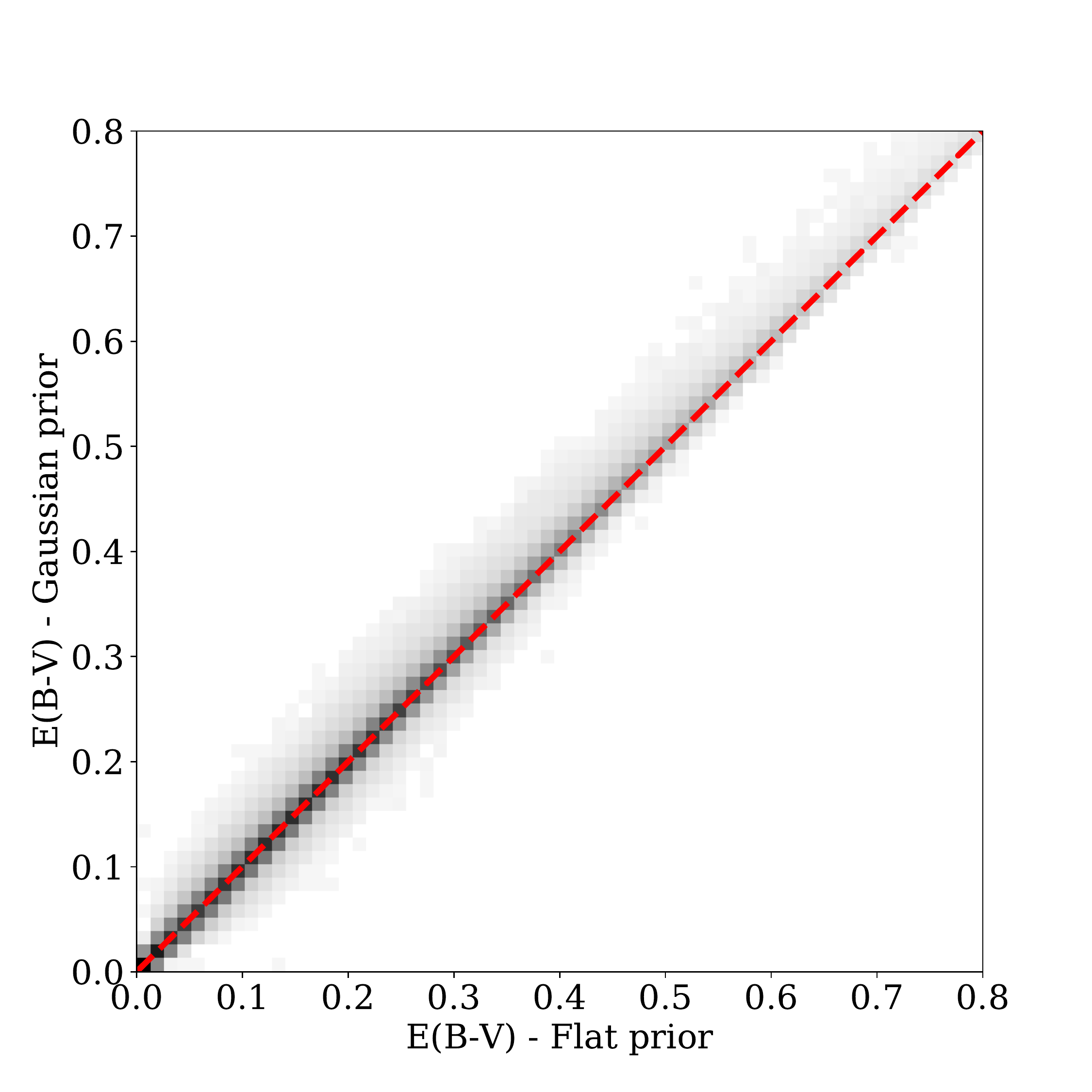}
    \end{minipage}
    \caption{Comparison between the results obtained with IZI with a flat (x-axis) and a Gaussian (y-axis) prior on the ionisation parameter, for metallicity, ionisation parameter and gas extinction, respectively. The red dashed line represents the one-to-one relation.} 
    \label{fig:priorVSnopriorall}
\end{figure*}

The differences between our two IZI runs (with a Gaussian and a flat prior on the ionisation parameter) can be appreciated from the three panels of Fig.~\ref{fig:priorVSnopriorall}. These figures compare the best-fit values (i.e. the median of the PDF computed by IZI, see Sec.~\ref{sec:izi}) of metallicity, ionisation parameter and gas extinction obtained without taking into consideration the \siii\ lines (x-axis) and imposing a Gaussian prior on log($q$) based on S3S2 (y-axis), respectively.
The metallicity distributions of the two runs are consistent at values 12+log(O/H)~$>8.9$, but below this threshold metallicities inferred by the IZI run with the Gaussian prior tend to be higher than those inferred by IZI with the flat prior.
The distribution of ionisation parameters of all considered spaxels is found to be wider when taking into account the prior, since the best-fit values tend to be lower for log($q$)~$<7$ and higher for log($q$)~$>7.2$. Finally, as expected, the prior on the ionisation parameter has no effect on the gas extinction.

\begin{figure*}
\begin{minipage}{1\columnwidth}
	\includegraphics[width=1\columnwidth]{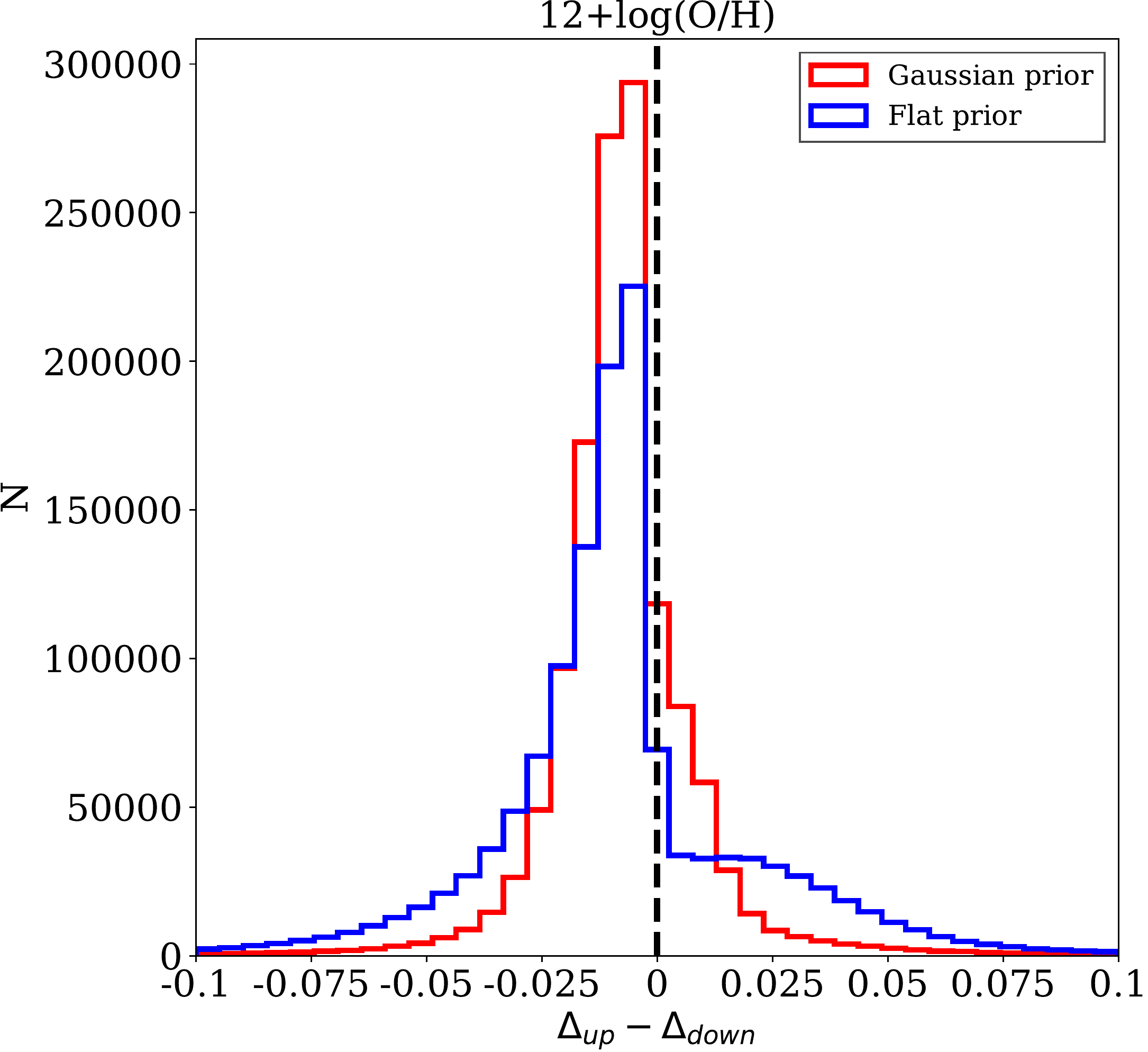}
\end{minipage}
\begin{minipage}{1\columnwidth}
	\includegraphics[width=1\columnwidth]{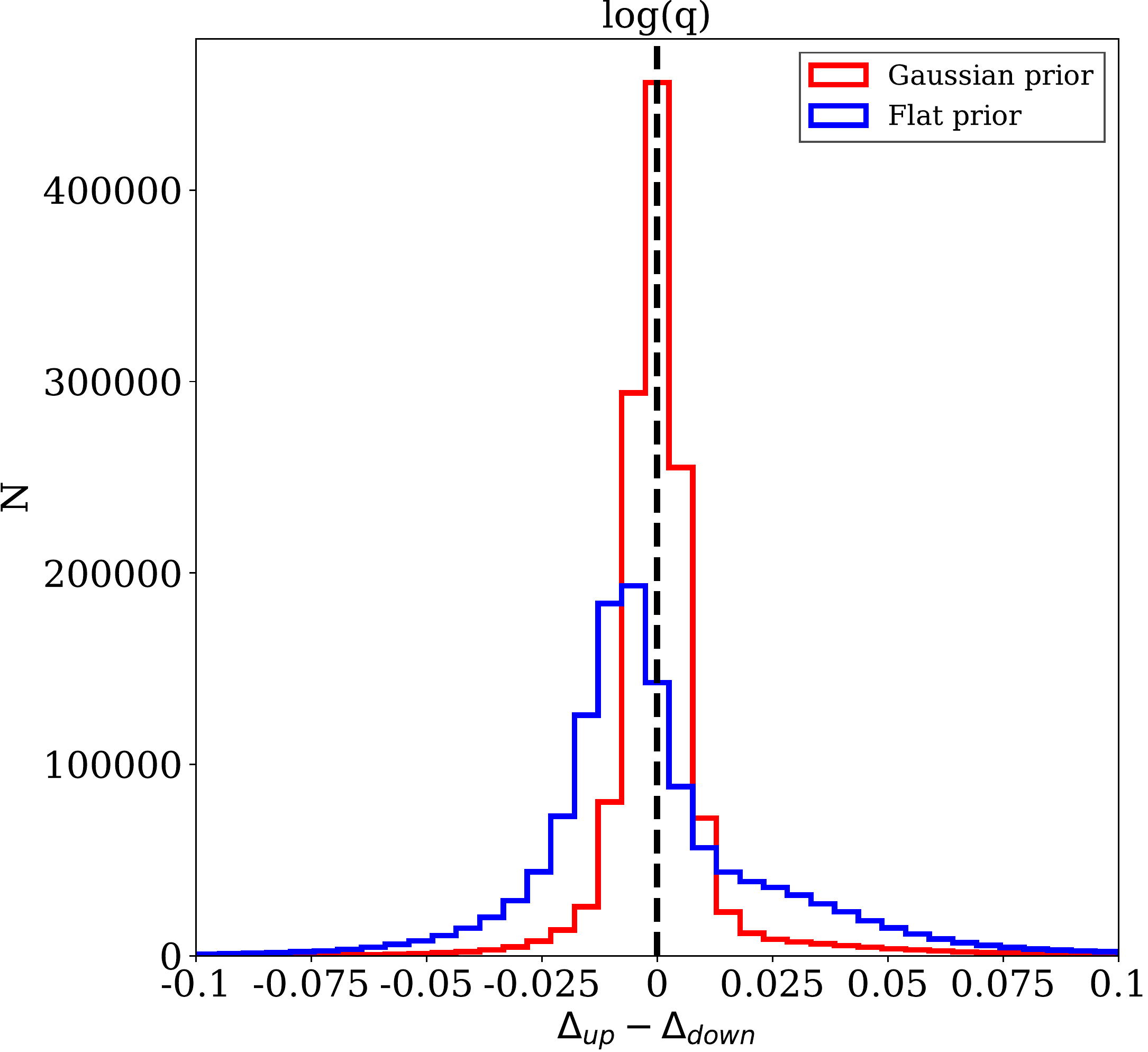}
\end{minipage}
    \caption{Distribution of the difference between $\Delta_{up}=84^{\mathrm{th}}-50^{\mathrm{th}}$ and $\Delta_{down}=50^{\mathrm{th}}-16^{\mathrm{th}}$ for the metallicity (left panel) and ionisation parameter (right panel), in the case with a flat (in blue) and a Gaussian (in red) prior on the ionisation parameter, respectively), for all the spaxels of all the galaxies.}
    \label{fig:priorVSnoprior}
\end{figure*}

Coming back to Fig.~\ref{fig:priorVSnopriormet}, the histogram of the best-fit values of the metallicity obtained by the IZI run with the Gaussian prior on the ionisation parameter is shown in red. Note that the bimodality discussed in Sec.~\ref{sec:applyizi} disappears when the Gaussian prior is imposed. Indeed, in the upper panel of Fig.~\ref{fig:tests}, the degeneracy is broken introducing the Gaussian prior based on the Diaz calibration (shown in red as well), in which a higher metallicity (8.8 vs 8.7) and a higher ionisation parameter (7.1 vs 7.0) solution is preferred.
  
As explained in Sec.~\ref{sec:izi}, the errors provided by IZI are obtained by computing the $16^{\mathrm{th}}$, $50^{\mathrm{th}}$ and $84^{\mathrm{th}}$ percentiles of the PDFs, and are defined as $\Delta_{up}=84^{\mathrm{th}}-50^{\mathrm{th}}$ and $\Delta_{down}=50^{\mathrm{th}}-16^{\mathrm{th}}$.
The difference between $\Delta_{up}$ and $\Delta_{down}$ indicates how much these PDFs differ from a symmetric distribution. In general this difference would be greater than zero if the distribution is tailed towards higher values with respect to its median value, considered as the best-fit, and vice-versa.
Fig.~\ref{fig:priorVSnoprior} shows the distribution of the difference between $\Delta_{up}$ and $\Delta_{down}$ for the metallicity (left panel) and ionisation parameter (right panel), in the case with a flat (blue) and a Gaussian (red) prior on the ionisation parameter, for all the spaxels considered in this work. In the case with the flat prior, these distributions are shifted towards negative values both for 12+log(O/H) and log($q$) $\Delta_{up}$ - $\Delta_{down}$, meaning that the majority of the spaxels have an asymmetric PDF tailed towards lower values with respect to the best-fit. Moreover, these distributions show a cut-off around 0 (this is more visible for 12+log(O/H)), with a considerable tail at positive values to $\sim 0.075$, implying that some spaxels have an asymmetric PDF tailed towards higher values with respect to the best-fit. This leads to the bimodality discussed above.

In the case with the Gaussian prior, the log(q) $\Delta_{up}$ - $\Delta_{down}$ distribution is almost Gaussian, centered on zero, mirroring the prior itself. The metallicity distribution remains slightly shifted towards negative values (i.e. high values of metallicity), but shows a much less conspicuous tail compared to the case with the flat prior, indicating a general significant reduction in the effect of the 12+log(O/H) $-$ log($q$) degeneracy.

\begin{figure*}
        	\includegraphics[width=1\columnwidth]{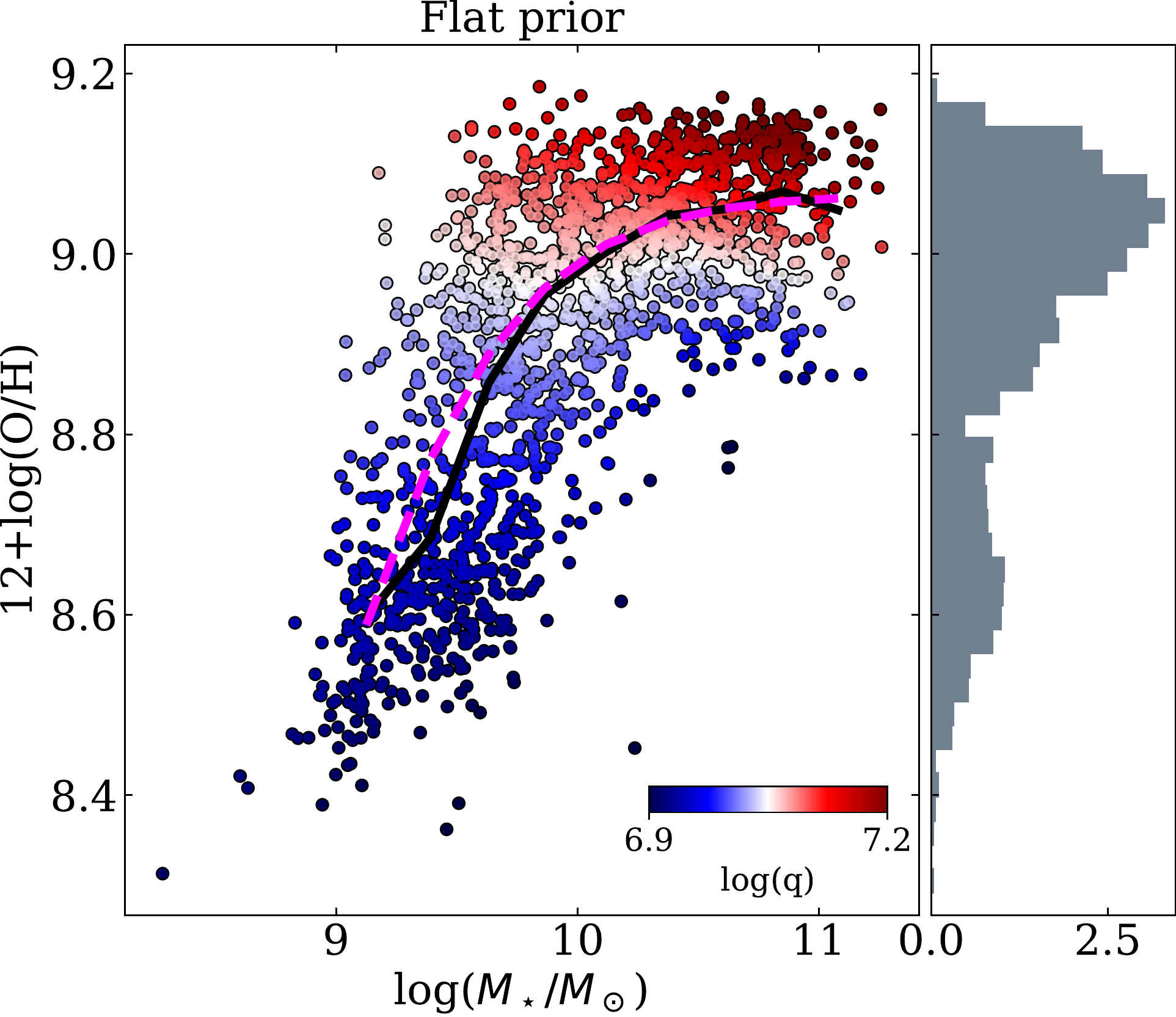}
        	\includegraphics[width=1\columnwidth]{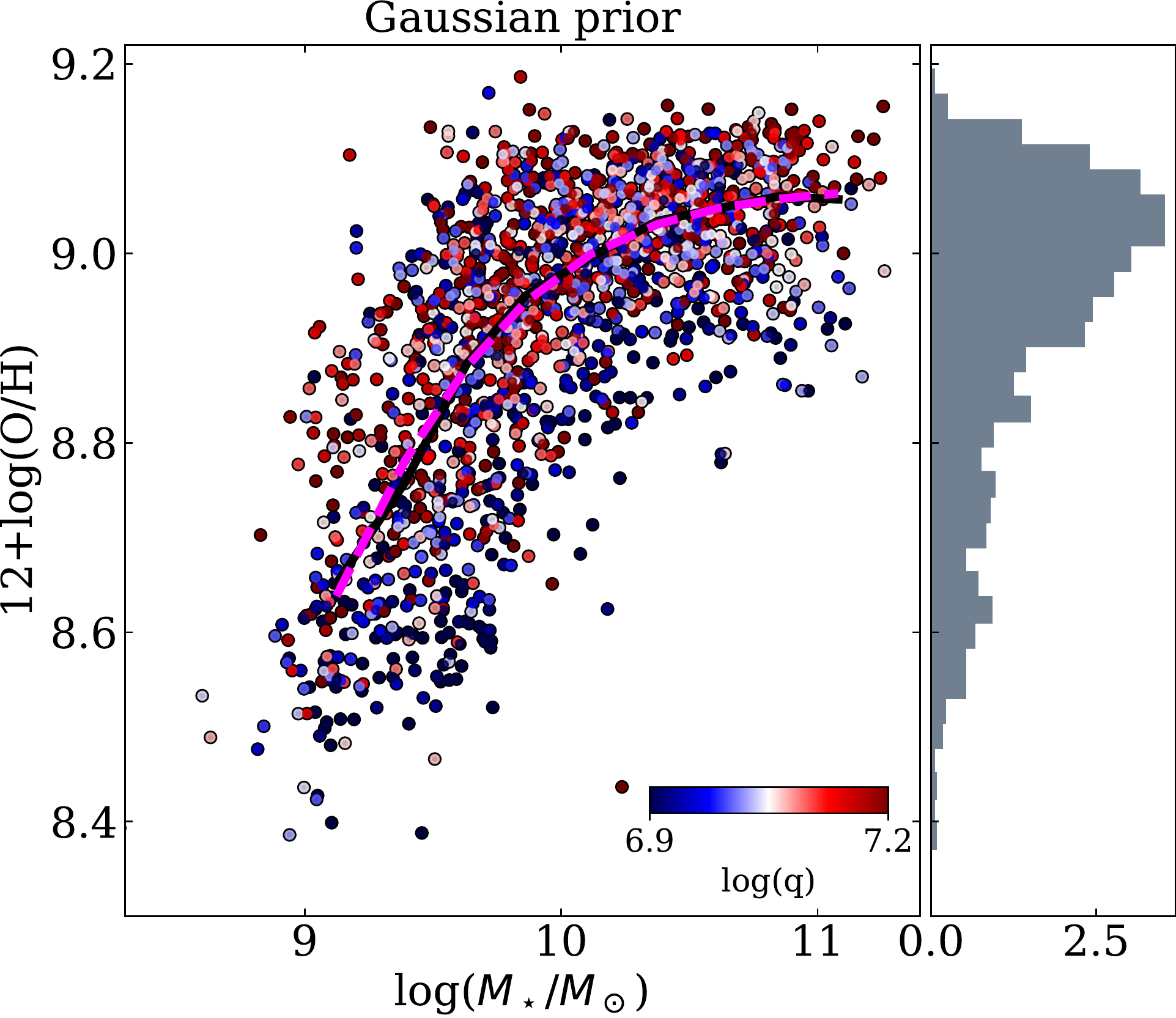}
    \caption{The mass-metallicity relation for the MaNGA star-forming galaxies considered in this work without using the 
    \siii\ lines (left) and by using the \siii\ lines via an ionisation parameter prior as discussed in the text (right). Metallicities are computed for each galaxy at 0.8~\re. The  colour-coding indicates the ionisation parameter (at 0.8~\re). A histogram of the metallicity distributions for all galaxies is shown in grey on the right. In the case where the \siii\ lines are not used (left panel) there is a striking correlation at fixed mass between the metallicity and ionisation parameter. To visually highlight this correlation the colour-coding has been smoothed according to the LOESS recipe of \citet{cappellari2013b}. The solid black lines show the median metallicity in bins of $M_\star$, while the dashed magenta lines are the best-fit using the parametrisation proposed in \citet{curti2019}.}
    \label{fig:massmet}
\end{figure*} 

Fig.~\ref{fig:massmet} shows the metallicity inferred for each galaxy at 0.8~\re, obtained by interpolating the metallicity profile as a function of galactocentric distance. This characteristic metallicity is shown as a function of the total stellar mass, colour-coded on the basis of the ionisation parameter (also at 0.8~\re) for the case with the flat and the Gaussian prior on the ionisation parameter, respectively. We present the corresponding metallicity distributions as grey histograms on the right. We comment here on this well-known scaling relation between mass and metallicity (MZR, \citealt{lequeux1979, tremonti2004}) to highlight the importance of breaking the degeneracy between metallicity and ionisation parameter when investigating secondary trends.

As the left panel of Fig.~\ref{fig:massmet} shows, in the case with the flat prior we find the hint of a bimodality in the metallicity distribution, shown in the grey histogram, with values around 12+log(O/H) $\sim$ 8.8 being less probable. This feature creates a steeper slope for the mass-metallicity relation in the range $9.5< \rm log(M_{\star}/M_\odot)<10$, with respect to the case with the prior, and in agreement with the even larger bimodality in the mass-metallicity relation derived by \citet{blanc2019} with IZI and the \citealt{levesque2010} models.  

More strikingly, at fixed mass there is a strong correlation between metallicity and ionisation parameter (a smoothing has been applied to the colour-coding for ease of visualisation, using the LOESS recipe from \citealt{cappellari2013b}). This correlation disappears when a Gaussian prior based on \siii/\sii\ is used (right panel of Fig.~\ref{fig:massmet}). We argue, therefore, that this secondary dependence of metallicity on log($q$) seen in the case with the flat ionisation parameter prior is solely due to the degeneracy between the two parameters. 

In Tab.~\ref{tab:tab1} we report the best-fit parameter values for the MZR shown in Fig.~\ref{fig:massmet}, using the new parametrisation proposed in \citet{curti2019} (C19):
\begin{equation}\label{eq:c19}
    {\rm 12+log(O/H)} = Z_0 - \gamma/\beta \times {\rm log}\Bigg(1+\Bigg(\frac{M}{\rm M_0}\Bigg)^{-\beta}\Bigg) 
\end{equation}
In this equation, $Z_0$ is the metallicity at which the relation saturates, quantifying the asymptotic upper metallicity limit, while $M_0$ is the characteristic turnover mass above which the metallicity asymptotically approaches the upper metallicity limit ($Z_0$). At stellar masses $M_\star < M_0$, the MZR reduces to a power law of index $\gamma$. In Eq.~\ref{eq:c19}, $\beta$ quantifies how ``fast'' the curve approaches its saturation value.
The parameters that we obtain for the two different MZRs, shown on the left and right of Fig.~\ref{fig:massmet}, are consistent within the errors, and the scatter around the best-fit relations are 0.12~dex and 0.10~dex, respectively.
 
\begin{table}
\label{tab:tab1}
\caption{Best-fit values for the parameters of the MZR shown in the right panel of Fig.~\ref{fig:massmet} derived with IZI, putting a Flat and a Gaussian prior on the ionisation parameter based on \siii/\sii, assuming the new parametrisation
proposed in \citet{curti2019} (Eq.~\ref{eq:c19}).} 
\label{table:1}      
\centering                          
\begin{tabular}{c c c c}        
\hline
\hline                 
\multicolumn{4}{c}{MZR C19 - Flat prior} \\
\hline
\hline
$Z_0$ & $M_0/$M$_\odot$ & $\gamma$ & $\beta$ \\  
  $9.07^{+0.06}_{-0.04}$ & $9.3^{+0.1}_{-0.2}$  & $1.3^{+0.4}_{-0.4}$ & $1.1^{+0.5}_{-0.4}$ \\
\hline  
\multicolumn{4}{c}{MZR C19 - Gaussian prior} \\
\hline
\hline
$Z_0$ & log($M_0/$M$_\odot$) & $\gamma$ & $\beta$ \\    
   $9.06^{+0.05}_{-0.04}$ & $9.3^{+0.1}_{-0.2}$  & $1.1^{+0.5}_{-0.4}$ & $1.2^{+0.5}_{-0.4}$ \\      
\hline
\hline
\end{tabular}
\end{table}


In order to show the reliability of our method, in App.~\ref{sec:quality_test}, we show the comparison between the observed value of \ha, \nii$\lambda$6584, \oii, \oiii$\lambda$5007, \sii$\lambda$6717, \sii$\lambda$6731 and \siii$\lambda$9532 and the fluxes of the the best-fitting model predicted by IZI (all normalised to \hb), taking into account D13 models and a Gaussian prior on the ionisation parameter, as a function of the Petrosian effective radius and stellar mass, for all the spaxels of all the galaxies taken into account. Overall, the stellar-mass- and radius-dependent variations lie in the range [-0.1,0.1]~dex, meaning that they are all consistent within the assumed uncertainty of 0.1 dex (0.01 dex for \ha) for the photoionisation models. As expected, this does not hold for the \siii$\lambda$9532 emission line, that is largely overestimated by IZI, with differences between observed and predicted fluxes up to 0.8 dex.

\section{Metallicity, ionisation parameter and gas extinction in MaNGA galaxies}
\label{sec:results}

In this section we show the results obtained with the IZI run which makes use of the Gaussian prior on the ionisation parameter, as described in Sec.~\ref{sec:method}. 
In all the plots, only the spaxels classified as star-forming according to both the \nii- and the \sii-BPT diagrams with S/N(\ha)~$>15$ and in which the prior based on the S3S2\, line ratio can be used  are shown.

\subsection{Example maps for low- and high-mass galaxies}

\begin{figure*}
    \begin{minipage}{1\columnwidth}
    \vspace{-0.5cm}
        \centering
    	 \includegraphics[width=0.5\columnwidth]{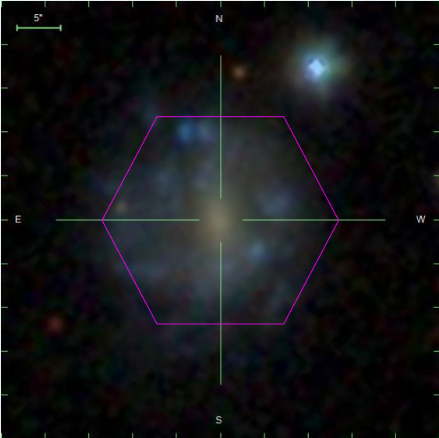}
    \end{minipage}    \hfill
    \begin{minipage}{1\columnwidth}
    \vspace{-0.5cm}
       \centering
        	\includegraphics[width=0.7\columnwidth]{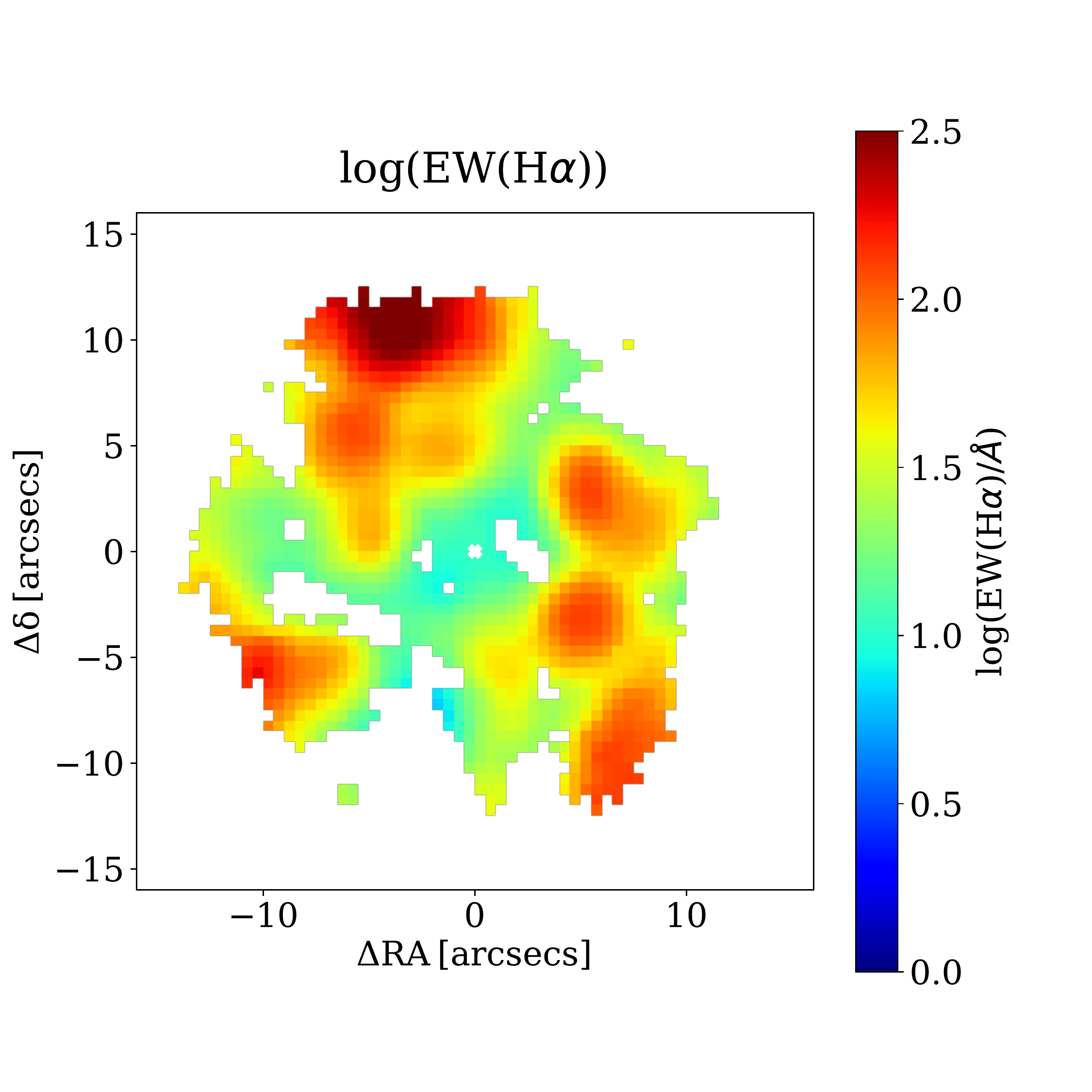}
    \end{minipage} \hfill
    \begin{minipage}{1\columnwidth}
    \vspace{-0.5cm}
       \centering
        	\includegraphics[width=0.7\columnwidth]{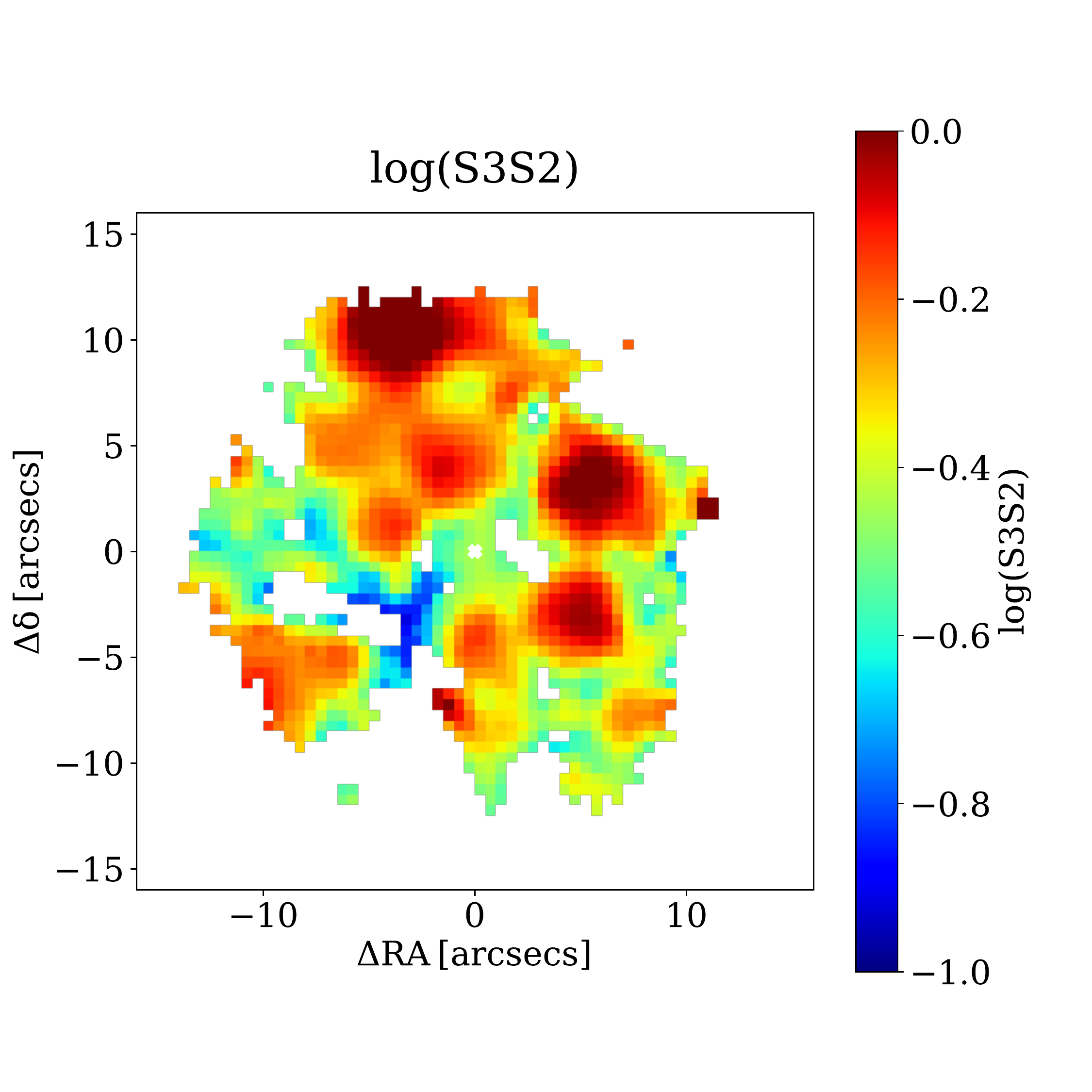}
    \end{minipage} \hfill
    \begin{minipage}{1\columnwidth}
    \vspace{-0.5cm}
        \centering
        	\includegraphics[width=0.7\columnwidth]{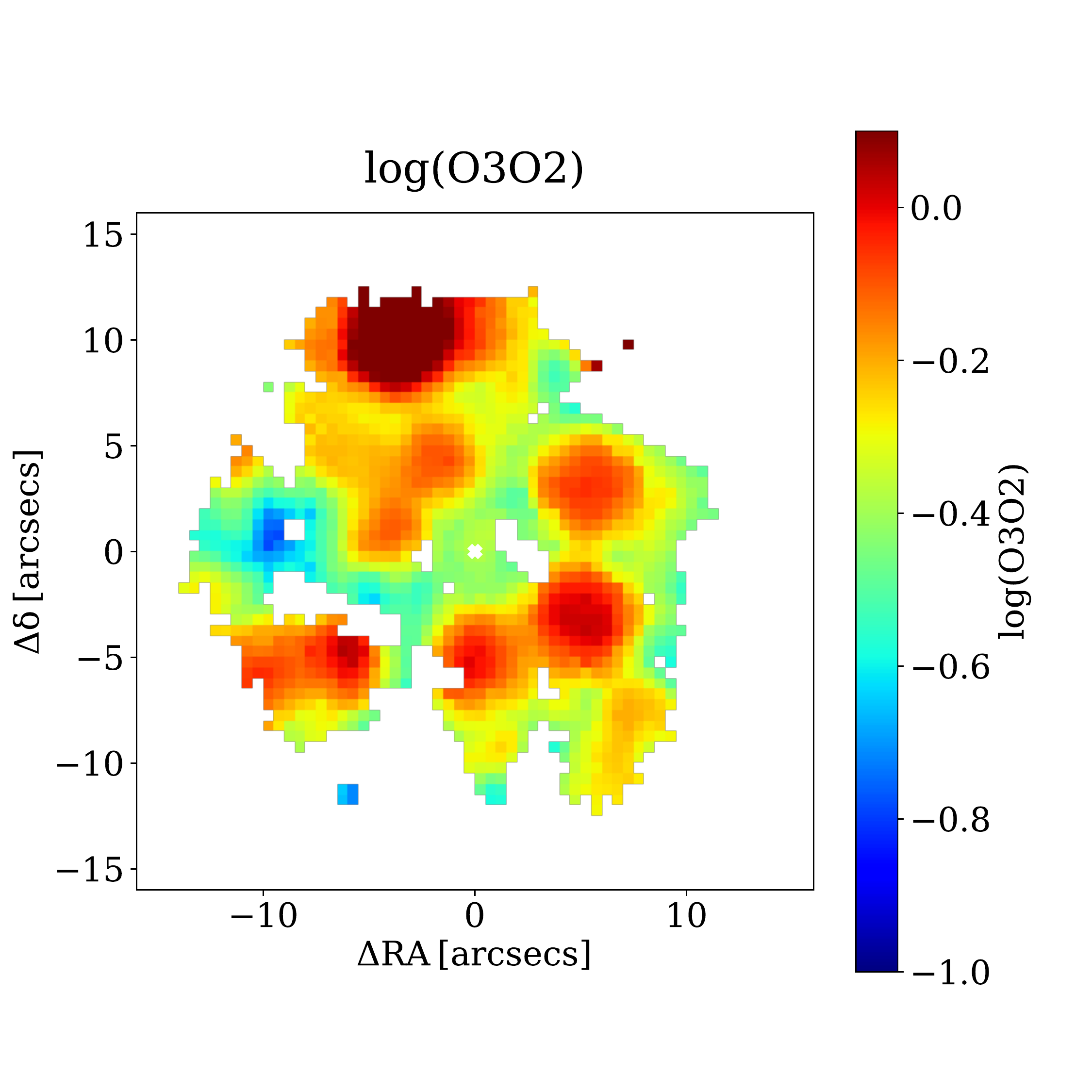}
    \end{minipage} \hfill
    \begin{minipage}{1\columnwidth}
    \vspace{-0.5cm}
       \centering
        	\includegraphics[width=0.7\columnwidth]{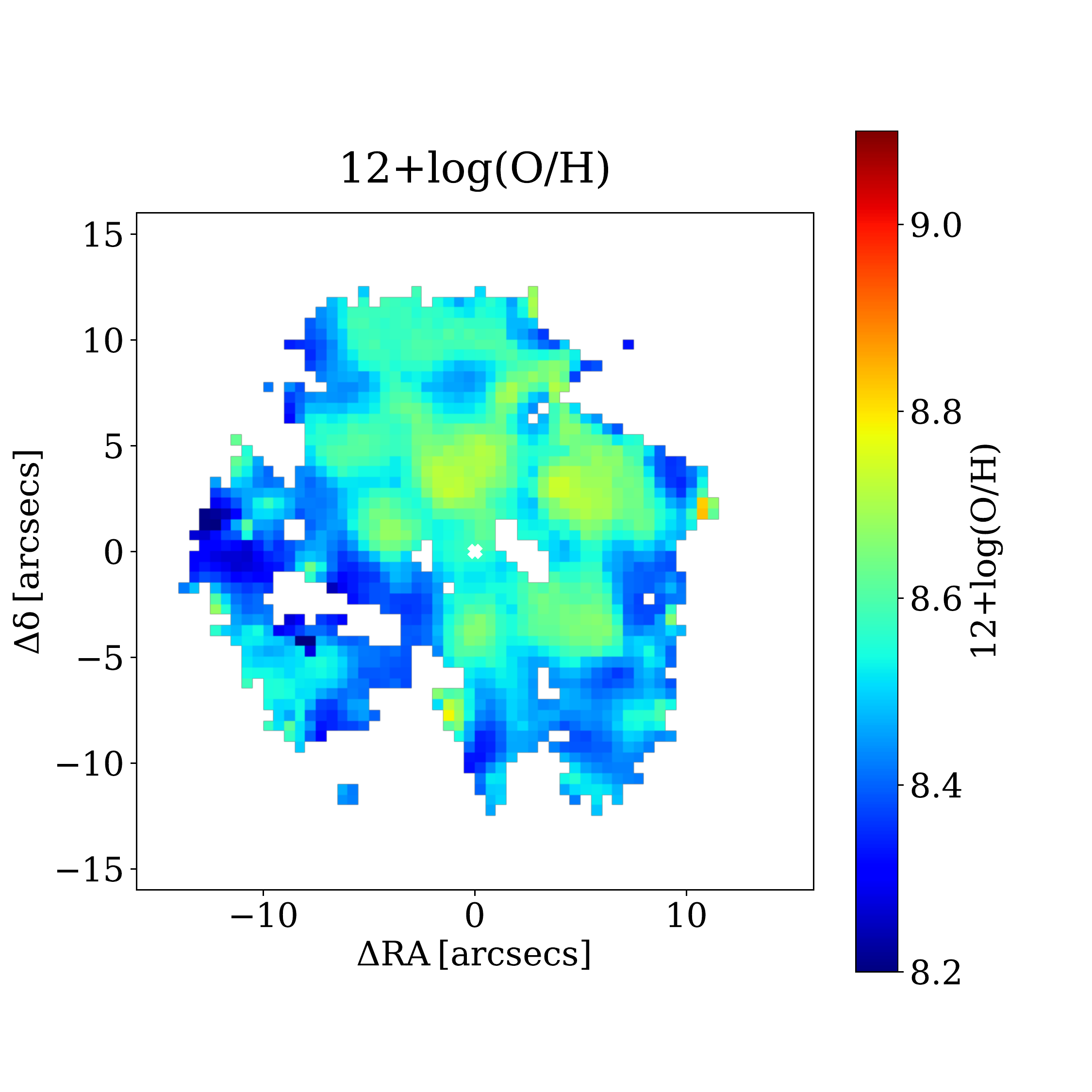}
    \end{minipage} \hfill
    \begin{minipage}{1\columnwidth}
    \vspace{-0.5cm}
        \centering
        	\includegraphics[width=0.7\columnwidth]{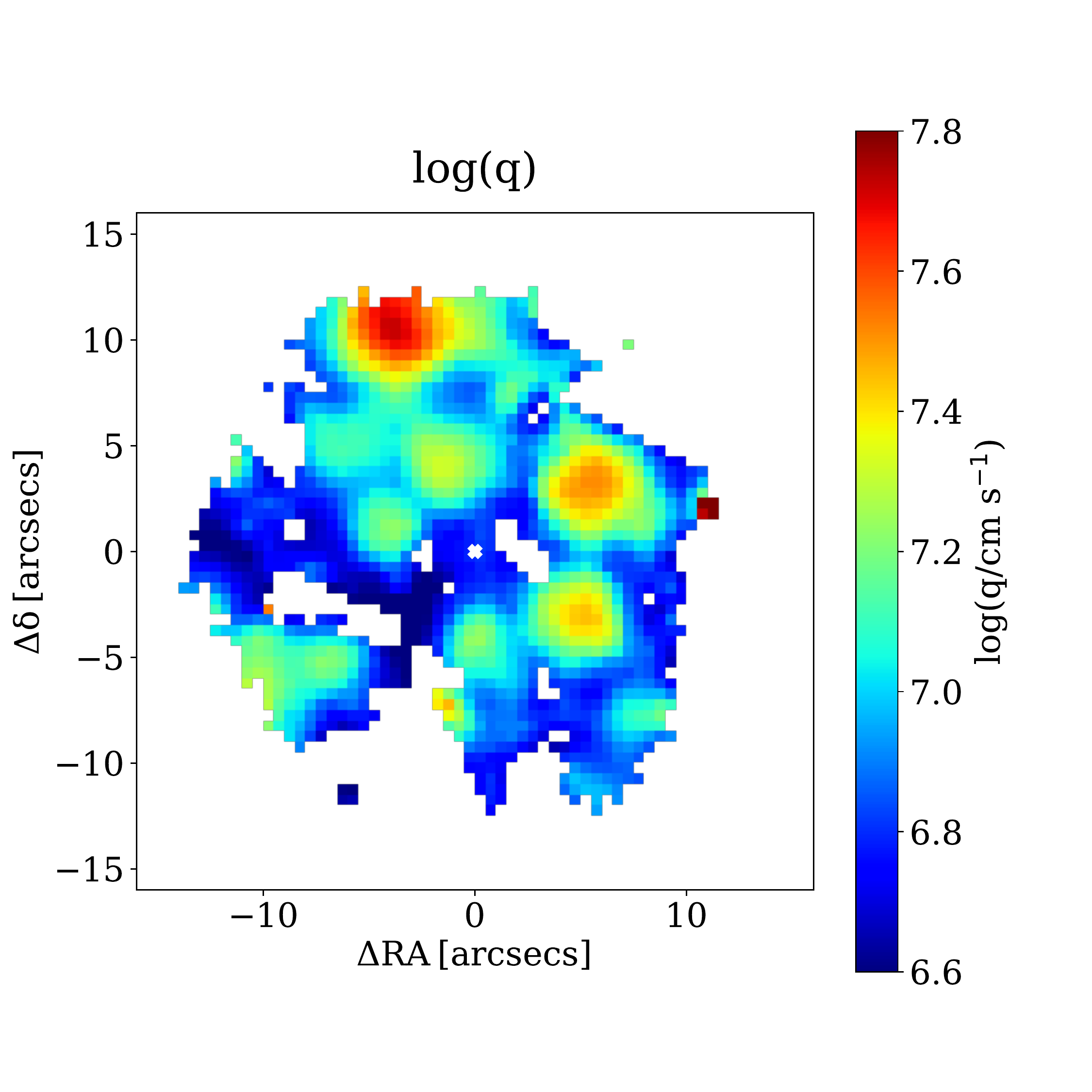}
    \end{minipage} \hfill
    \begin{minipage}{1\columnwidth}
    \vspace{-0.5cm}
            \centering
        	\includegraphics[width=0.7\columnwidth]{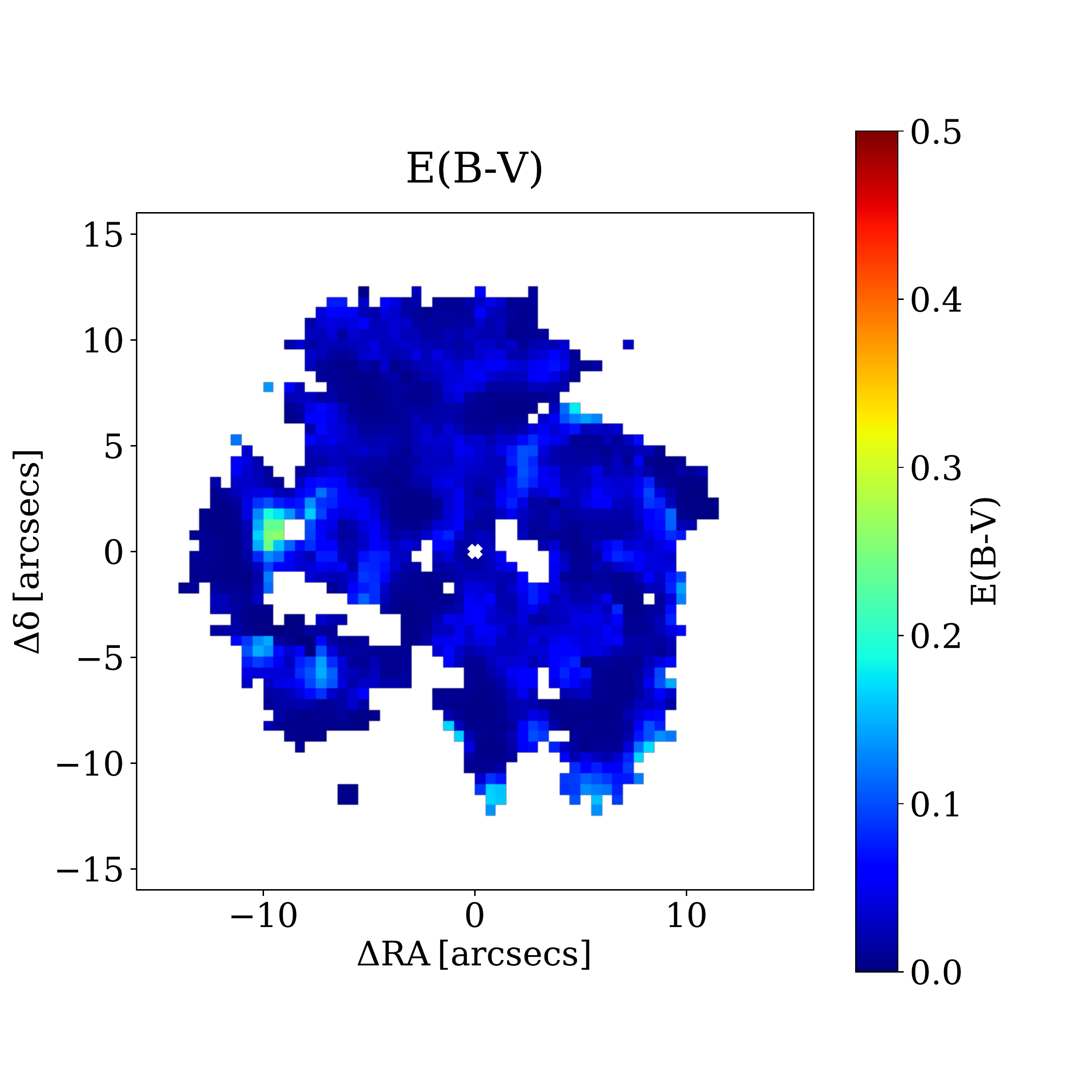}
    \end{minipage} \hfill
    \begin{minipage}{1\columnwidth}
    \vspace{-0.5cm}
        \centering
        	\includegraphics[width=0.7\columnwidth]{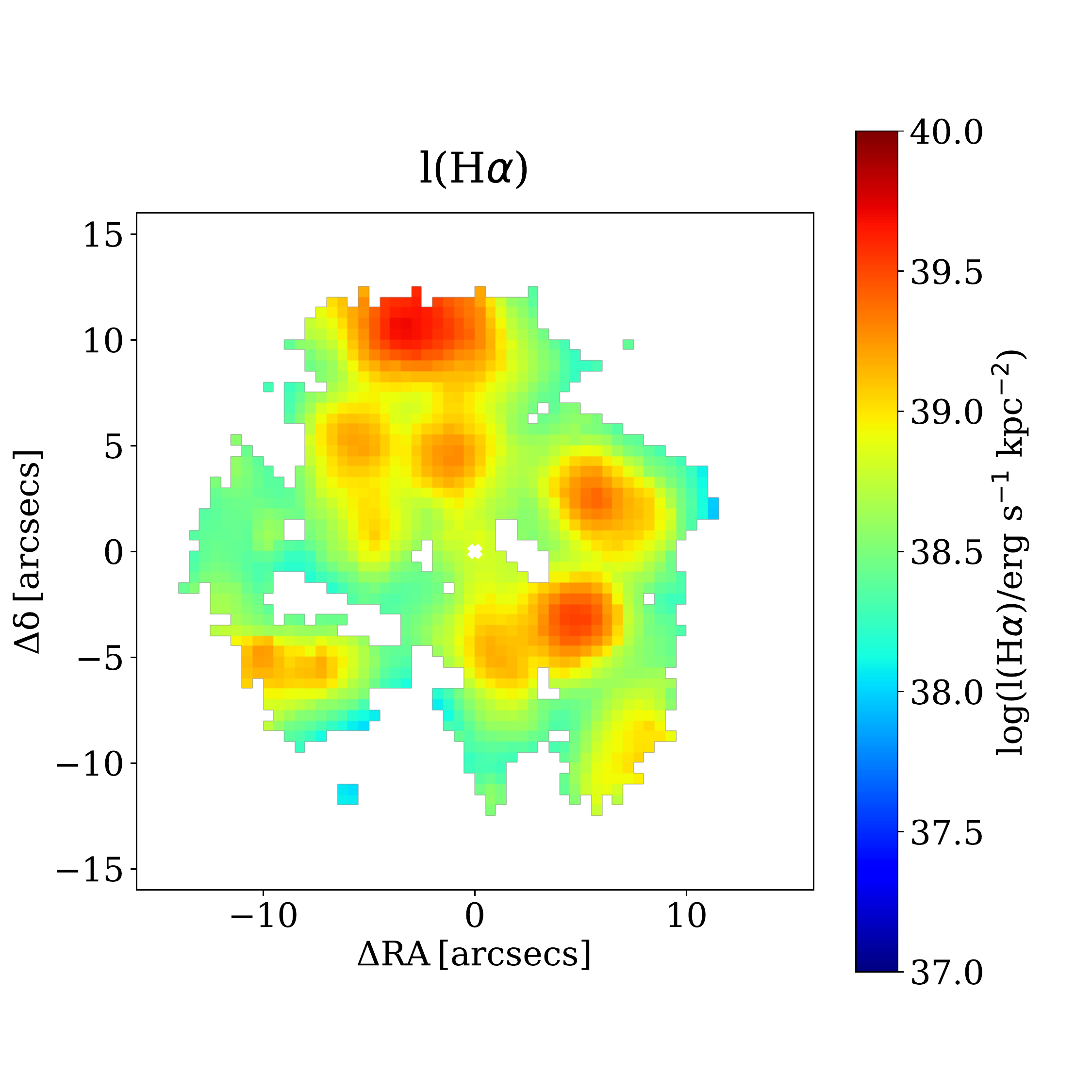}
    \end{minipage}    
    
    \caption{Maps for the galaxy 8147-9102, with a stellar mass $\rm log(M_{\star}/M_\odot) = 9.10$. The $g-r-i$ image composite from SDSS with the MaNGA hexagonal FoV overlaid, equivalent width of \ha, S3S2, O3O2, IZI metallicity 12+log(O/H), IZI ionisation parameter log($q$), IZI gas extinction $E$(B-V) and \ha\,luminosity per spaxel $L$(\ha) maps are reported, respectively. East is to the left.}
        	\label{fig:lowmassgal}
\end{figure*}

\begin{figure*}
        \begin{minipage}{1\columnwidth}
        \vspace{-0.5cm}
        \centering
    	 \includegraphics[width=0.5\columnwidth]{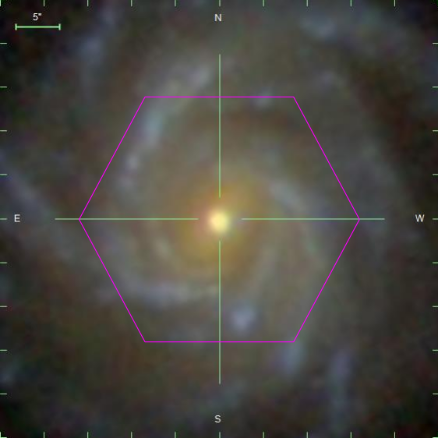}
   \end{minipage}    
       \begin{minipage}{1\columnwidth}
       \vspace{-0.5cm}
       \centering
        	\includegraphics[width=0.7\columnwidth]{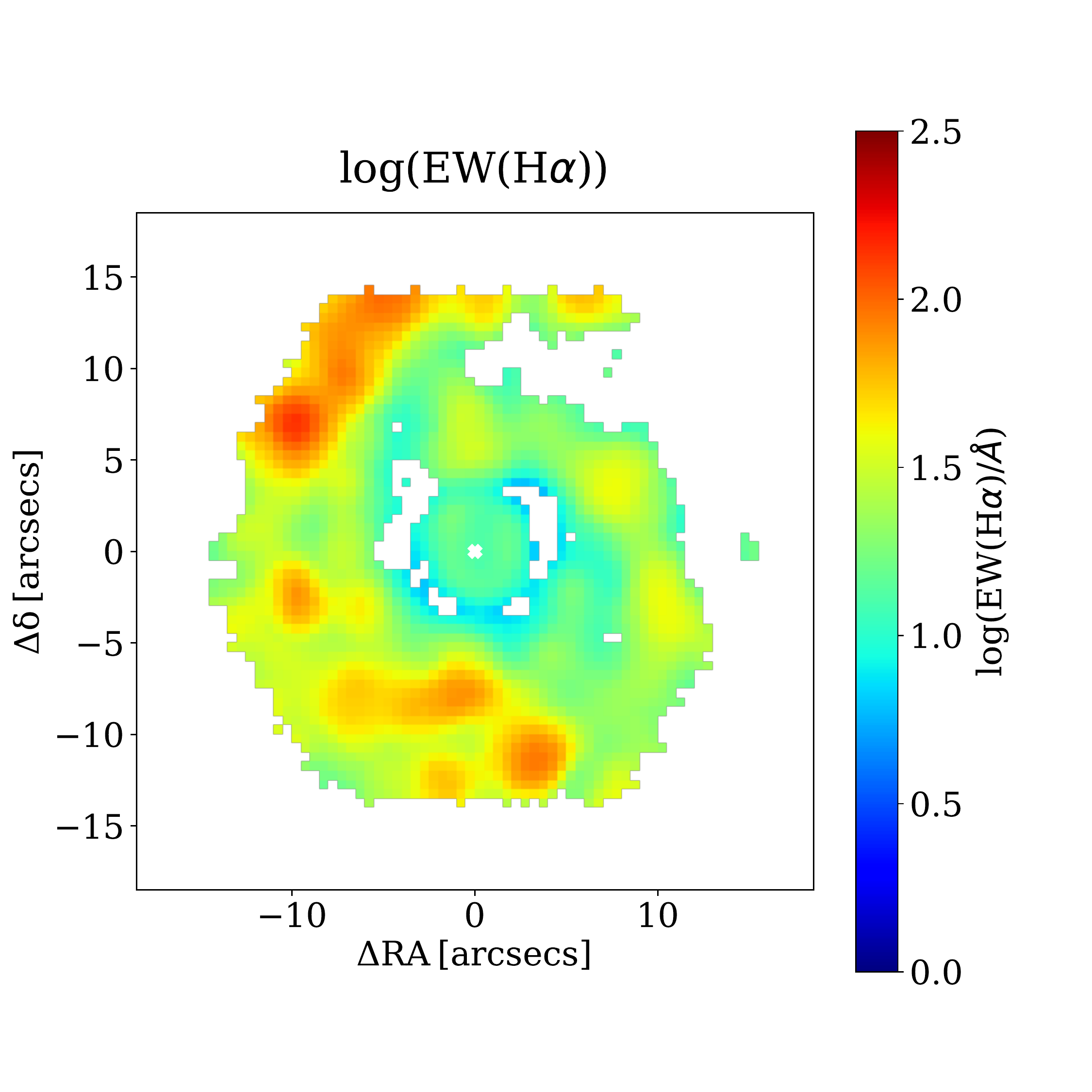}
    \end{minipage} 
       \begin{minipage}{1\columnwidth}
       \vspace{-0.5cm}
       \centering
        	\includegraphics[width=0.7\columnwidth]{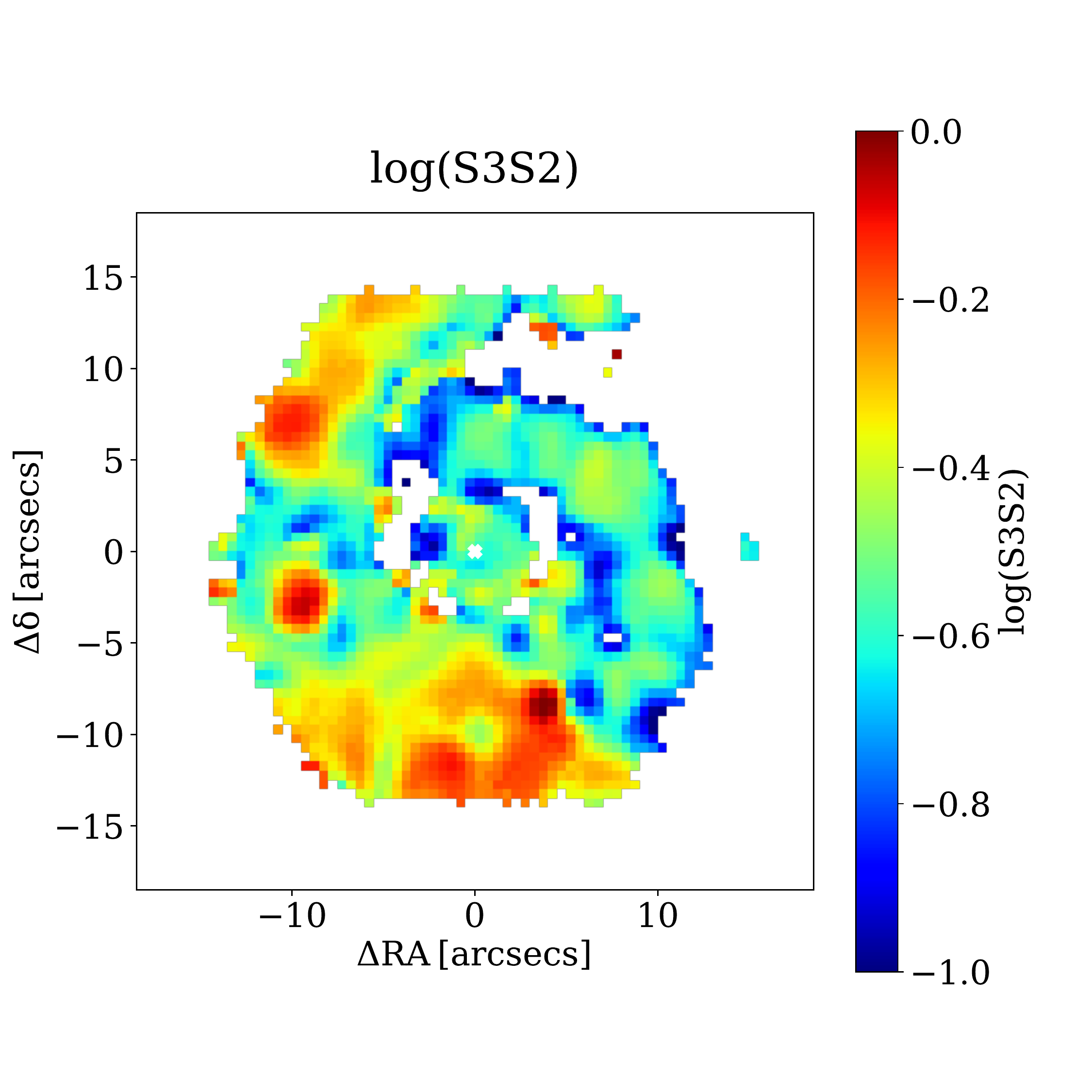}
    \end{minipage}     
    \begin{minipage}{1\columnwidth}
    \vspace{-0.5cm}
    \centering
        	\includegraphics[width=0.7\columnwidth]{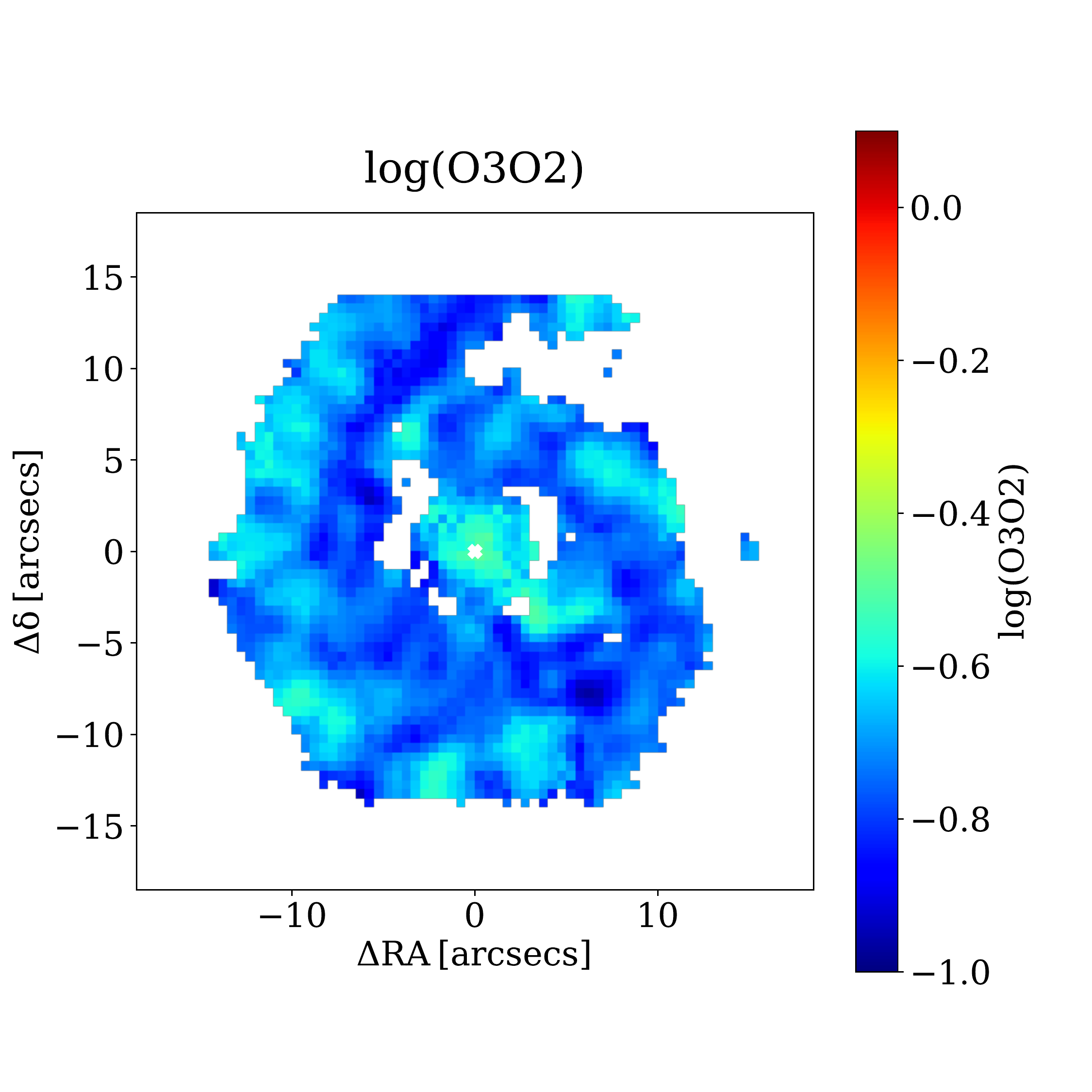}
    \end{minipage} 
       \begin{minipage}{1\columnwidth}
       \vspace{-0.5cm}
       \centering
        	\includegraphics[width=0.7\columnwidth]{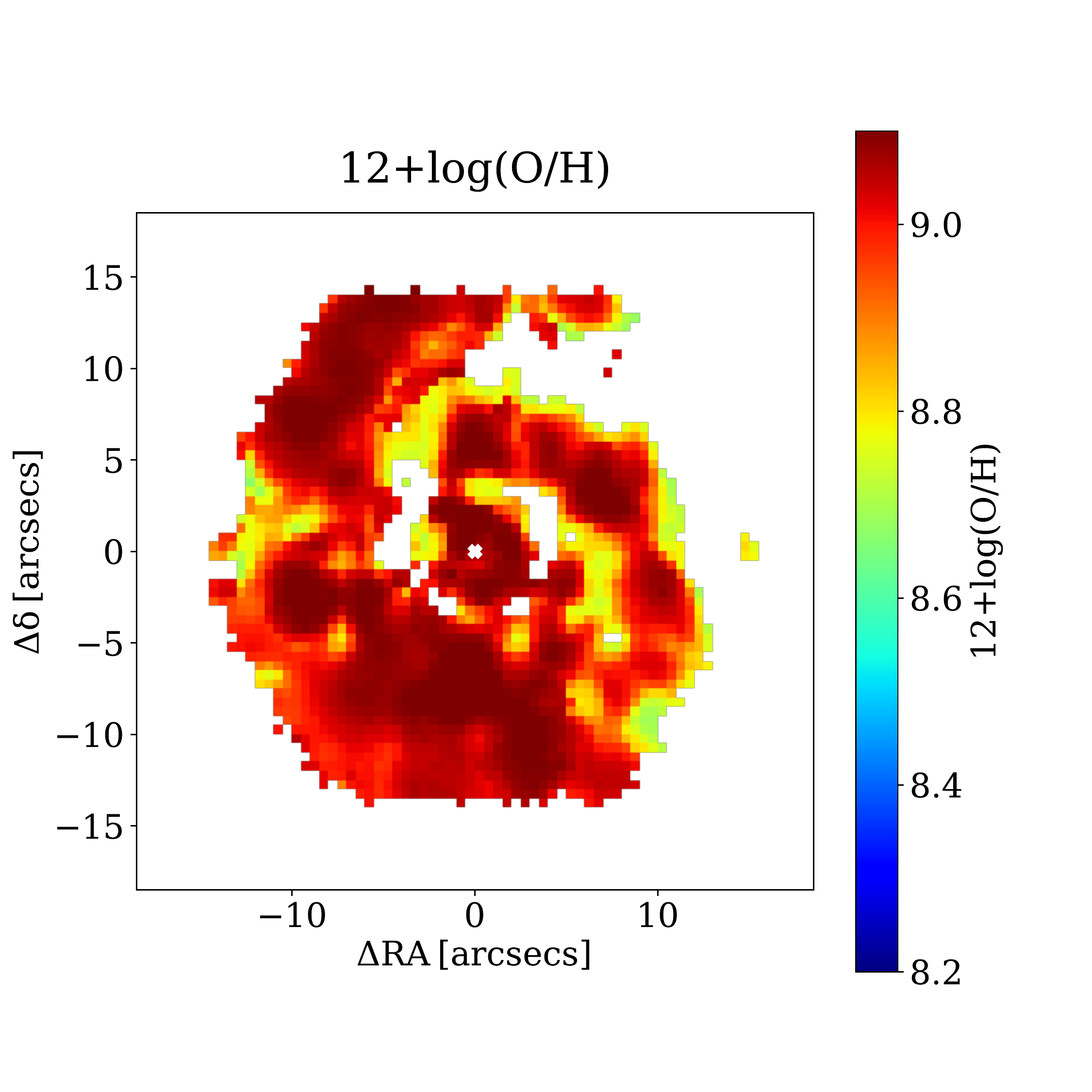}
    \end{minipage} 
        \begin{minipage}{1\columnwidth}
        \vspace{-0.5cm}
        \centering
        	\includegraphics[width=0.7\columnwidth]{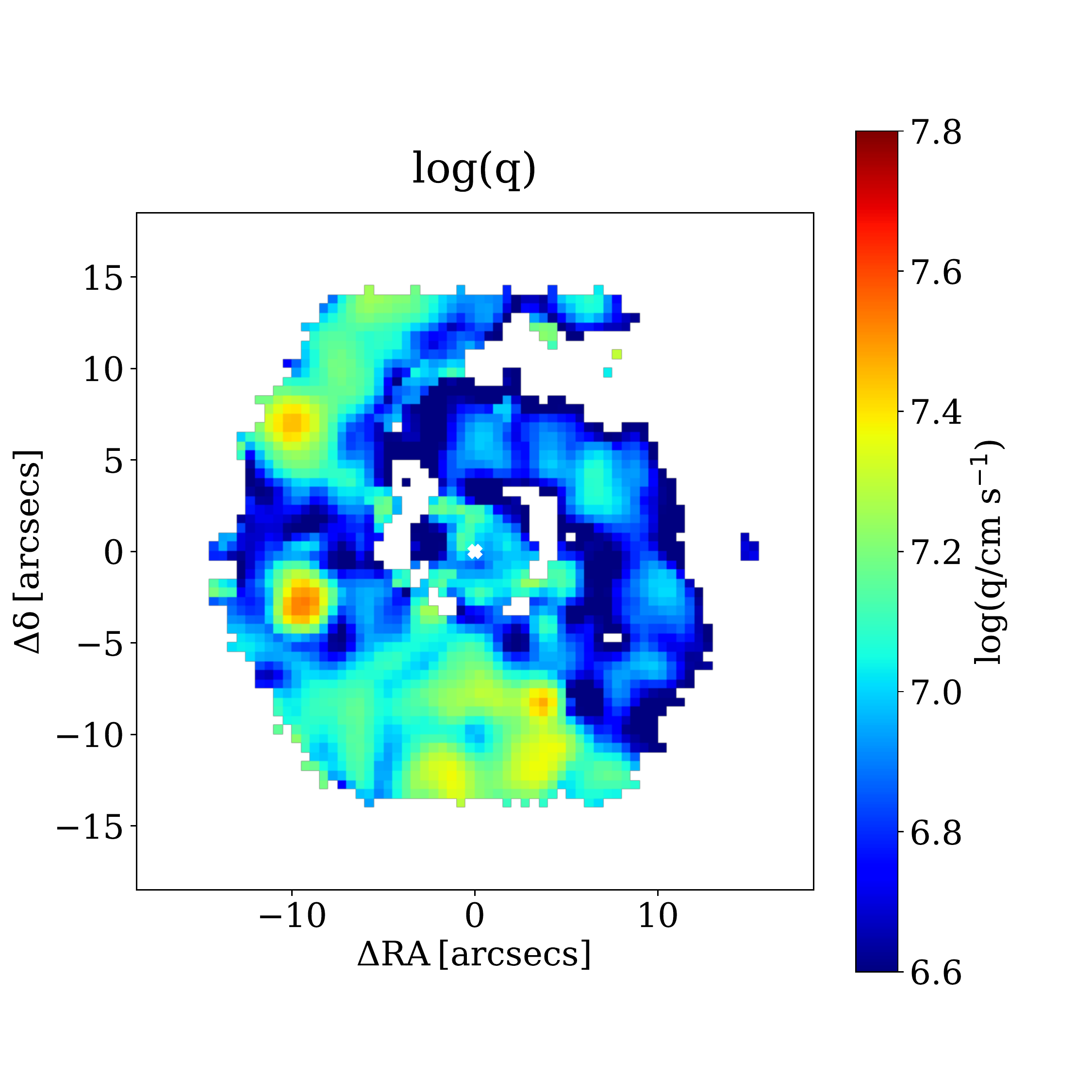}
    \end{minipage} 
            \begin{minipage}{1\columnwidth}
            \vspace{-0.5cm}
            \centering
        	\includegraphics[width=0.7\columnwidth]{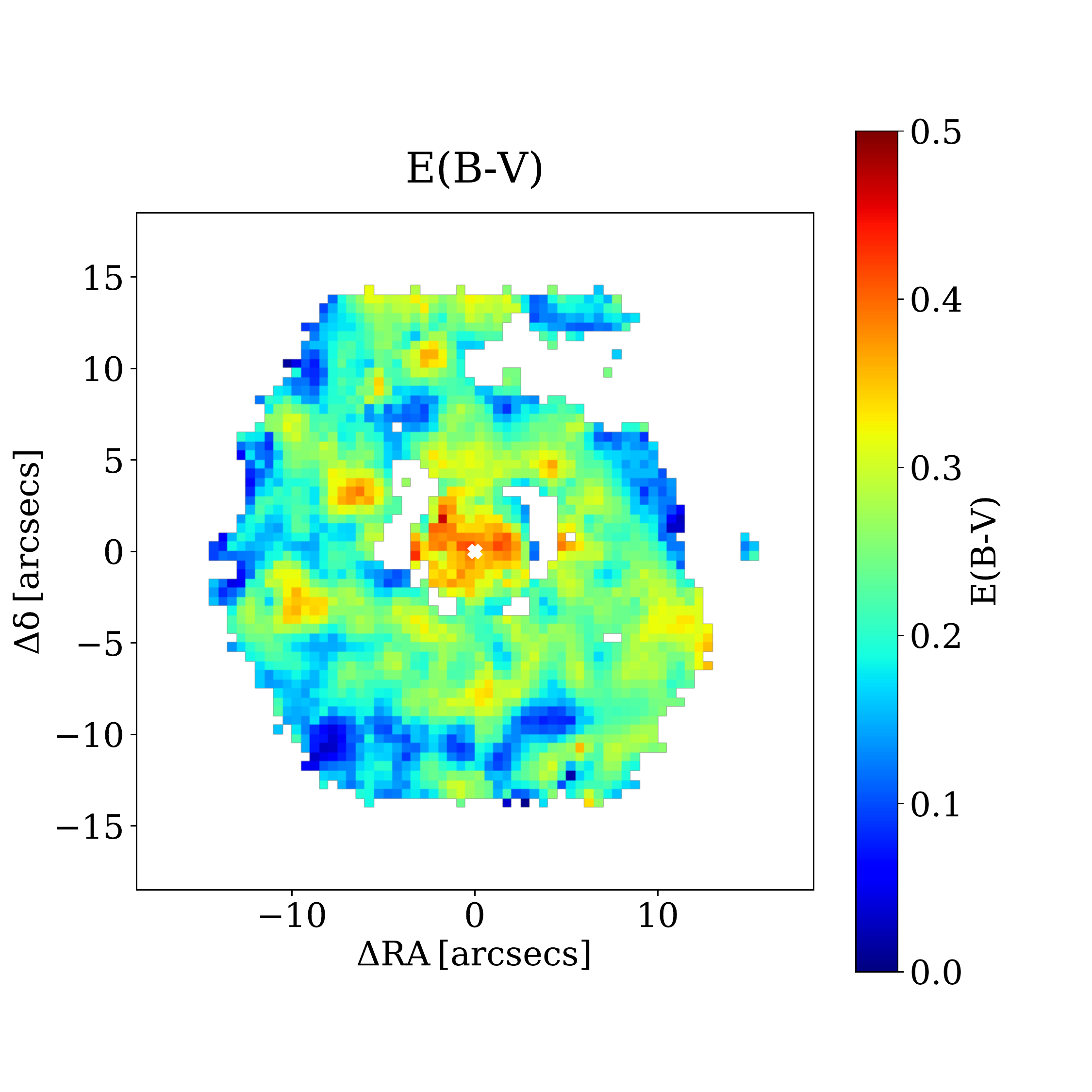}
    \end{minipage} \hfill
        \begin{minipage}{1\columnwidth}
        \vspace{-0.5cm}
        \centering
    	\includegraphics[width=0.7\columnwidth]{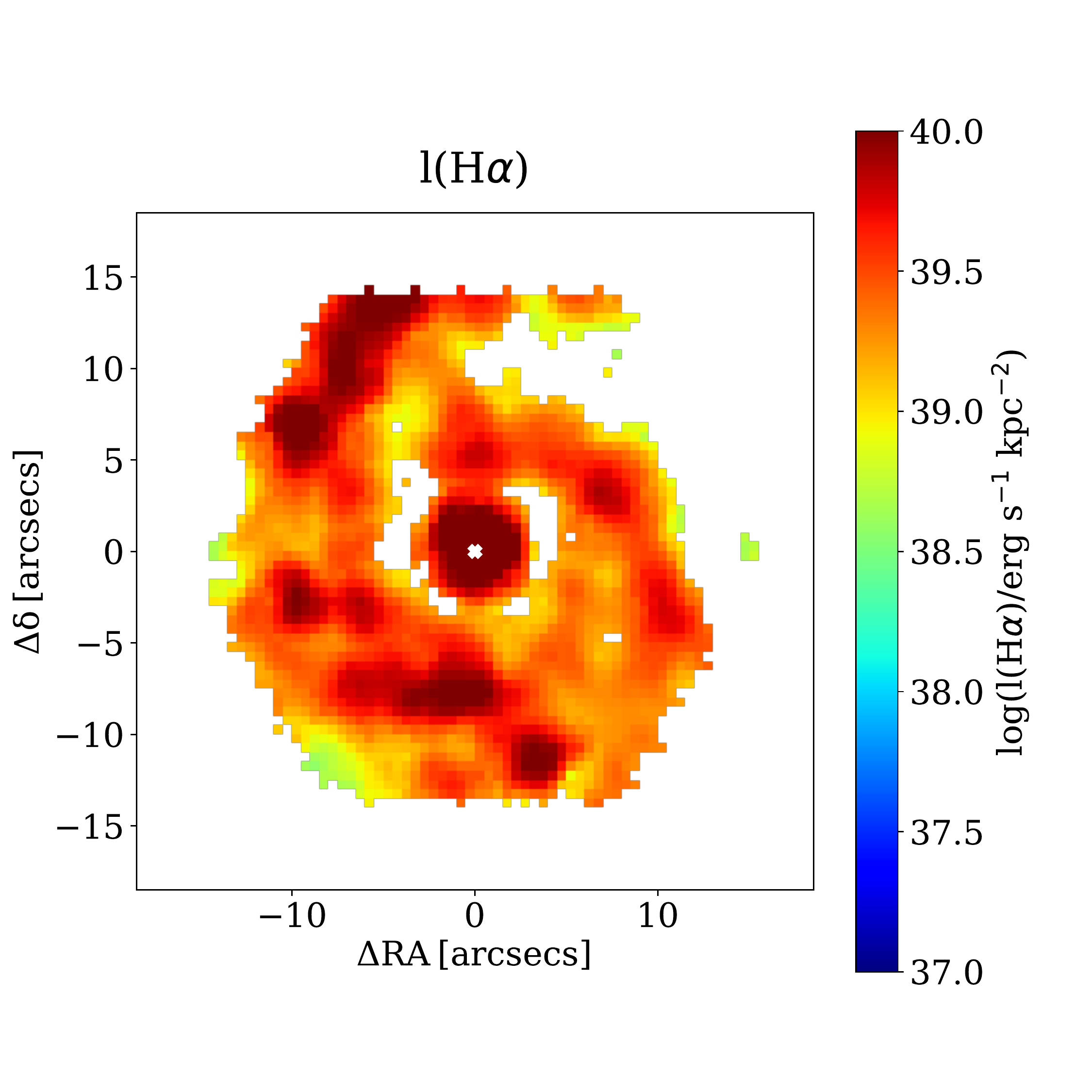}
    \end{minipage}    
    \caption{Same as Fig.~\ref{fig:lowmassgal} for the galaxy 9041-12701, with a stellar mass $\rm log(M_{\star}/M_\odot) = 10.93$.}
        	\label{fig:highmassgal}
\end{figure*}

Fig.~\ref{fig:lowmassgal} and Fig.~\ref{fig:highmassgal} show an example of a low-mass (8147-9102, $\rm log(M_{\star}/M_\odot) = 9.10$) and a high-mass (9041-12701, $\rm log(M_{\star}/M_\odot) = 10.93$) galaxy in our sample. 
The quantities displayed are: $g-r-i$ image composite from SDSS with the MaNGA hexagonal FoV overlaid, the logarithm of equivalent width of \ha\, [$EW$(\ha)], the log(S3S2) and log(O3O2) line ratios, the IZI best-fit metallicity 12+log(O/H), ionisation parameter log($q$) and gas extinction $E$(B-V), and the logarithm of the resulting luminosity of \ha \, per spaxel corrected for extinction [$l$(\ha)]. 

In the low-mass galaxy, the S3S2 line ratio map, used as a prior for the ionisation parameter, and the O3O2 line ratio map look very similar. Interestingly, both the S3S2 and O3O2 line ratios tend to trace the regions with higher values of $EW$(\ha) and $l$(\ha), and thus show a clumpy distribution.

On the other hand, in the high-mass galaxy the S3S2 and O3O2 maps look qualitatively different. The S3S2 line ratio map is clumpy with an enhancement in the eastern direction, while the O3O2 line ratio is higher in the center, but shows a roughly flat profile over the FoV. 
The similarity between S3S2 line ratios and $EW$(\ha) persists, while there is no clear correlation between S3S2 and $l$(\ha).
Indeed, the central regions in which $l$(\ha) is enhanced are characterised by lower values of S3S2. 
The discrepancy between S3S2 and O3O2 in the high mass galaxy, could be due to the strong dependence of O3O2\, on metallicity (e.g. Fig.~1, \citealt{kewley2002}), as already mentioned in Sec.~\ref{sec:intro}. 
Specifically, in the low-mass galaxy the highest metallicities (up to 12+log(O/H)~$\sim8.8$) are found in several clumps all around the field of view. The high-mass galaxy, in contrast, displays higher metallicities (with values up to 12+log(O/H)~$\sim9.1$) and an almost regular metallicity gradient, with some deviations only in the inter-arm regions. In this high-metallicity regime the O3O2 line ratio is only marginally sensitive to log($q$) \citep{kewley2002}, and thus the higher values of the O3O2 line ratio observed in the centre can be explained by the metallicity enhancement. 

Interestingly, in these two galaxies the ionisation parameter does not show a monotonic and smooth radial profile, but, similarly to the S3S2 line ratio map, is enhanced in `structures' which follow the $EW$(\ha) maps, concentrated along the spiral arms visible in the $g-r-i$ images. 

Finally, the gas extinction is low across the entire FoV of the lower-mass galaxy, with values below $E$(B-V)~$\sim 0.1$, but shows a radial gradient in the high-mass galaxy, with values up to $E$(B-V)~$\sim 0.5$ in the central regions.

\subsection{The shape of the metallicity gradients}
In the following, we analyse the radial profiles and gradients of metallicity, ionisation parameter and gas extinction for all the 1795 galaxies, keeping in mind that some of these quantities may have trends not well reproduced by a simple radial average,
as shown in the example maps.

To determine the radial gradients, we used the deprojected distance of each spaxel derived by \citet{belfiore2017a}, taking into account the inclination from the measured semi-axis ratio, assuming a constant oblateness of $0.13$. Following the approach of \citet{sanchez2014}, \citet{ho2015} and \citet{belfiore2017a}, we adopt as a scale length to normalise the gradients the elliptical Petrosian effective radius (henceforth \re), which is the most robust measure of the photometric properties of MaNGA galaxies provided by the NSA catalogue.
The median radial profile of each quantity is computed for every galaxy in the range $0.5-2$~\re\, (in bins of 0.2~dex) as the median of the quantity measured in the spaxels lying in each bin. Then the median profiles are divided in eight bins of stellar mass, in the range $\rm log(M_{\star}/M_\odot)=9-11$ (in bins of 0.25~dex). 
The radial range is chosen to minimise the effects of inclination and beam-smearing on the metallicity gradients \citep{belfiore2017a}, and because of significant deviations from a linear fit are observed out of this range \citep{sanchez2014,sanchez-menguiano2016}.
The shaded regions in the following figures represent the $16^{\mathrm{th}}$ and $84^{\mathrm{th}}$ percentiles of the distribution in each stellar mass bin, divided by $\sqrt{N}$, where N is the number of galaxies lying in the bin. For each mass bin, a profile is computed only if more than 100 galaxies have a valid measured radial profile.

\begin{figure*}
    \includegraphics[width=0.9\columnwidth]{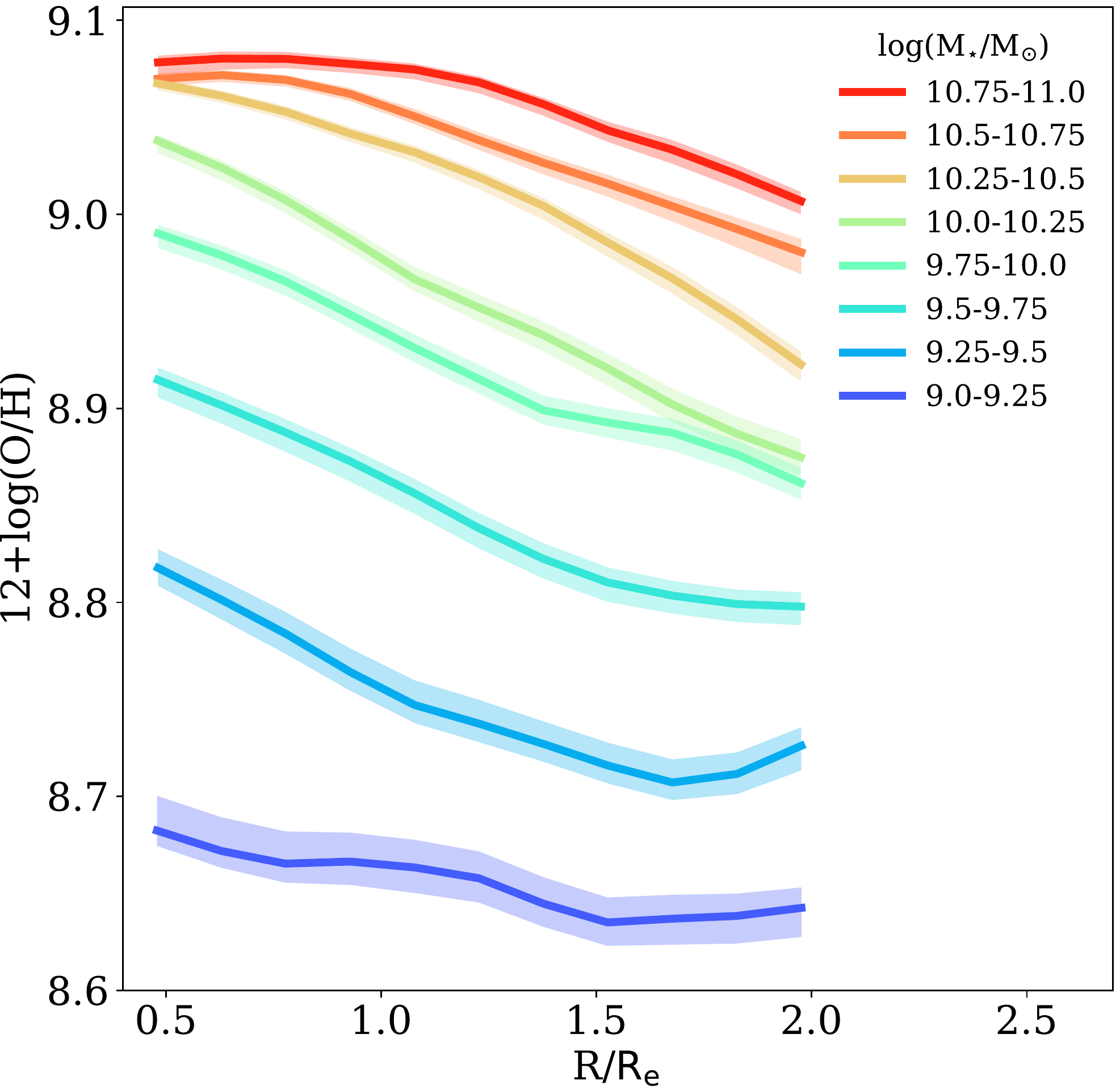}
    \includegraphics[width=1.0\columnwidth]{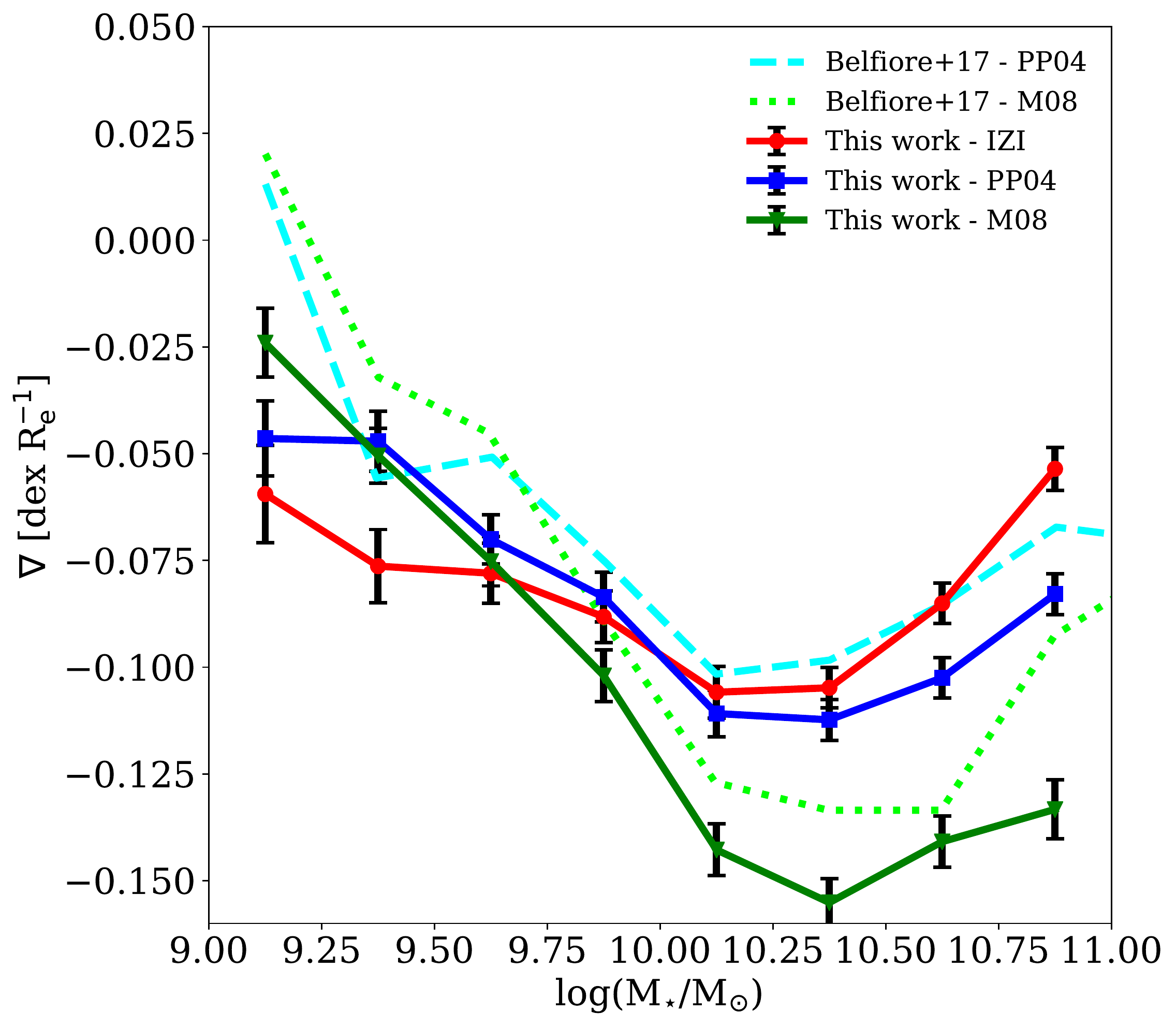}
    \caption{\textit{Left}: Oxygen abundance 12+log(O/H) as a function of the radius (in units of \re), colour-coded as a function of the stellar mass $\rm log(M_{\star}/M_\odot)$, as reported in the legend.
    The shaded regions represent the upper and lower errors of the radial gradients, obtained by calculating the $16^{\mathrm{th}}$ and $84^{\mathrm{th}}$ of the distribution in each stellar mass bin taking into account the number of galaxies lying in the bin. \textit{Right}: Metallicity gradient estimated in the radial range 0.5--2~\re\, from a linear fit of the corresponding radial profiles in different mass bins as a function of stellar mass (in red). The metallicity gradients estimated with M08 and PP04 calibrations are reported in green and blue, respectively, while those taken by \citet{belfiore2017a} are reported in lime and cyan.}
    \label{fig:metgradientizi}
\end{figure*}

Fig.~\ref{fig:metgradientizi} displays the metallicity radial profiles computed with IZI for galaxies of different stellar mass (stellar mass bins as reported in the legend). 
The main observed features are the following:
\begin{itemize}
    \item The mean oxygen abundance radial profiles are negative at all stellar masses in the range [-0.05, -0.10]~dex~\re$^{-1}$.
    \item The metallicity profile shape of galaxies with $\rm log(M_{\star}/M_\odot)>10.25$ is characterised by a flattening in the central regions ($R < 0.5$~\re);
    \item Galaxies with $\rm log(M_{\star}/M_\odot)<10$ show a mild flattening of the radial metallicity profile in the outer regions ($R>1.5$~\re).
    \item The average slope shows a weak non-linear trend with stellar mass.
\end{itemize}

The negative abundance gradients are consistent with infall models of galaxy formation, that predict that spiral discs build up through accretion of material, which leads to an inside-out growth \citep{matteucci1989, molla1996, boissier1999, belfiore2019b}. In this scenario, accretion brings gas into the inner regions of the discs, where high density leads to efficient star formation. The fast reprocessing of gas in the inner regions leads to a population of old, metal-rich stars in a high-metallicity gaseous environment. The outer regions, on the other hand, are characterised by younger, metal-poor stars surrounded by less enriched material (e.g., \citealt{dave2011, gibson2013, prantzos2000, pilkington2012}). The vertical offset between the different profiles is due to the mass-metallicity relation \citep{tremonti2004,curti2019}.

A decrease or a nearly flat distribution of the abundance in the innermost region of discs was first observed by \citet{belley1992}, while several works reported a flattening in the gradient in the outer regions (e.g. \citealt{martin1995, vilchez1996, roy1997}). This behaviour deviates from the pure inside-out scenario and could be due to the presence of radial migration (e.g. \citealt{minchev2011, minchev2012}).
Furthermore, the flattening of the metallicity gradient in the central region of the most massive spiral galaxies, found also in other works using CALIFA (e.g. \citealt{zinchenko2016}) and MaNGA data (e.g. \citealt{belfiore2017a}), can be a consequence of the metallicity saturating in the central most metal rich regions and can readily be explained by classical inside-out chemical evolution models \citep{belfiore2019b}. One should also consider the possibility that contamination from the central LIER-like emission in massive galaxies may still contaminate the measured line fluxes (e.g., \citealt{maiolino2019}).

In Fig.~\ref{fig:metgradientizi}, right panel, we show the metallicity profile slopes, estimated from a linear fit to the median profile in each mass bin, as red circles as a function of stellar mass.
We observe a slight steepening with stellar mass in the range $\rm log(M_{\star}/M_\odot)=9-10.25$, going from -0.06~dex~\re$^{-1}$ to -0.1~dex~\re$^{-1}$, and then a flattening down to -0.04~dex~\re$^{-1}$ towards $\rm log(M_{\star}/M_\odot)=11$.

These mass trends may appear in contrast with the conclusions from \citet{belfiore2017a}, who analysed gas-phase metallicity gradients using 550 star-forming MaNGA galaxies from a previous SDSS data release (data release 13). The metallicity gradients in \citet{belfiore2017a} are nearly flat for low-mass galaxies ($\rm log(M_\star/M_\odot) =9$) and become progressively steeper (more negative) for more massive galaxies until slopes $\sim-0.15$~dex~\re$^{-1}$ at $\rm log(M_{\star}/M_\odot)=10.25$. A steepening of the metallicity gradient as a function of stellar mass - though weaker than the one observed by \citet{belfiore2017a} - is also inferred by \citet{poetrodjojo2018}, using data from the SAMI survey and an iterative process based on R23 and O3O2 to estimate 12+log(O/H) and log($q$) (see \citealt{kobulnicky2004} for more details).
This steepening is not per-se in contrast with the inside-out galaxy formation scenario \citep{belfiore2019b}, but may also point towards redistribution of metals in the early stages galaxy formation (see e.g. \citealt{maiolino2019}).

While the current sample of galaxies is larger than the one considered in \citet{belfiore2017a} and we make use of a different S/N selection criterion, the main discrepancy between the current work and \citet{belfiore2017a} is the method applied to infer metallicities. Indeed, in  \citet{belfiore2017a} two strong-line ratio diagnostics were used, namely a calibration of R23 from \citet{maiolino2008} (M08) and the well-established O3N2 calibration from \citealt{pettini2004} (PP04). 
To allow for a fairer comparison with \citet{belfiore2017a}, in Fig.~\ref{fig:metgradientizi} we show the average metallicity gradients obtained applying the PP04 (blue squares) and M08 (green triangles) calibrations to the data presented in this work. The cyan dashed and the lime dotted lines are the results obtained by \citet{belfiore2017a} using PP04 and M08, respectively, and are fairly consistent with the gradients obtained with the current data set. At low stellar masses [$\rm log(M_{\star}/M_\odot)<9.75$], both the PP04 and M08 calibration find flatter gradients than IZI, in accordance to the findings in \citet{belfiore2017a}. 
It is reassuring that some of the qualitative features (e.g. the flattening in the central regions of massive galaxies) are found for different choices of calibrator. It seems, moreover, that the slope of the metallicity gradients for low-mass galaxies was found to be too shallow in \citet{belfiore2017a}. Further work may be warranted to better understand the chemical abundance distribution on the low-mass end, especially in light of recent work pointing out the diversity of metallicity gradients exhibited by low-mass star-forming galaxies \citep{bresolin2019}.

\subsection{The ionisation parameter on resolved scales}
\label{sec:ionu}
The ionisation parameter changes within galaxies are still surprisingly poorly understood, despite the importance of determining q for correctly utilising ISM diagnostics. 
\citet{dopita2006a} argued that in the local universe ionisation parameter gradually decreases with $M_{\star}$ (see also \citealt{brinchmann2008}). Some studies presented a possible correlation between ionisation parameter and star formation rate (SFR, e.g. \citealt{dopita2014, kaplan2016}), while others argue for a better correlation with specific SFR ($\rm sSFR = SFR/M_{*}$, e.g., \citealt{kewley2015,bian2016,kaasinen2018}). 
On the other hand, \citet{ho2015} investigated $\sim50$ galaxies from the CALIFA survey, finding indications of a smooth increase in the ionisation parameter from their centres to the outskirts, which they interpret as a radial change in the properties of the ionising radiation. \citet{kaplan2016}, instead, analysed the ionisation parameter distribution in 8 nearby galaxies with the VENGA survey \citep{blanc2013}, finding a peak in the central parts and regions of localised enhancements in the outer disc of their galaxies. 
Finally, \citet{poetrodjojo2018}, analysying 25 face-on star-forming spiral galaxies from the SAMI survey, found that the ionisation parameter does not have clear radial or azimuthal trends, showing a range of different distributions ranging from weak gradients, to flat or clumpy distributions.
All these authors estimated log($q$) using the O3O2 diagnostic following the iterative method proposed by \citet{kobulnicky2004}, apart from \citet{dopita2014}, who used the method developed by \citet{dopita2013}, \textsf{pyqz}, that performs a two-dimensional fit to a given diagnostic grid, estimating metallicity and ionisation parameters for a given set of diagnostics

\begin{figure*}
\begin{minipage}{1\columnwidth}
        \includegraphics[width=1\columnwidth]{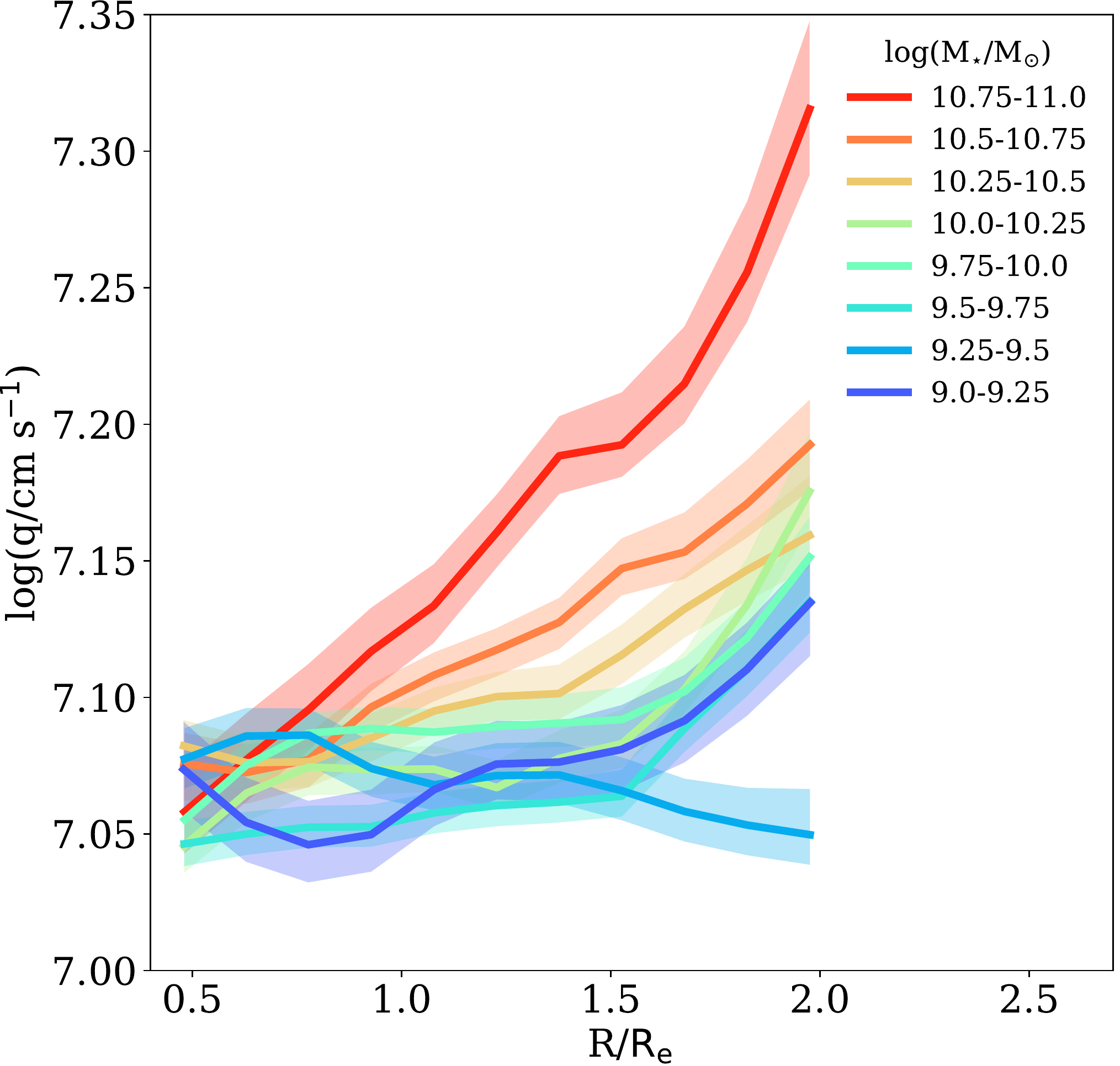}
        \end{minipage}
\begin{minipage}{1\columnwidth}
        \includegraphics[width=1\columnwidth]{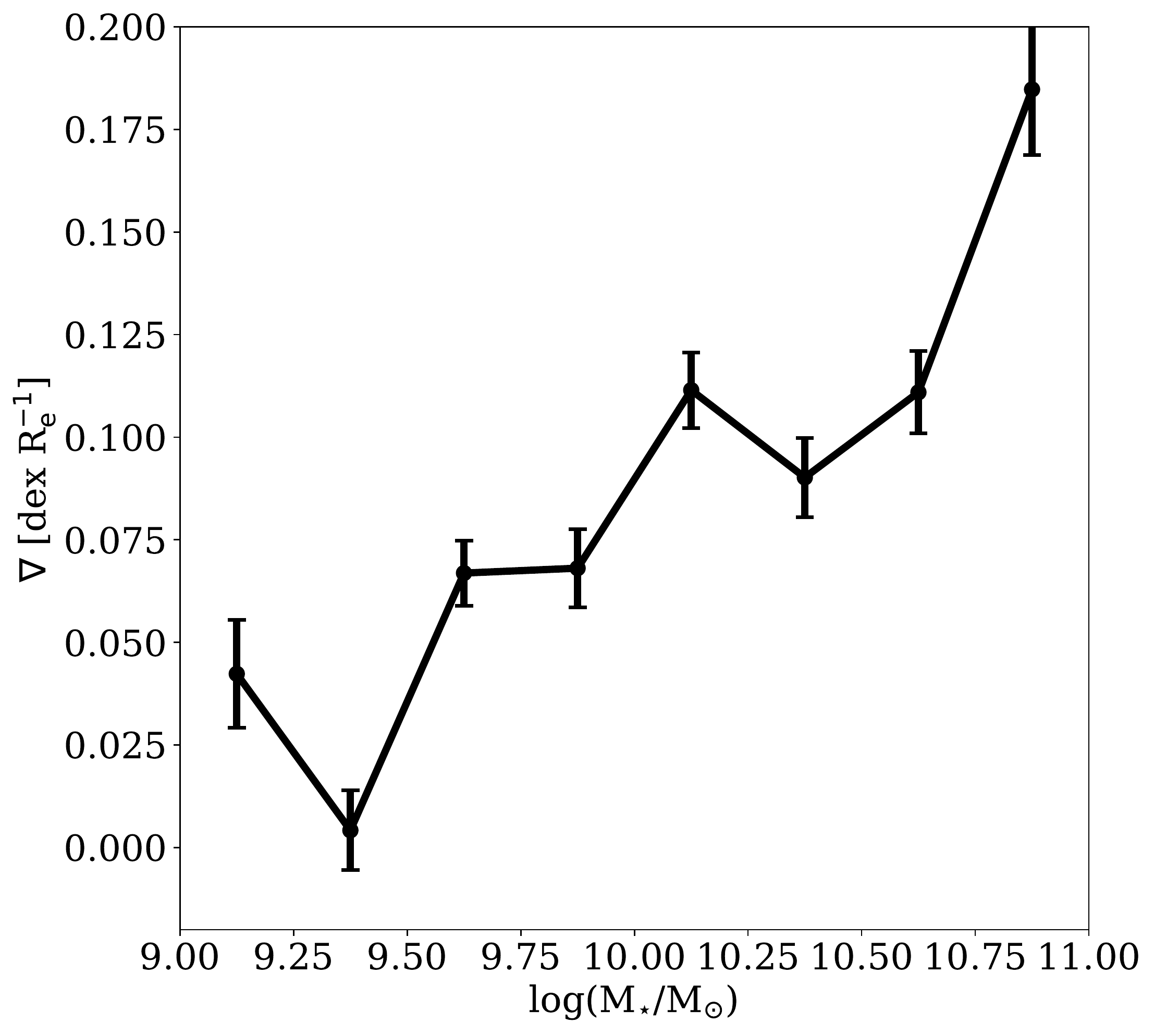}
        \end{minipage}
    \caption{\textit{Left}: Ionisation parameter radial distribution (in units of \re), colour-coded as a function of the stellar mass $\rm log(M_{\star}/M_\odot)$, as reported in the legend.
    The shaded regions represent the $16^{\mathrm{th}}$ and $84^{\mathrm{th}}$ pecentiles of the distribution in each stellar mass bin taking into account the number of galaxies lying in the bin. \textit{Right}: log($q$) radial profile slopes as a function of stellar mass in units of dex \re$^{-1}$.}
    \label{fig:qgradientizi}
\end{figure*}

The left panel of Fig.~\ref{fig:qgradientizi} shows the radial profiles of ionisation parameter estimated by IZI, colour-coded as a function of stellar mass bins as reported in the legend. The right panel shows the slope of the radial profiles as a function of stellar mass.
We highlight two main features from these figures:
\begin{itemize}
    \item the ionisation parameter radial profiles are approximately flat at low stellar masses (${\rm log}(M_{\star}/M_\odot<10$), with slopes becoming more positive and steeper at increasing $M_{\star}$, up to 0.15~dex~\re$^{-1}$ at ${\rm log}(M_{\star}/M_\odot)=11$;
    \item in the central regions, galaxies of all masses show similar values of log($q$) around ${\rm log}(q/{\rm cm\,s}^{-1})\sim 7.05-7.1$, while at large radii, higher stellar mass galaxies tend to have higher average values of log($q$) (with values up to ${\rm log}(q/{\rm cm\,s}^{-1})\sim7.3$).
\end{itemize}

The steepening with stellar mass and increasingly positive slopes of log($q$) profiles echo the $EW$(\ha) radial distributions as a function of stellar mass (see Fig.~3, \citealt{belfiore2018}). 
These are found to be flat for low mass galaxies, becoming increasingly positive towards higher stellar masses. 
As discussed in \citet{leitherer2005}, the equivalent width of the strongest hydrogen recombination lines, such as $EW$(\ha), can be very powerful age indicators, measuring the ratio of the young, ionising stars over the old, non-ionising population (see also \citealt{kewley2015,kaasinen2018}). Therefore, $EW$(\ha) is a good proxy for the sSFR, that is defined as the number of massive young (O,B) stars with respect to the total number of formed stars. 
A similarity between log($q$) and $EW$(\ha) was already suggested by analysing Fig.~\ref{fig:lowmassgal} and Fig.~\ref{fig:highmassgal}.

\begin{figure*}
       \begin{minipage}{1\columnwidth}
        	\includegraphics[width=1\columnwidth]{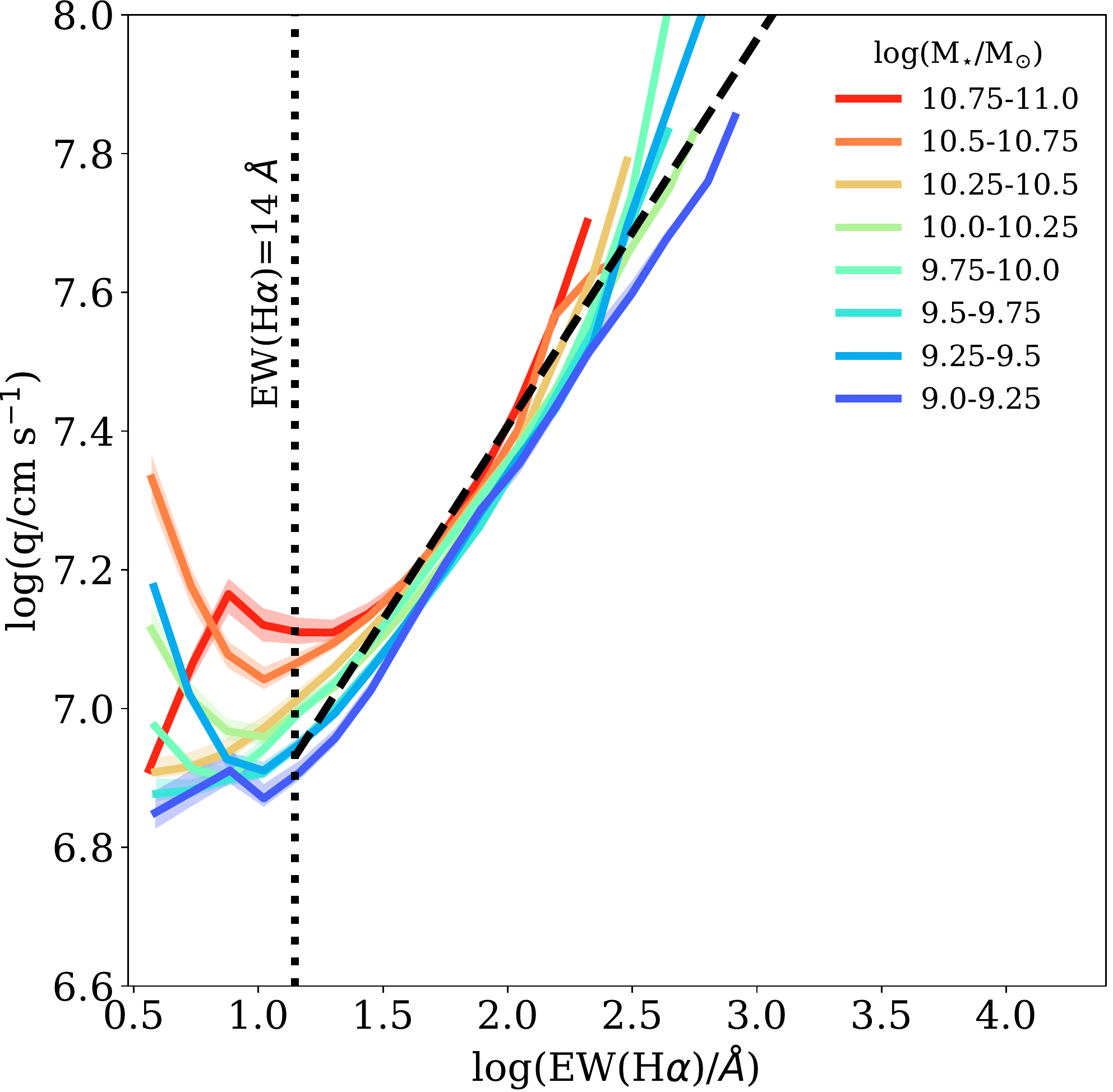}
    \end{minipage} 
        \begin{minipage}{1\columnwidth}
        	\includegraphics[width=1\columnwidth]{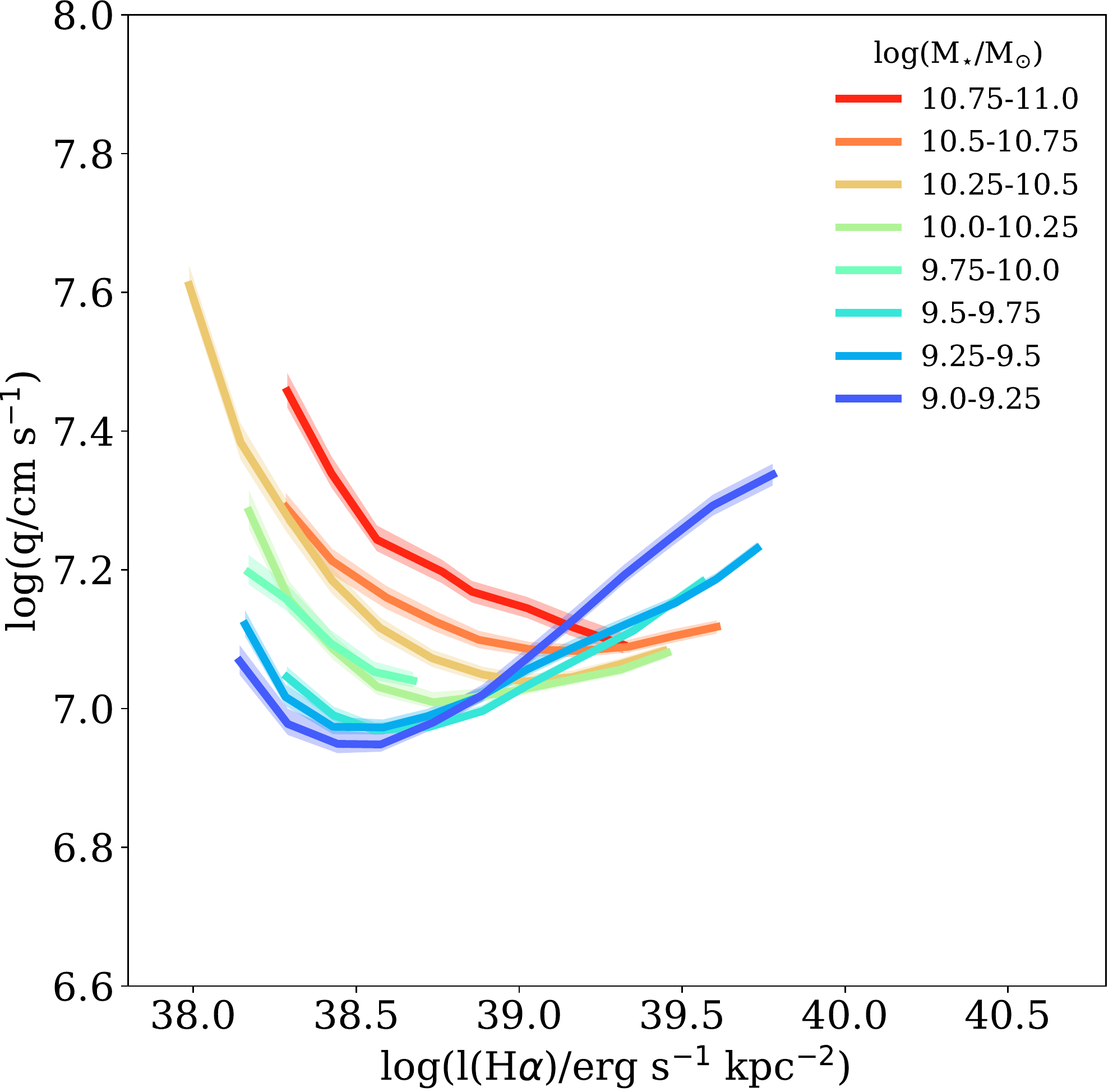}
    \end{minipage} 
        \caption{\textit{Left}: Ionisation parameter log($q$) as a function of \ha\, equivalent width [$EW$(\ha)/\AA] dividing all the spaxels used in this work in bins of 0.15 dex, and then separating the galaxies in bins of stellar mass, as reported in the legend. A nearly-universal power-law relation is evident between the two quantities for $EW$(\ha) $>$ 14 \AA\ (labelled with a dotted vertical line).
        \textit{Right}: Same as the left panel but for \ha\ luminosity per spaxel [$l$(\ha)/erg s$^{-1}$]. No universal relation is found in this case.}
        	\label{fig:ionucorr}
\end{figure*}

In light of this, we investigated the dependence of the ionisation parameter on $EW$(\ha), finding a strong correlation at all stellar masses, as shown in the left panel of Fig.~\ref{fig:ionucorr}, obtained dividing all the spaxels of each galaxy in $EW$(\ha) bins of 0.15 dex, and then separating the galaxies in bins of stellar mass $M_\star/{\rm M_\odot}$, as reported in the legend.
For $EW$(\ha)~$>14$~\AA \,(above the threshold of a possible DIG contamination; \citealt{lacerda2018}), log($q$) increases with $EW$(\ha) following a nearly universal relation for galaxies of different masses. 
We quantify this relation with a linear slope in the log-log plane as 
\begin{equation}
{\rm log}(q) = 0.56\times {\rm log}(EW(H\alpha)) +6.29    
\end{equation}
with a scatter of 0.12 dex (dashed line in Fig.~\ref{fig:ionucorr}).
This relation could be useful to constrain log($q$) in order to calculate metallicity from a limited set of emission lines (e.g. \nii\, and \ha \,in high-$z$ galaxies).
We also test the presence of a correlation between log($q$) and $l$(\ha). $l$(\ha) traces the number of ionising photons produced by young and massive stars, and thus the current star formation on timescales of $\sim10$~Myr \citep{calzetti2005,calzetti2012}.
The right panel of Fig.~\ref{fig:ionucorr} shows the ionisation parameter as a function of $l$(\ha), dividing all the spaxels in this work in $l$(\ha) bins of 0.15 dex, and then separating the galaxies in bins of stellar mass $M_\star/{\rm M_\odot}$.
No universal relation is found between log($q$) and $l$(\ha). Indeed, log($q$) increases with $l$(\ha) at the lowest stellar masses ($M_\star/{\rm M_\odot}=9$), while it decreases at the highest stellar masses ($M_\star/{\rm M_\odot}=11$).

Interestingly, \citet{pellegrini2019} found a good correlation between ionisation parameter and the age of the stellar population (i.e. with $EW$(\ha)), that they explain in terms of a correlation between q and the hardness of the spectrum, without finding any correlation between q and SFR, in agreement with our findings.

We note that, as evidenced in Fig.~\ref{fig:qgradientizi}, the majority of spaxels have log($q$) in the range log($q$/cm~s$^{-1}) = 7.05-7.15$, while the correlation with $EW$(\ha) observed in in Fig.~\ref{fig:ionucorr} only starts to appear for all masses for log($q$)~$ > 7.2$.
This indicates that the correlation is driven by spaxels corresponding to the brightest \hii\ regions.
The flat log($q$) radial profiles are the result of averaging a very clumpy distribution in log($q$) with a large scatter, as highlighted in Fig.~\ref{fig:lowmassgal} and Fig.~\ref{fig:highmassgal}.
Therefore, the radial averages used here to derive radial trends might not be ideal to understand the variations of $q$ across the galaxy discs, that seem to be dominated by sSFR variations traced by $EW$(\ha).

\begin{figure*}
\label{fig:obsgradientss}
    \begin{minipage}{1\columnwidth}
        	\includegraphics[width=1.\columnwidth]{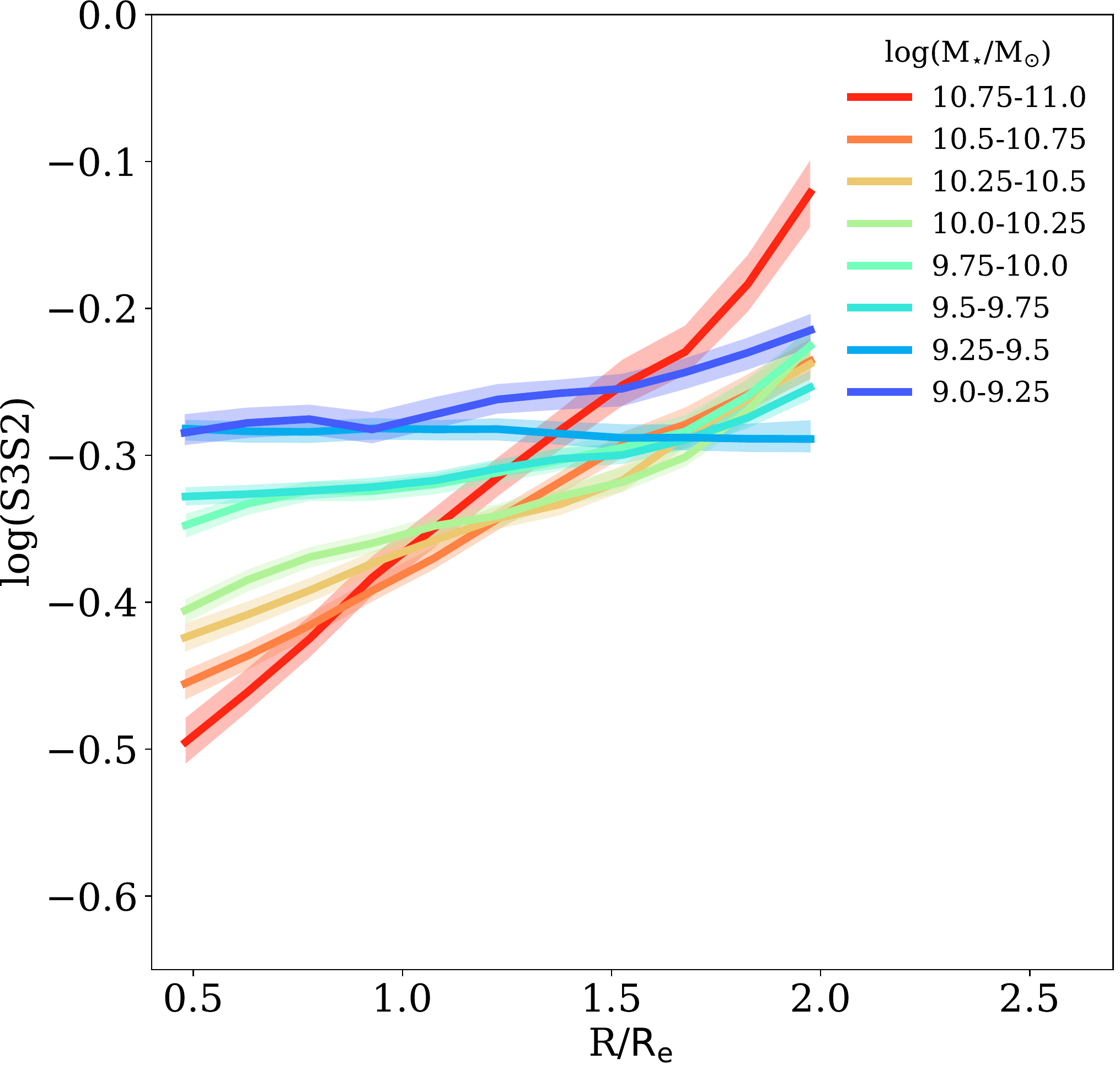}
    \end{minipage} 
        \begin{minipage}{1\columnwidth}
        	\includegraphics[width=1.\columnwidth]{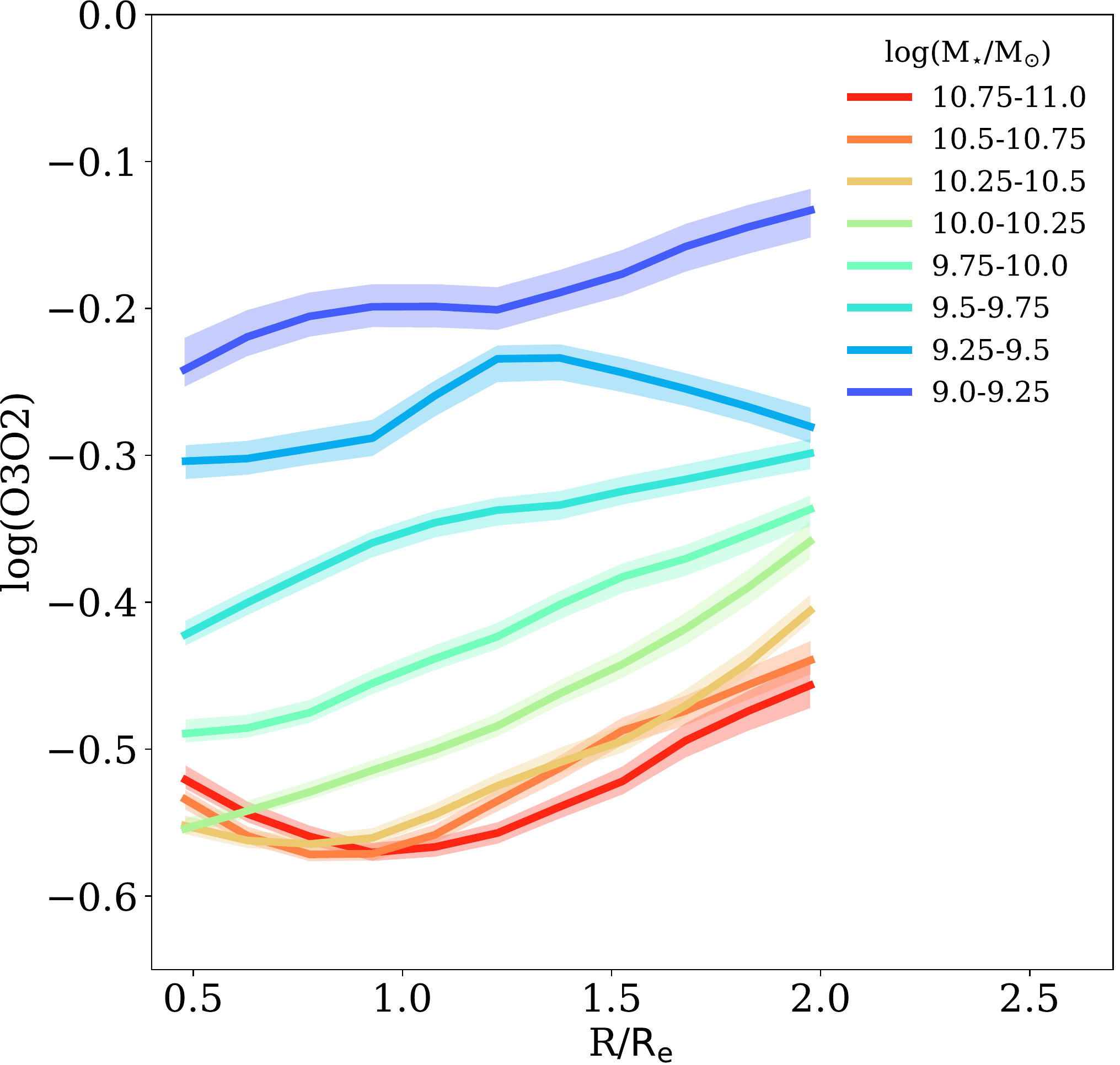}
    \end{minipage}
    \caption{S3S2\, and O3O2\, radial distributions, in bins of stellar mass $\rm M_\star/M_\odot$, as reported in the legend, respectively.}
\end{figure*}

\begin{figure}
        	\includegraphics[width=1.\columnwidth]{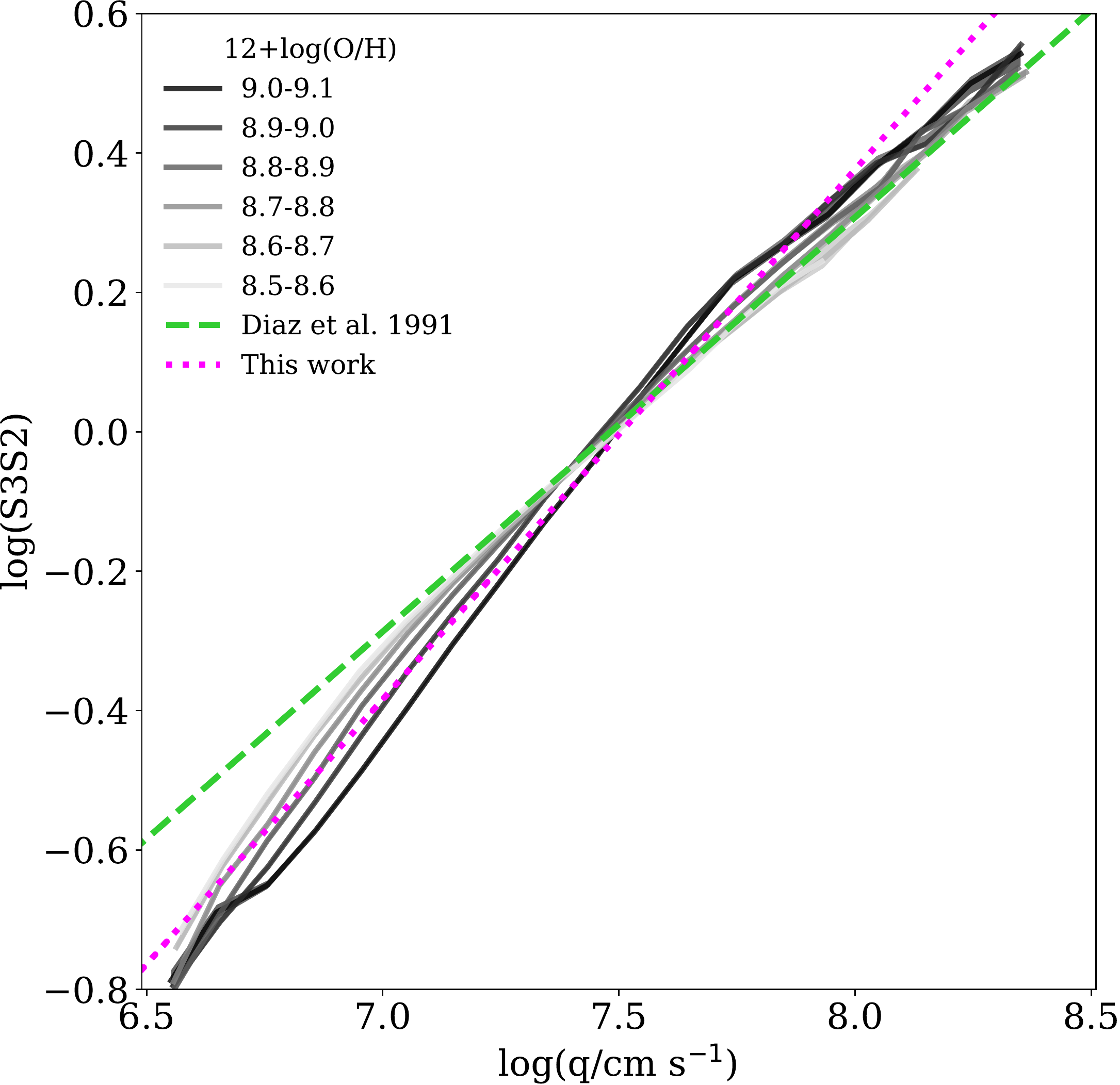}
    \caption{Distribution of log(S3S2) as a function of log($q$) inferred with IZI is shown in shades of grey, while its linear fit is given by the dotted magenta line. The green dashed line illustrates the D91 calibration, used as a prior.}
    \label{fig:s3s2q}
\end{figure}

The left and right panels of Fig.~\ref{fig:obsgradientss} show log(S3S2) and log(O3O2) radial distributions in stellar mass bins, derived analogously to Fig.~\ref{fig:qgradientizi}. 
These line ratios are both proxies for the ionisation parameter, and consistently show that low-mass galaxies are characterised by higher values of log(S3S2) and log(O3O2) and by flatter radial profiles. High-mass galaxies, on the other hand, tend to show positive slopes.
Interestingly, we used S3S2 line ratios as a prior for log($q$), so the log($q$) radial distribution should closely follow the same trend. However, comparing the right panels of Fig.~\ref{fig:qgradientizi} and Fig.~\ref{fig:obsgradientss} the log($q$) distribution obtained with IZI shows higher values for higher mass galaxies.
On the one hand, this ``flip" with stellar masses could be due to the residual dependence of these line ratios on metallicity.
Up to log($q$/cm~s$^{-1}$)~$=7.5$ (where the majority of spaxels lies) the log($q$) values retrieved by IZI are slightly lower with respect to the prediction of Eq.~\ref{eq:diaz}, and the discrepancy depends on metallicity. Interestingly, this happens even though the input prior is not metallicity-dependent. IZI, however, will predict a metallicity dependence of the relation between log(S3S2) and ionisation parameter as a consequence of the information provided by the other emission lines.
Above log(q/cm~s$^{-1}$)~$=7.5$, log($q$) retrieved by IZI is fairly consistent with D91. 
This leads to a ``re-calibrated'' relation between log(S3S2) and log($q$), highlighted by the magenta dotted line in Fig.~\ref{fig:s3s2q} and given by
\begin{equation}\label{eq:recal}
    \rm log(S3S2) = 0.76 \times log(q) -5.70,
\end{equation}
with a scatter of 0.11 dex, obtained using all the selected spaxels.
On the other hand, this residual dependence of log(S3S2) on metallicity after adding a prior on the ionisation parameter could also point to the fact that IZI still suffers from a certain degree of degeneracy between the 12+log(O/H) and log($q$).
Another possibility is that the difference between the \siii/\sii\, derived value for $q$ and the IZI output could be due to residual problems of the photoionisation models used in reproducing the observed line ratios. Therefore, a new generation of photoionisation models, designed to reproduce the sulfur emission lines as well, is required to better asses this issue.

\begin{figure}
 \centering
	\includegraphics[width=1\columnwidth]{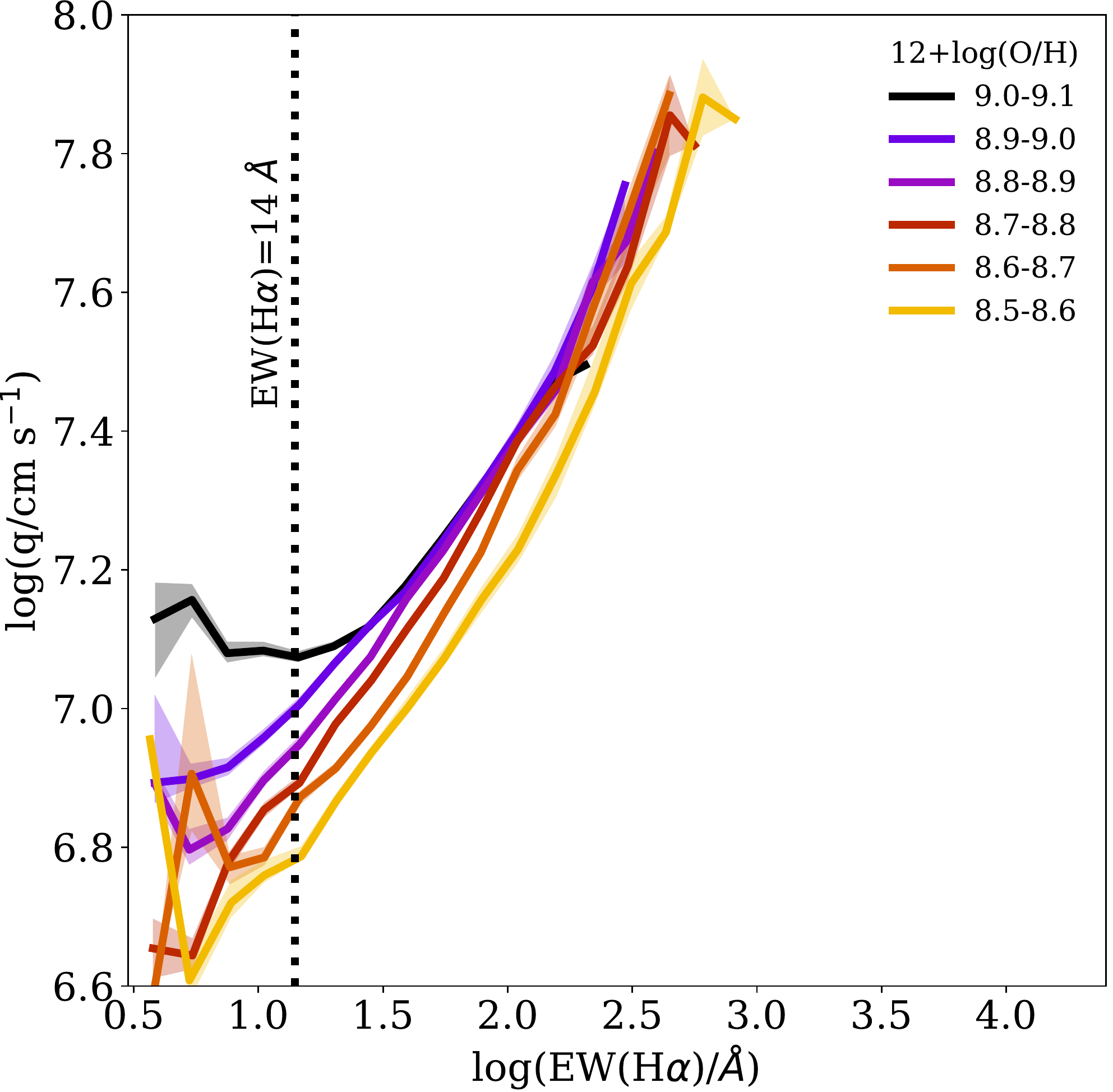}
    \caption{Ionisation parameter log($q$) as a function of log($EW$(\ha)) dividing all the spaxels used in this work in bins of 0.15 dex, and then separating the galaxies in bins of metallicity, as reported in the legend.}
    \label{fig:qmetrelation}
\end{figure}

Finally, Fig.~\ref{fig:qmetrelation} shows the relation between $EW$(\ha) and ionisation parameter in bins of metallicity, obtained by IZI, indicating that at fixed $EW$(\ha) there is a clear correlation between log($q$) and 12+log(O/H). However, this correlation disappears going towards the highest value of $EW$(\ha) ($EW$(\ha)>150~\AA).
A correlation between the two quantities is in contrast with the theoretical relation presented in \citet{dopita1986} and \citet{dopita2006a} [log($q$/cm~s$^{-1}$)~$\propto$~12+log(O/H)$^{-0.8}$].
However, \citet{ho2015}, \citet{kaplan2016} and \citet{poetrodjojo2018} do not find any clear radial trend between ionisation parameter and gas metallicity, both when considering regions at different galactocentric distances and in the 2-D maps, while \citet{dopita2014} found a positive correlation between the two quantities. 
\citet{dopita2014} explained the positive correlation between 12+log(O/H) and log($q$) as a consequence of the positive correlation that they find between SFR density and ionisation parameter. They conclude that the correlation between SFR density and log($q$) is mainly caused by geometrical effects (i.e. overlapping between \hii\, regions or non-spherical geometries), but is also due to the presence of dense gas in the vicinity of \hii\, regions. On the other hand, \citet{cresci2017} investigated the physical properties of the ionised gas in the prototypical \hii\, galaxy He~2-10, finding that the central extreme star forming knots are highly enriched with super solar metallicity and characterised by a large ionisation parameter, highlighting again a correlation between the two quantities.

\subsection{Gas extinction}
\label{sec:ebv}
\begin{figure*}
        \includegraphics[width=1\columnwidth]{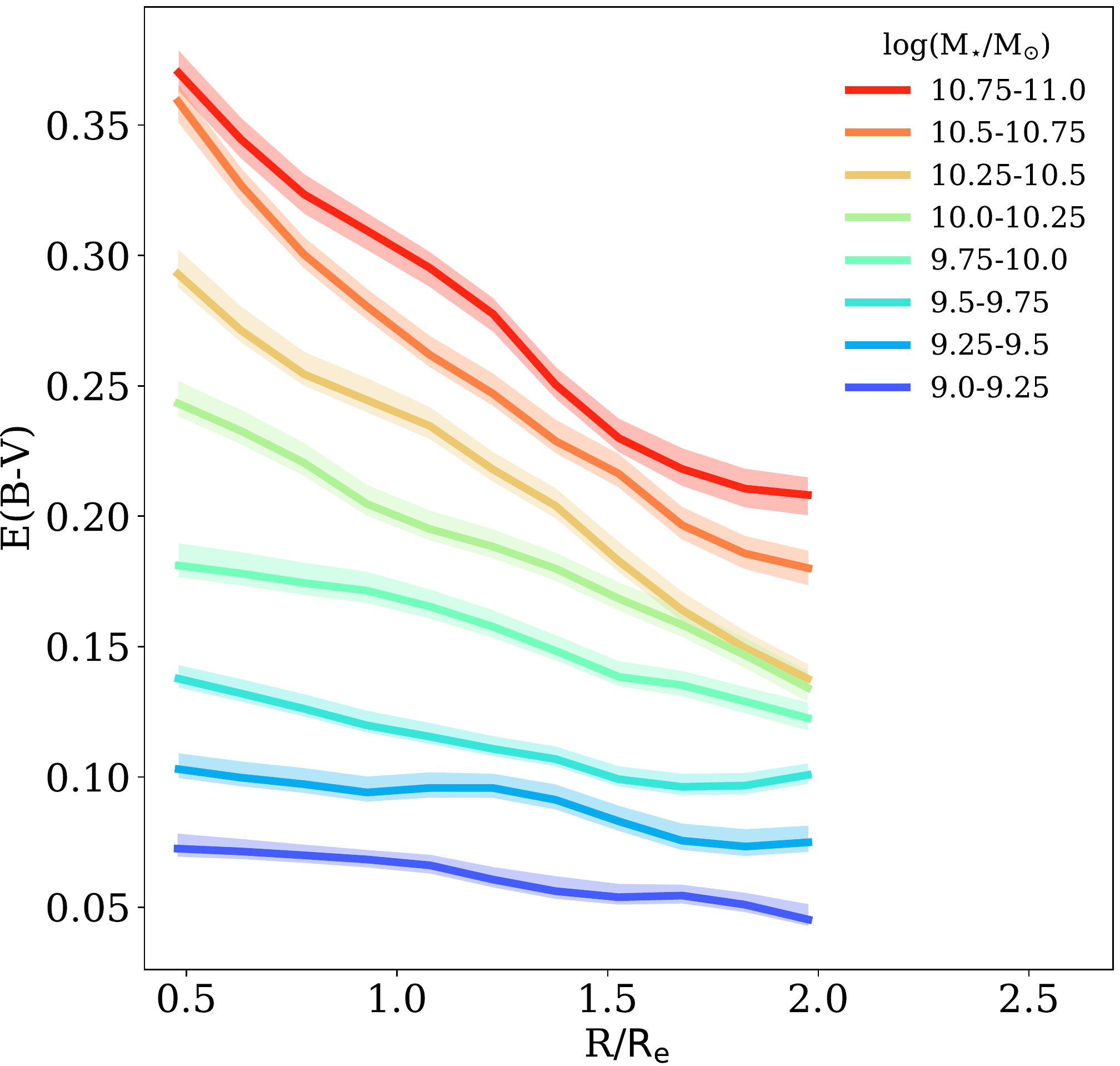}
        \includegraphics[width=1\columnwidth]{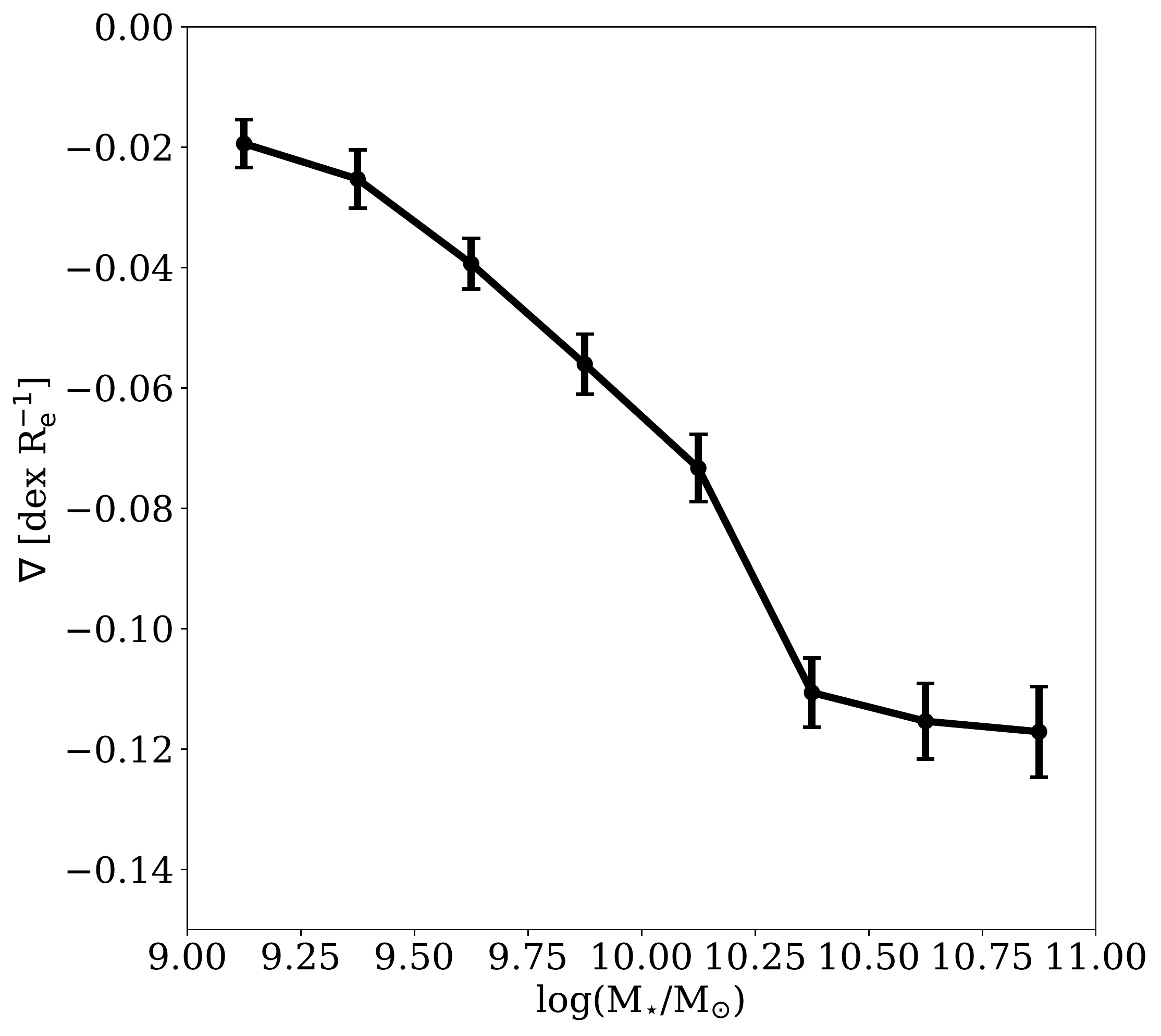}
    \caption{\textit{Left}: Gas extinction radial distribution (in units of \re), colour-coded as a function of the stellar mass $\rm log(M_{\star}/M_\odot$, as reported in the legend. The shaded regions represent the $16^{\mathrm{th}}$ and $84^{\mathrm{th}}$ percentiles of the distribution in each stellar mass bin taking into account the number of galaxies lying in the bin. \textit{Right}: Slope of the radial profiles of $E$(B-V) in units of mag/\re\ as a function of stellar mass.}
    \label{fig:ebvgradientizi}
\end{figure*}

The left panel of Fig.~\ref{fig:ebvgradientizi} illustrates the radial profiles of gas extinction $E$(B-V) estimated by IZI, colour-coded as a function of stellar mass bins as reported in the legend, while the right panel shows the radial profile as a function of stellar mass. We report on these profiles only briefly in this work, since they will be the subject of future work by our team.
Here we only highlight two main features of these figures:
\begin{itemize}
    \item the $E$(B-V) radial profiles show a strong dependence on stellar mass, with slopes around $-0.02$~dex~\re$^{-1}$ at $\rm log(M_{\star}/M_\odot)=9$ and increasingly negative and steeper slopes in the range up to $\sim-0.13$~dex~\re$^{-1}$ at larger stellar masses;
    \item $E$(B-V)~$\sim0.07$ for the lowest mass galaxies and reaches values as high as $E$(B-V)~$\sim0.4$ in the central regions of the more massive galaxies.
\end{itemize}

Indeed our findings in this section are not surprising, since more massive galaxies usually have larger dust reservoirs, especially in the central regions \citep{bell2000}. 
Since dust is formed from metals, a correlation between dust and the gas-phase oxygen abundance is expected and observed both in the local universe (e.g., \citealt{heckman1998, zahid2013}) and at high redshift (e.g., \citealt{reddy2010}). 
A follow-up study of the relation between gas extinction and other galaxy properties will be presented in a forthcoming paper.

\begin{figure}
 \centering
	\includegraphics[width=0.5\textwidth]{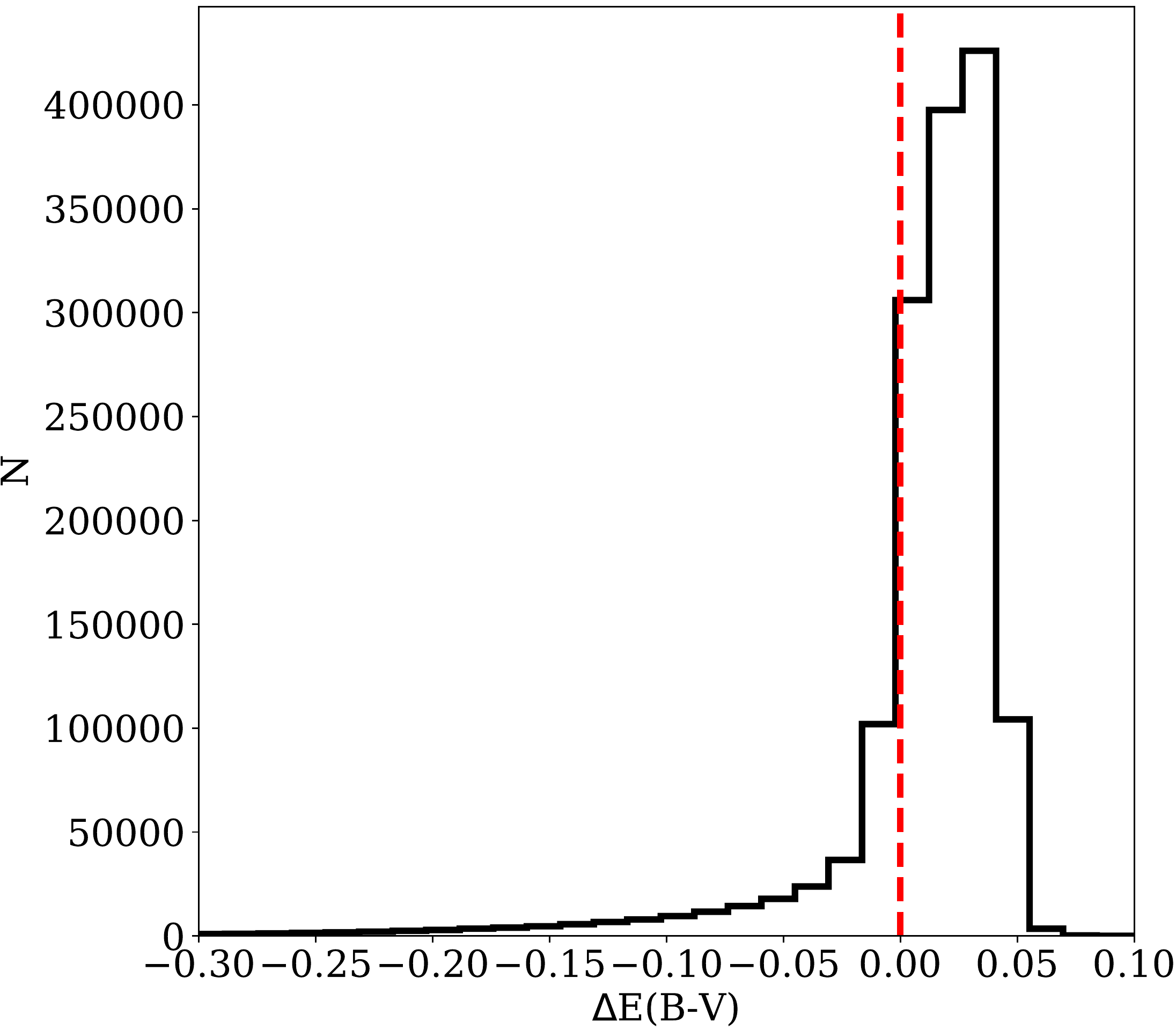}
    \caption{Histogram of the difference between the $E$(B-V) obtained by assuming a fixed \ha/\ha\, line ratio of 2.86 and the one retrieved by IZI.}
    \label{fig:ebvizi_vs_standard}
\end{figure}

Fig.~\ref{fig:ebvizi_vs_standard} shows the histogram of the difference between gas extinction obtained by assuming a fixed \ha/\hb\, line ratio of 2.86 and the one retrieved by IZI [$\Delta E$(B-V)]. The two quantities are pretty consistent, but in general IZI tends to retrieve slightly lower $E$(B-V) values, in agreement with \citet{brinchmann2004}. 
 
\section{Conclusions}
\label{sec:conclusion}
In this work we investigated the gas phase metallicity, ionisation parameter and gas extinction for a sample of 1795 local star forming galaxies, spanning the stellar mass range $10^9-10^{11}$~M$_\odot$, by exploiting integral field spectroscopy from the SDSS-IV MaNGA DR~15.
We selected star forming galaxies following the classification scheme proposed in \citet{belfiore2016}, further excluding spaxels which are not classified as star forming according to both the \nii- and the \sii-BPT diagrams. 
Moreover, we excluded spaxels with S/N(\ha)~$<15$, which assures that the main optical emission lines (i.e. \oii$\lambda\lambda$3726,29, \hb, \oiii$\lambda\lambda$4959,5007, \nii$\lambda\lambda$6548,84, \ha, \sii$\lambda$6717, \sii$\lambda$6731 and \siii$\lambda\lambda$9069,9532) are generally detected with $S/N>1.5$ without introducing metallicity biases.
We characterise self-consistently the gradients of metallicity, ionisation parameter and gas extinction with a method that consists in an update of the software tool IZI \citep{blanc2015}, which compares an arbitrary set of observed emission lines with photoionisation model grids.
Our revised version takes as input observed (as opposed to de-reddened) fluxes and simultaneously estimates the dust extinction ($E$(B-V)).

In the following we summarise our main findings.
\begin{itemize}
    \item We confirm the existence of a discrepancy between models and observations of the \siii\ lines already reported in the literature that persists with observations of a large sample of local galaxies and latest-generation photoionisation models (based on both the \textsc{mappings} and \textsc{cloudy} codes) (Fig.~\ref{fig:sulfurproblems}).
    \item We argue that the set of emission lines comprising \oii$\lambda\lambda$3726,29, \hb, \oiii$\lambda\lambda$4959,5007, \nii$\lambda\lambda$6548,84, \ha, \sii$\lambda$6717 and \sii$\lambda$6731 is not sufficient to break the degeneracy between metallicity and ionisation parameter with the current photoionisation models. We therefore used the \siii\ lines to add extra information to the fit performed with our revised version of IZI, taking into account a Gaussian prior based on the D91 calibration, that links S3S2 to $q$ (Sec.~\ref{sec:priorvsnoprior}). 
    \item The oxygen abundance radial profiles (in the range 0.5--2~\re) have negative slopes [$\rm -0.1\,dex\,R_e^{-1}$ to  $\rm -0.04\,dex\,R_e^{-1}$], and show a flattening in more massive systems.
    Galaxies with $\rm log(M_{\star}/M_\odot)>10.25$ show a flat gradient in the central regions ($R\sim0.5$~\re), while galaxies with $\rm log(M_{\star}/M_\odot)<10$ tend to have a flatter gradients in the external regions ($R>1.5$~\re) (Fig.~\ref{fig:metgradientizi}).
    \item The ionisation parameter gradients are approximately flat at low stellar masses ($\rm log(M_{\star}/M_\odot)<10$), and tend to steepen (more positive slopes) at increasing $\rm M_{\star}$. In galaxy outskirts, higher stellar mass galaxies tend to have higher average values of log($q$) than less massive objects. All galaxies, however, show a median value around log($q$)~$\sim 7.05-7.1$ in the central regions (Fig.~\ref{fig:qgradientizi}).
\item A tight correlation between log($EW$(\ha)) and log($q$) is observed at all stellar masses, expressed by a simple power-law relation for $EW$(\ha)~$>14$~\AA. A clear correlation between $l$(\ha) (i.e. SFR) and log($q$) is only found in low-mass galaxies, but this correlation does not hold for the whole galaxy sample (Fig.~\ref{fig:ionucorr}). 
    \item A correlation between metallicity and ionisation parameter is found at fixed $EW$(\ha) up to $EW$(\ha)~$<150$ (Fig.~\ref{fig:qmetrelation}). 
\item The gas extinction radial gradients strongly depend on stellar mass. $E$(B-V) slopes are approximately flat at $\rm log(M_{\star}/M_\odot)=9$ with values of $E$(B-V)~$\sim0.07$ mag. The profiles steepen towards higher stellar masses, and $E$(B-V) reaches values as high as $\sim0.4$ mag in the central regions of the more massive systems (Fig.~\ref{fig:ebvgradientizi}).
\end{itemize}

The work presented in this paper can be seen as a first step in the simultaneous study of 12+log(O/H) and $q$ within a large set of galaxies.
As such, it is affected by some limitations.
For example, at this stage we have not included in our analysis the variation of N/O ratio and the effective temperature of the stellar cluster.
Indeed, the nucleosynthetic origin of nitrogen is more complex than that of oxygen, and its nucleosynthesis is metallicity-dependent, implying a non-linear relation between O/H and N/O (see e.g. \citealt{belfiore2017a, schaefer2019}).
Furthermore, the effective temperature of the exciting stars is related to both metallicity and ionisation parameter since it increases as metallicity is lowered, but the mechanical energy flux in the stellar winds decreases towards lower metallicity, causing changes in the ionisation parameter (see e.g. \citealt{perez-montero2005, dopita2006a, dopita2006b, perez-montero2014}). Since these fundamental parameters of \hii\ regions are not independent, more physical \hii\ region models, such as WARPFIELD-EMP \citep{rahner2017,pellegrini2019}, that allows to model the time evolution of feedback in molecular clouds taking into account the physical processes regulating the emission from the clouds (e.g. stellar winds, radiation, supernovae, gravity, thermal conduction, cooling), may be helpful in reducing the size of the parameter space to be explored.

However, the main drawback of the current approach lies in the limitations of the current photoionisation models. Even though the model grids are generally able to reproduce the bulk properties of \hii\, regions in galaxies, we have shown here that S3S2 is not well reproduced by any of the state-of-the art model grids taken into account in this work. This fact severely limits the usefulness of S3S2 at present as a tracer of the ionisation parameter.

\section*{Acknowledgements}

\begin{footnotesize}
    We thank Nell Byler for sharing her \textsc{cloudy} models and for assistance with their use, and Kathryn Kreckel for sharing the PHANGS data shown in this work. We thank Sebastian Sanchez for support on the Pipe3D Value-Added Catalog for MaNGA DR15. We are grateful to Mike Dopita and Emiliy Levesque for making their photoionisation model grids publicly available. We acknowledge F. Mannucci for his precious advice and the anonymous referee for their comments and suggestions, which contributed to improve the quality of our work. MM and GC have been supported by the INAF PRIN-SKA 2017 programme 1.05.01.88.04. MB acknowledges FONDECYT regular grant 1170618. RM acknowledges ERC Advanced Grant 695671 "QUENCH" and support by the Science and Technology Facilities Council (STFC). 
    
	This work makes use of data from SDSS-IV.
	Funding for the Sloan Digital Sky Survey IV has been provided by the Alfred P. Sloan Foundation, the U.S. Department of Energy Office of Science, and the Participating Institutions. SDSS acknowledges support and resources from the Center for High-Performance Computing at the University of Utah. The SDSS web site is {\tt www.sdss.org}. 
	This research made use of Marvin, a core Python package and web framework for MaNGA data, developed by Brian Cherinka, Jos\'e S\'anchez-Gallego, and Brett Andrews \citep{Cherinka2018}.
	SDSS is managed by the Astrophysical Research Consortium for the Participating Institutions of the SDSS Collaboration including the Brazilian Participation Group, the Carnegie Institution for Science, Carnegie Mellon University, the Chilean Participation Group, the French Participation Group, Harvard-Smithsonian Center for Astrophysics, Instituto de Astrofísica de Canarias, The Johns Hopkins University, Kavli Institute for the Physics and Mathematics of the Universe (IPMU) / University of Tokyo, the Korean Participation Group, Lawrence Berkeley National Laboratory, Leibniz Institut f\"ur Astrophysik Potsdam (AIP), Max-Planck-Institut f\"ur Astronomie (MPIA Heidelberg), Max-Planck-Institut f\"ur Astrophysik (MPA Garching), Max-Planck-Institut f\"ur Extraterrestrische Physik (MPE), National Astronomical Observatories of China, New Mexico State University, New York University, University of Notre Dame, Observatório Nacional / MCTI, The Ohio State University, Pennsylvania State University, Shanghai Astronomical Observatory, United Kingdom Participation Group, Universidad Nacional Autónoma de México, University of Arizona, University of Colorado Boulder, University of Oxford, University of Portsmouth, University of Utah, University of Virginia, University of Washington, University of Wisconsin, Vanderbilt University, and Yale University.
	The MaNGA data used in this work is publicly available at {\tt http://www.sdss.org/dr15/manga/manga-data/}.
	
\end{footnotesize}

%
   \bibliographystyle{aa} 
   \bibliography{mingozzi_2019} 

\appendix

\section{Comparison with Pipe3D}
\label{app:pipe3d}

Pipe3D code \citep{sanchez2016, sanchez2018} provides a valid point of comparison to the DAP, since it performs an independent continuum subtraction and emission-line measurements. 
An overview of the data products released by these two pipelines is provided by \citet{aguado2019}. 
The output of Pipe3D code on MaNGA data is presented in the MaNGA Pipe3D value-added catalog (VAC)\footnote{\url{https://www.sdss.org/dr15/manga/manga-data/manga-pipe3d-value-added-catalog/}}.
The MaNGA VAC generated by the Pipe3D team uses MIUSCAT templates for spectral fitting, instead of the MILES models used by the DAP as reported in Sec.~\ref{sec:spectralfitting}.
Specifically, MIUSCAT is a set of simple stellar-population (SSP) models generated according to \citet{vazdekis2012}, that extend the wavelength range of MILES models to cover the range $3465-9469$~\AA.
The Pipe3D VAC therefore contains line fluxes for \siii$\lambda9069$. Intriguingly, the publicly-available VAC also contains line fluxes for the \siii$\lambda9532$, which lies outside the MIUSCAT wavelength coverage. It appears that Pipe3D in this case performs a extrapolation of the model continuum for $\sim100$~\AA\ (S. Sanchez, private communication).

\begin{figure*}
    \begin{minipage}{0.66\columnwidth}
                \includegraphics[width=1\columnwidth]{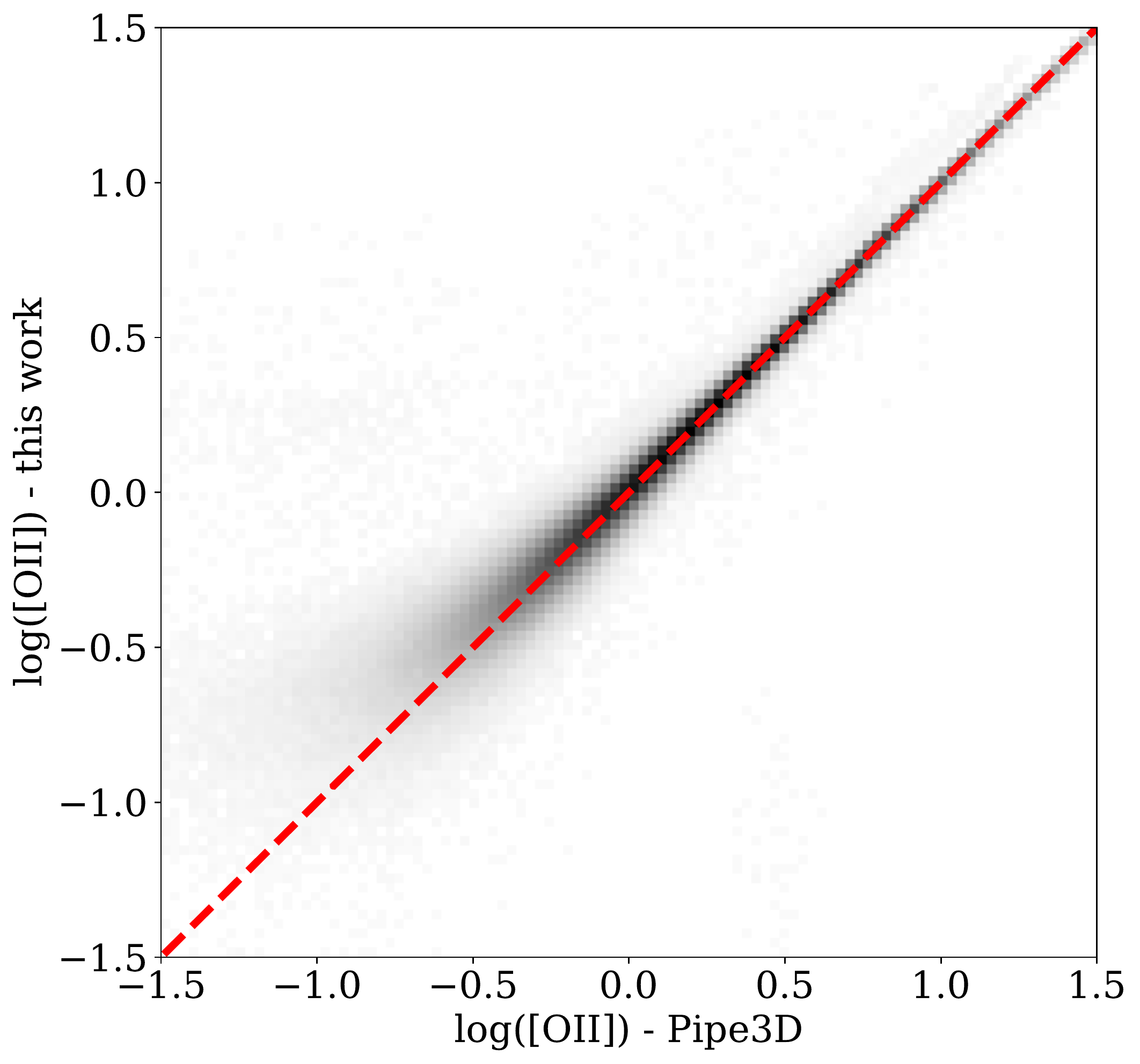}
    \end{minipage}
    \begin{minipage}{0.66\columnwidth}
                \includegraphics[width=1\columnwidth]{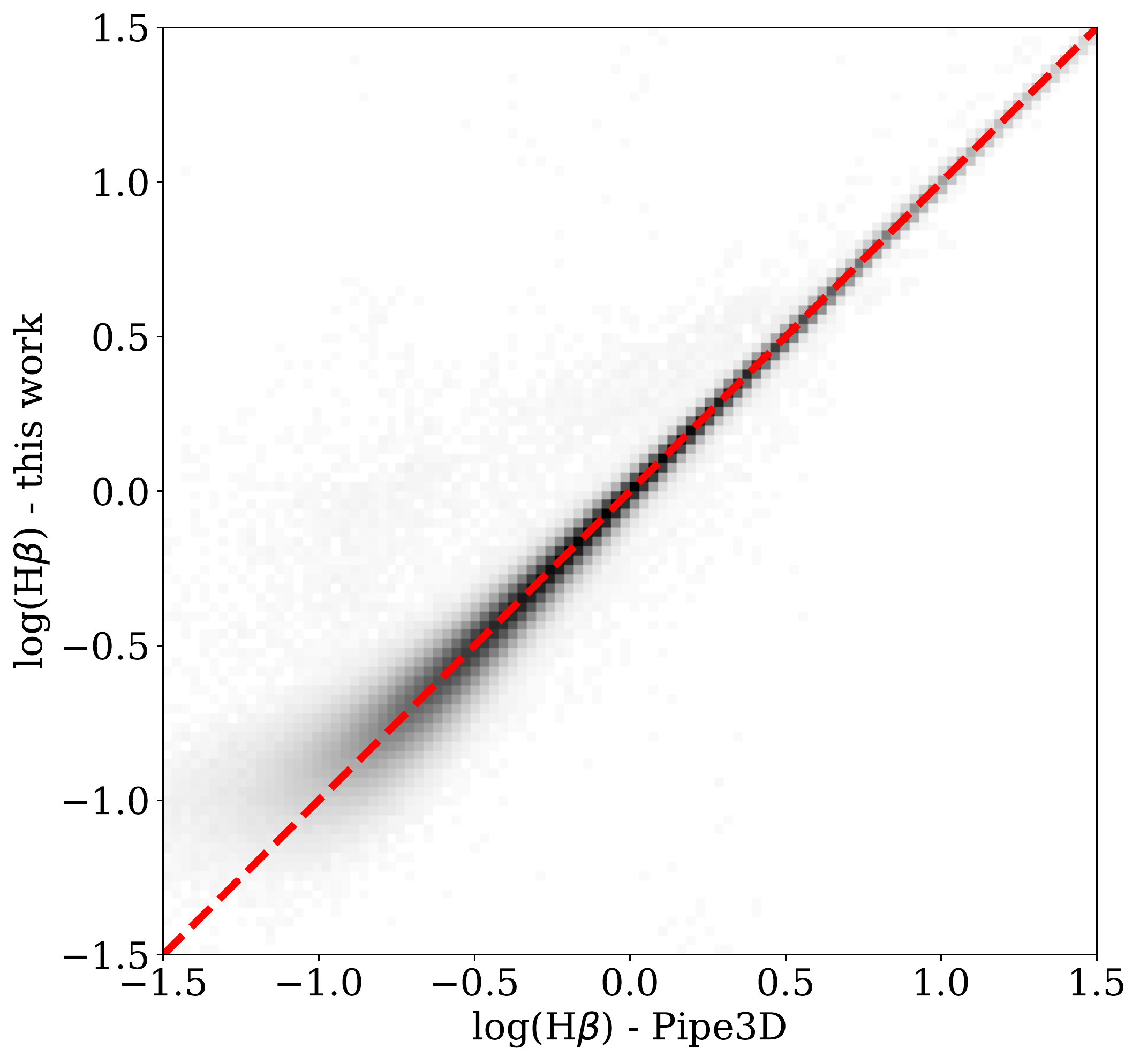}
    \end{minipage}
    \begin{minipage}{0.66\columnwidth}
                \includegraphics[width=1\columnwidth]{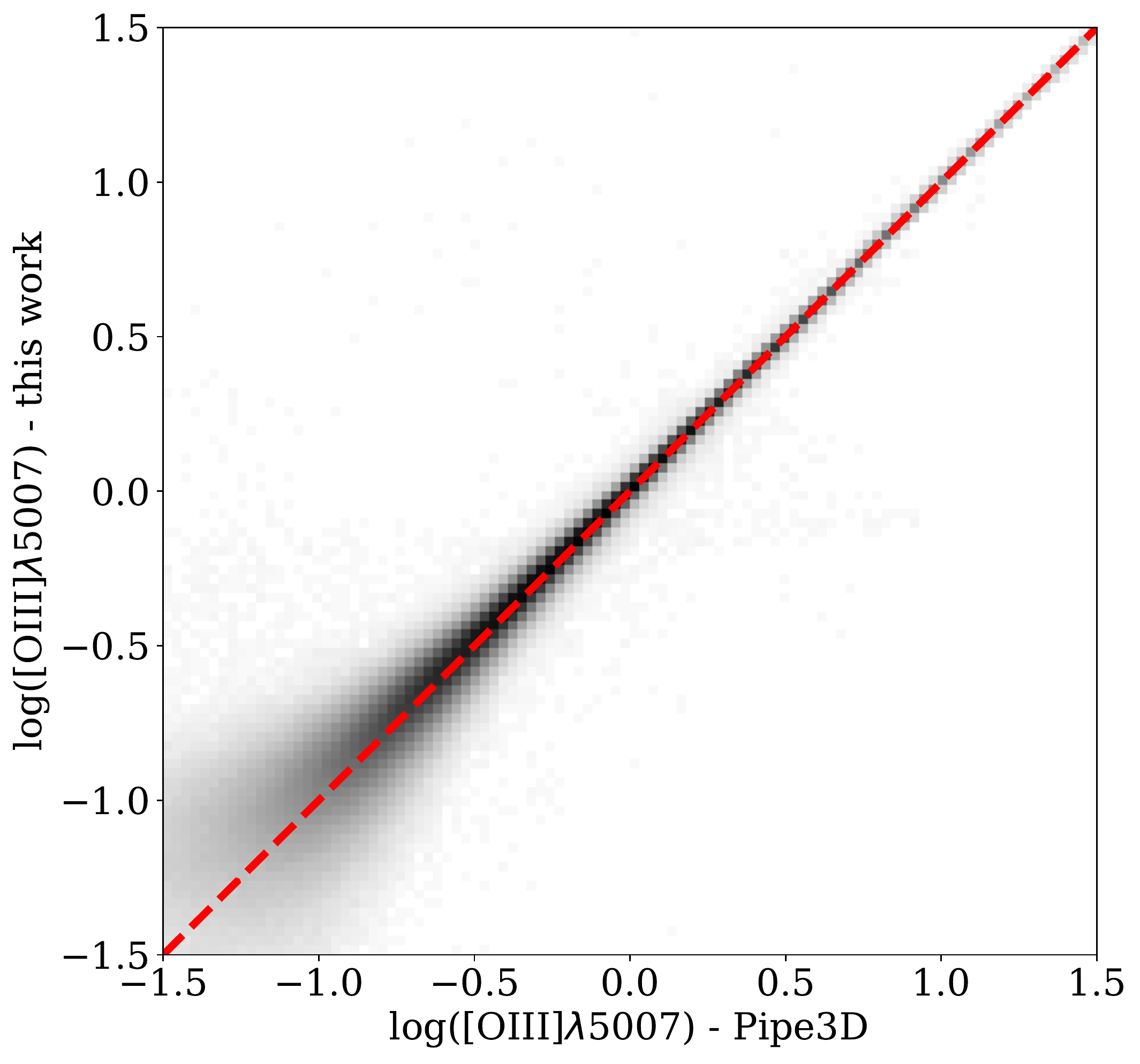}
    \end{minipage}
    
    \begin{minipage}{0.66\columnwidth}
                \includegraphics[width=1\columnwidth]{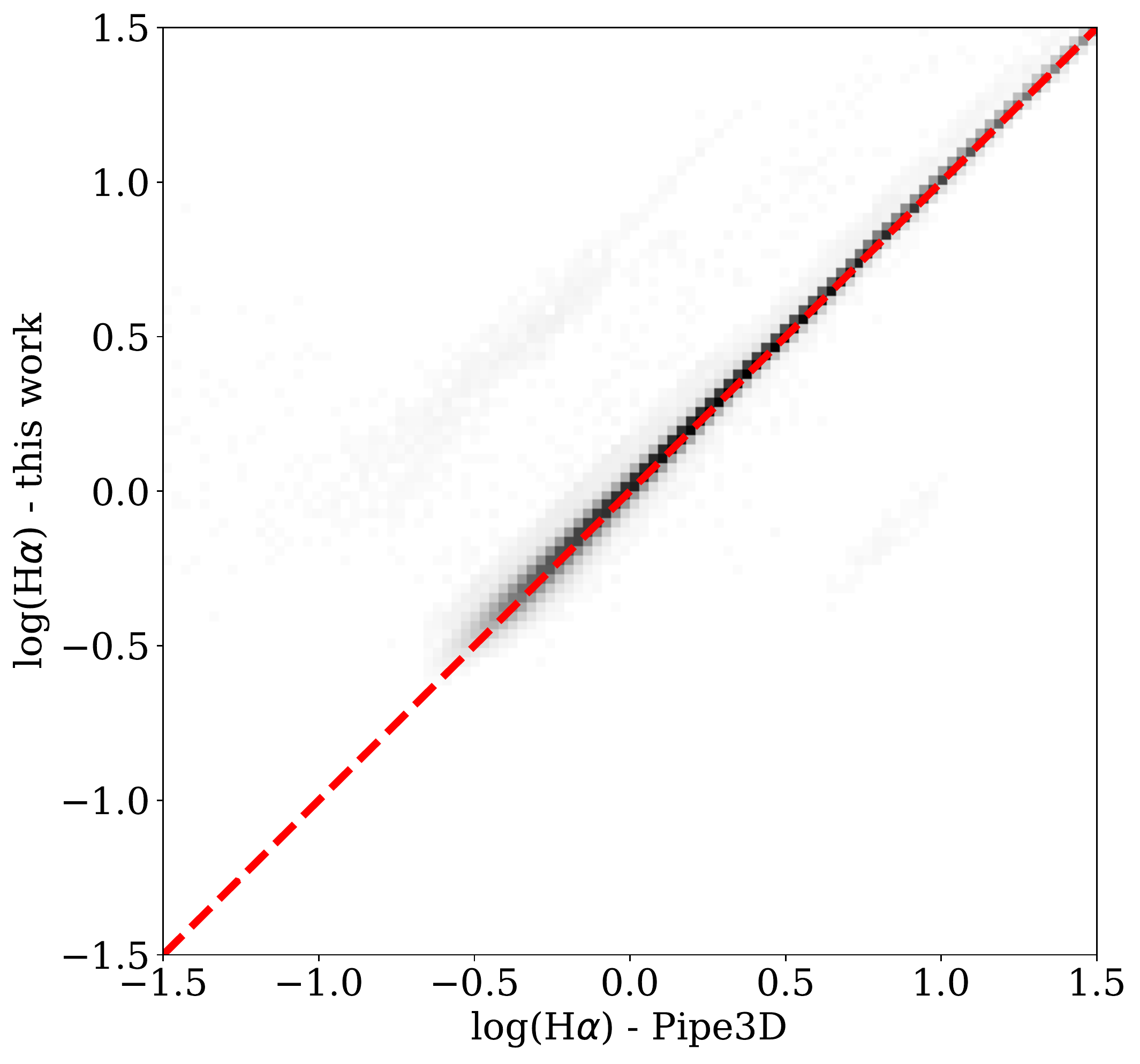}
    \end{minipage} 
    \begin{minipage}{0.66\columnwidth}
                \includegraphics[width=1\columnwidth]{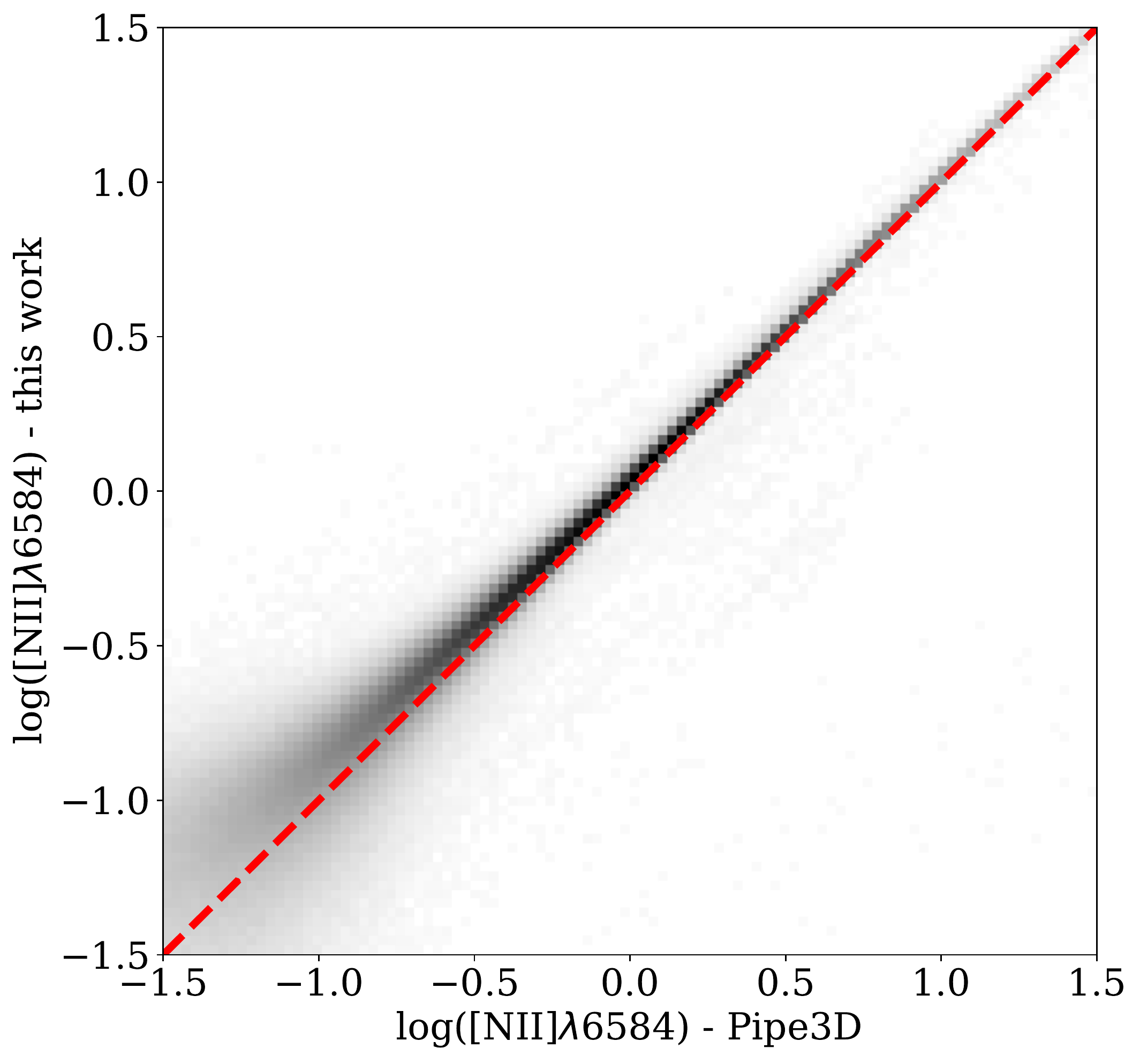}
    \end{minipage}    
    \begin{minipage}{0.66\columnwidth}
                \includegraphics[width=1\columnwidth]{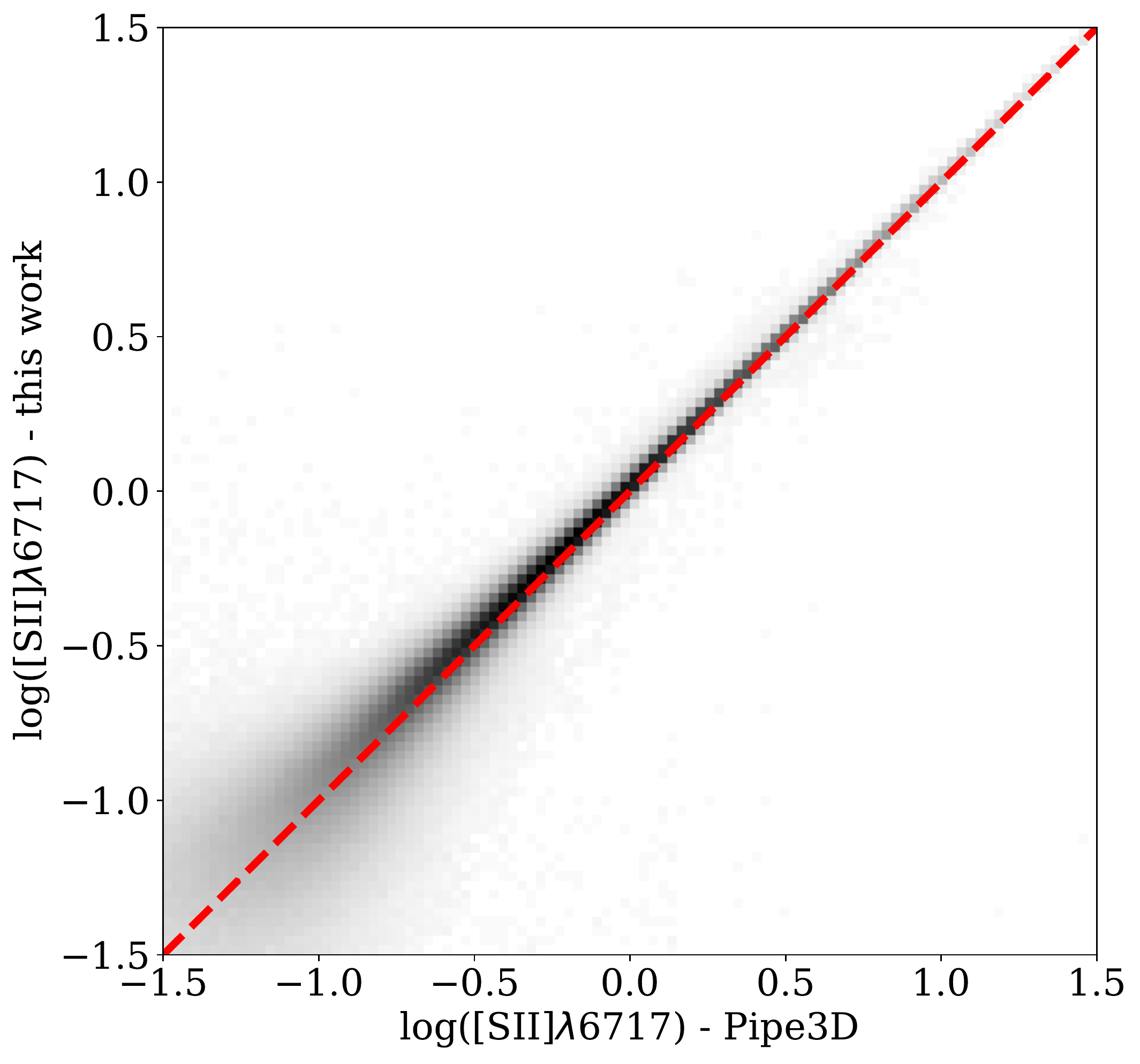}
    \end{minipage} 
    
    \begin{minipage}{0.66\columnwidth}
                \includegraphics[width=1\columnwidth]{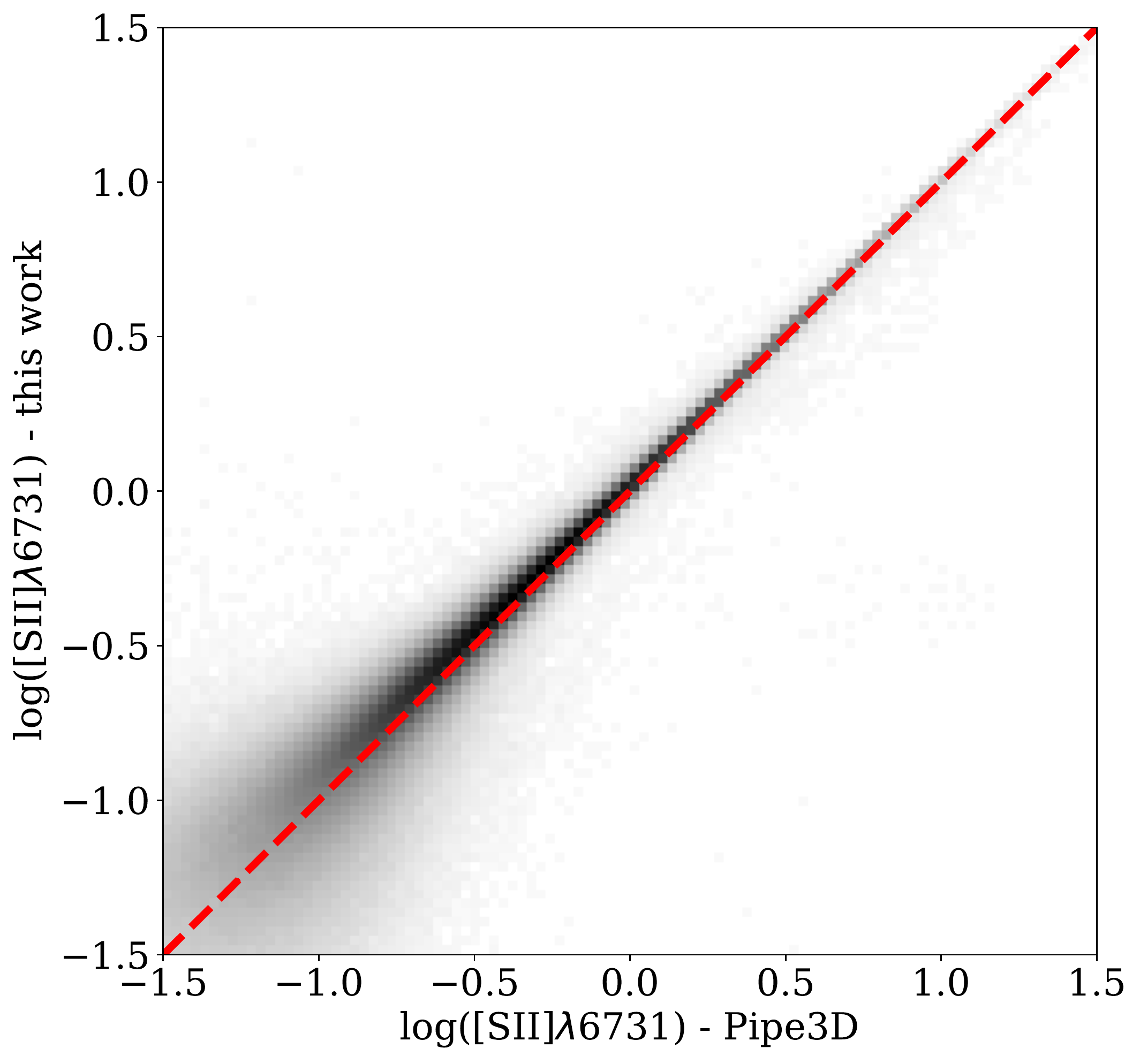}
    \end{minipage}   
    \begin{minipage}{0.66\columnwidth}
                \includegraphics[width=1\columnwidth]{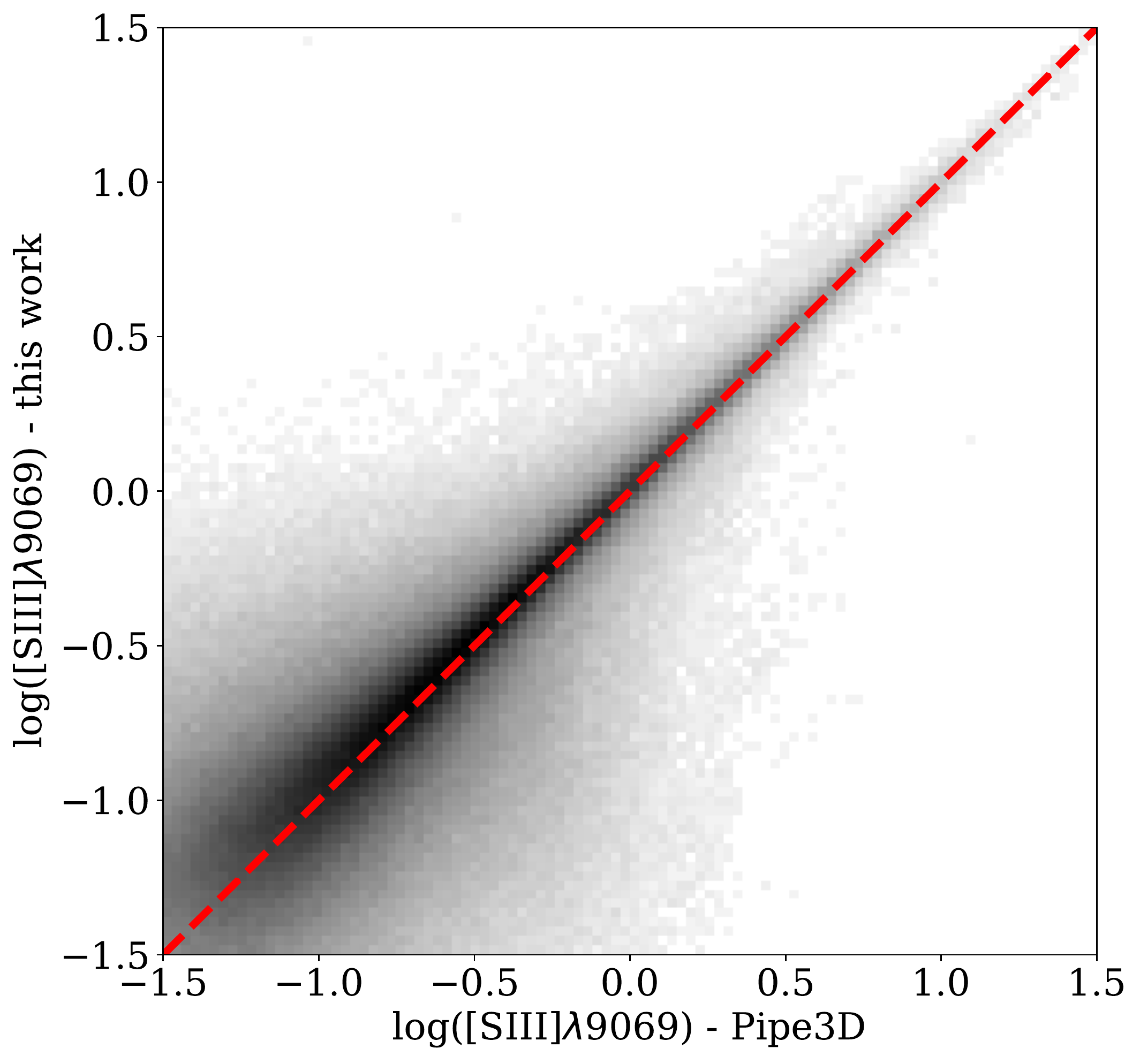}
    \end{minipage} 
    \begin{minipage}{0.66\columnwidth}
                \includegraphics[width=1\columnwidth]{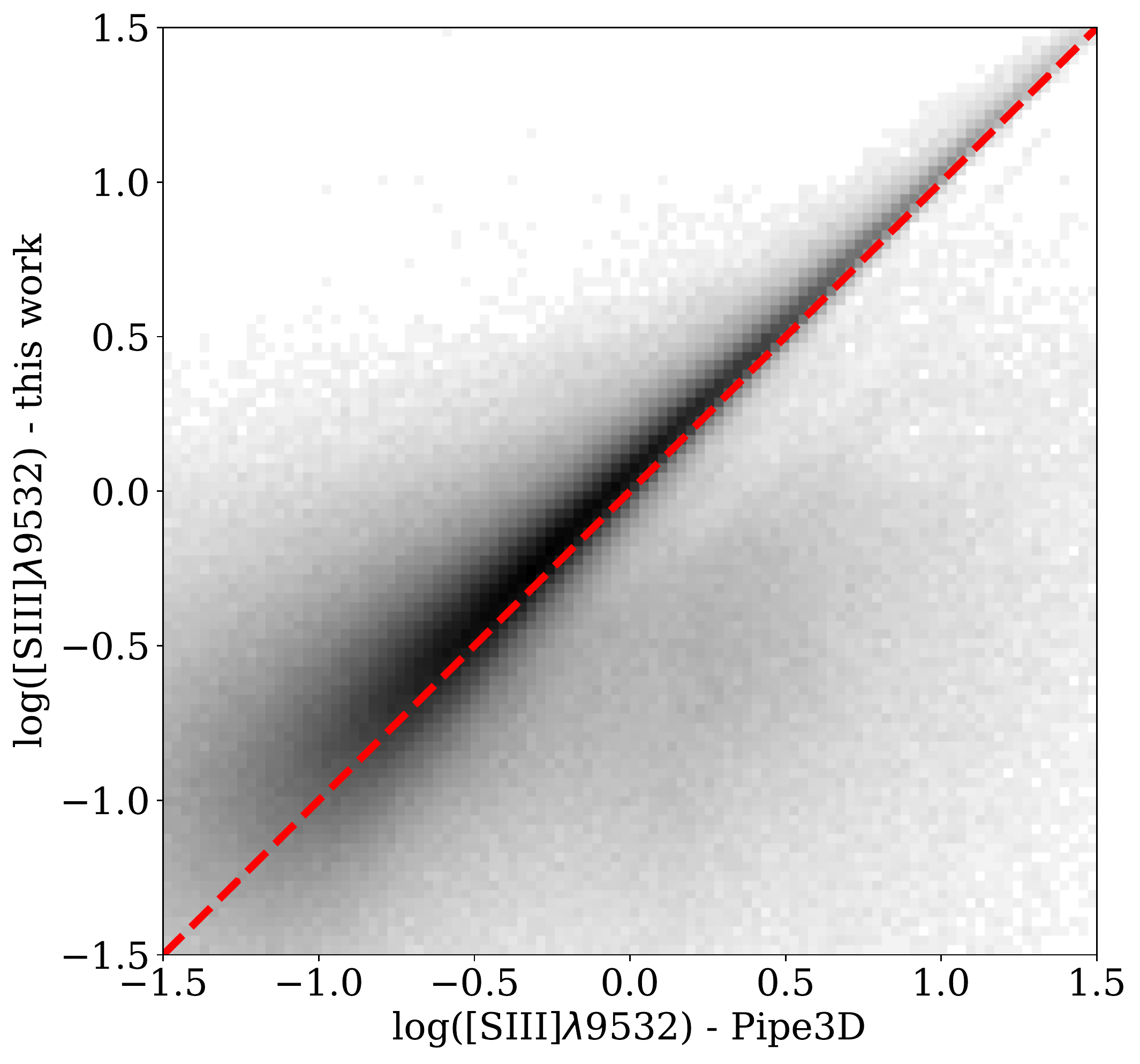}
    \end{minipage}     

        \caption{Spaxel-by-spaxel comparison between the emission line fluxes taken into account in this work (y-axis) and the results of Pipe3D (x-axis): namely \oii, \hb, \oiii$\lambda$5007, \ha, \nii$\lambda6584$, \sii$\lambda6717$, \sii$\lambda6731$, \siii$\lambda9069$, \siii$\lambda9532$. The fluxes are in units of $10^{-17}$~erg~s$^{-1}$\AA$^{-1}$~cm$^{-2}$ and are expressed in logarithm. The red dashed line represents the one-to-one line.}
        \label{fig:pipe3d}
   \end{figure*}

Fig.~\ref{fig:pipe3d} shows the comparison between the emission line fluxes taken into account in this work (y-axis) with the ones obtained with Pipe3D (x-axis) (see also Appendix~A, \citealt{belfiore2019}) for all the star forming spaxels characterised by a signal-to-noise S/N(\ha)~$>15$. In general there is good agreement between the two pipelines, since the majority of the measurements lie in the vicinity of the red dashed line one-to-one line. 
However, there is a larger spread in the comparison of the \siii\, lines with respect to the other transitions, especially for \siii$\lambda9532$. Overall, our \siii$\lambda9532$ measurements are slightly larger than those performed by Pipe3D, while in few cases Pipe3D measurements have values up to two orders of magnitude higher than the ones derived by us. There is good agreement for the \siii$\lambda9069$ emission line, instead. 
In contrast to Pipe3D, in our fitting procedure we fixed the flux ratio of the two \siii\, lines to the intrinsic value of 2.47 \citep{pyneb}, as explained in Sec~\ref{sec:spectralfitting}. Therefore, the fact that there is good agreement for the \siii$\lambda9069$ but not for the \siii$\lambda9532$ suggests that Pipe3D is underestimating the \siii$\lambda9532$ line flux in the majority of the spaxels, on top of showing a high failure rate of spaxels with clearly nonphysically \siii\ line ratios. We therefore strongly recommend against the use of the Pipe3D \siii$\lambda9532$ line fluxes.

\section{Signal-to-noise radial profiles}
\label{sec:snr_app}

\begin{figure*}
    \begin{minipage}{0.66\columnwidth}
                \includegraphics[width=1\columnwidth]{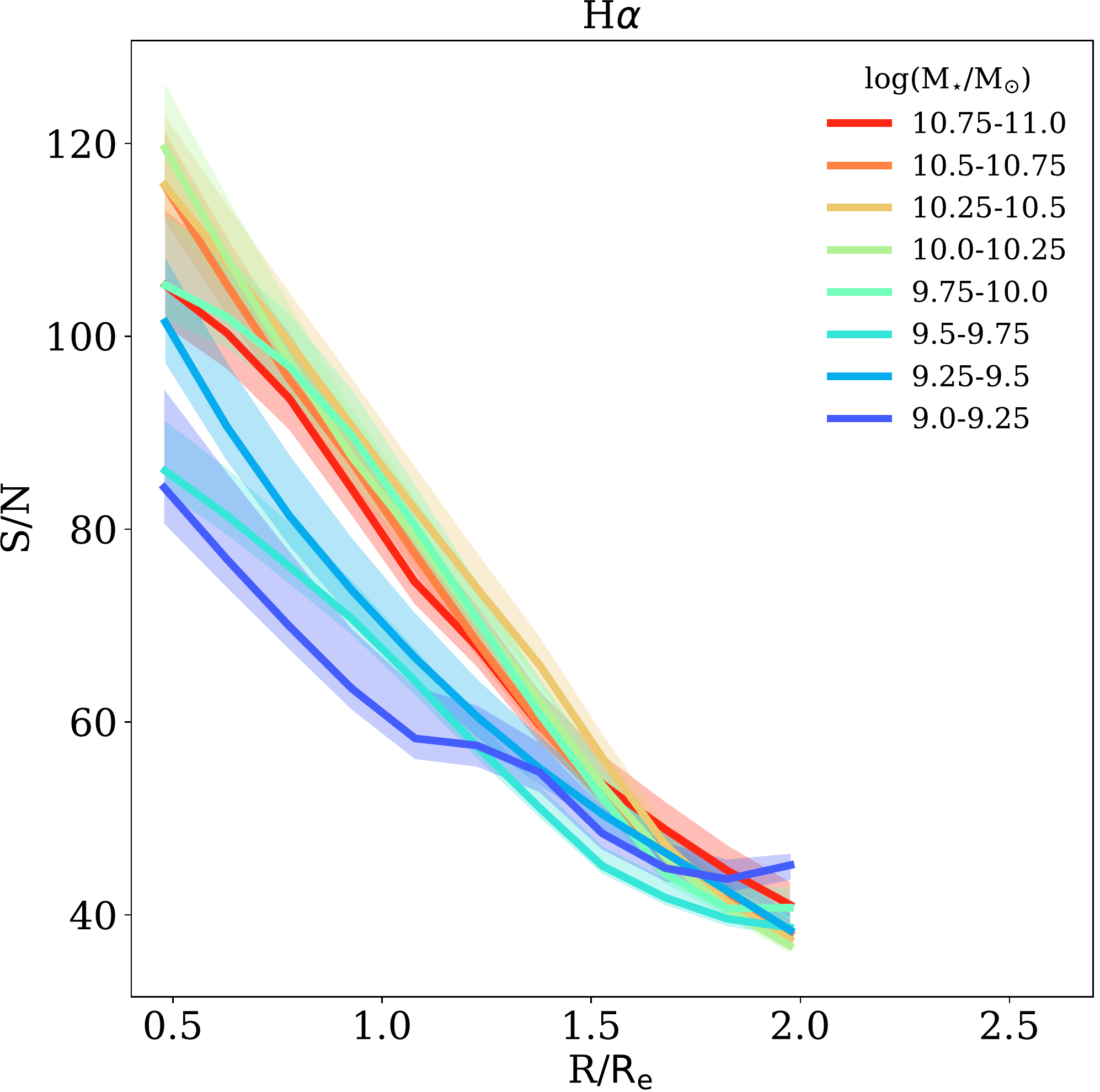}
    \end{minipage}
    \begin{minipage}{0.66\columnwidth}
                \includegraphics[width=1\columnwidth]{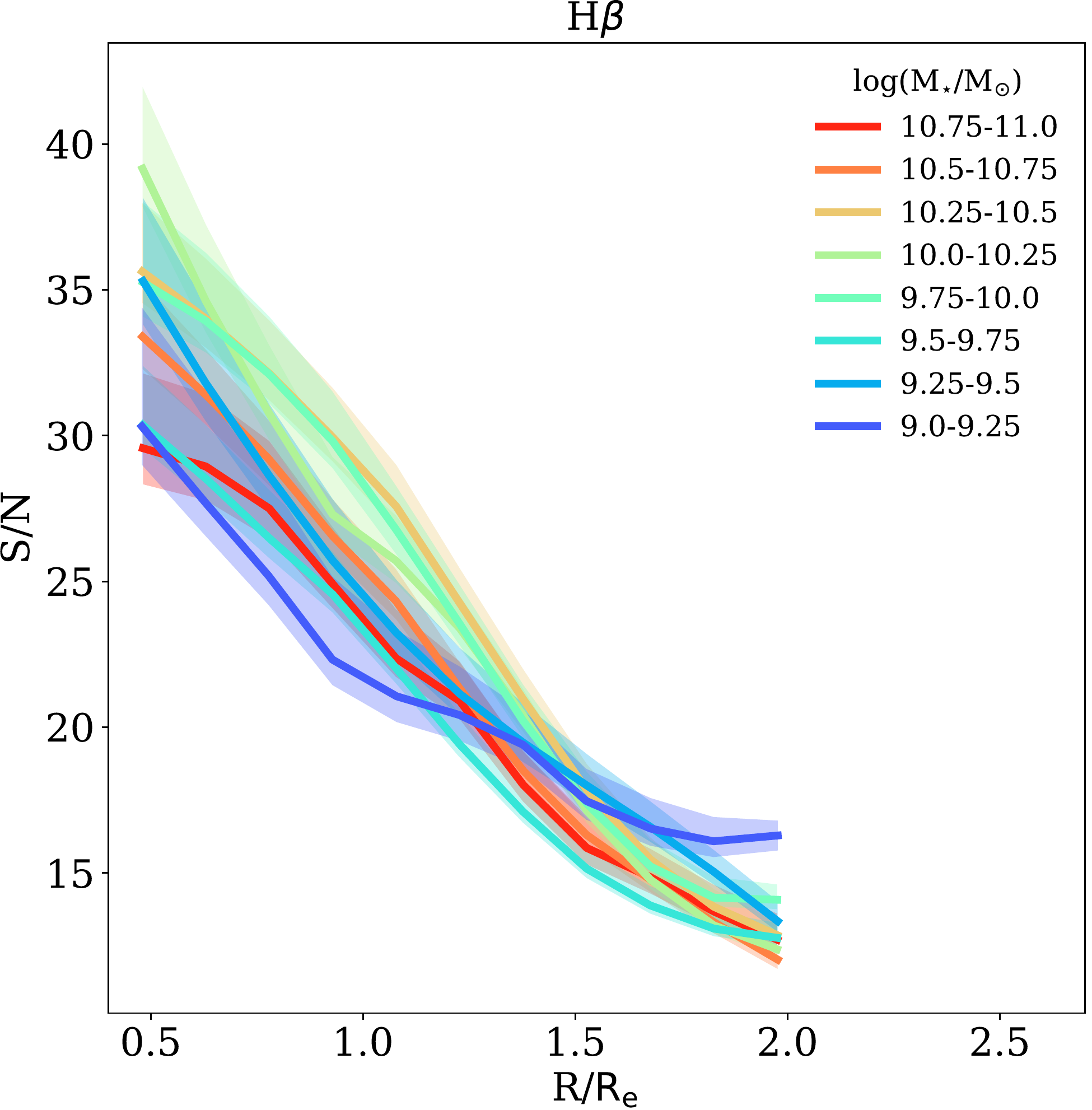}
    \end{minipage}
    \begin{minipage}{0.66\columnwidth}
                \includegraphics[width=1\columnwidth]{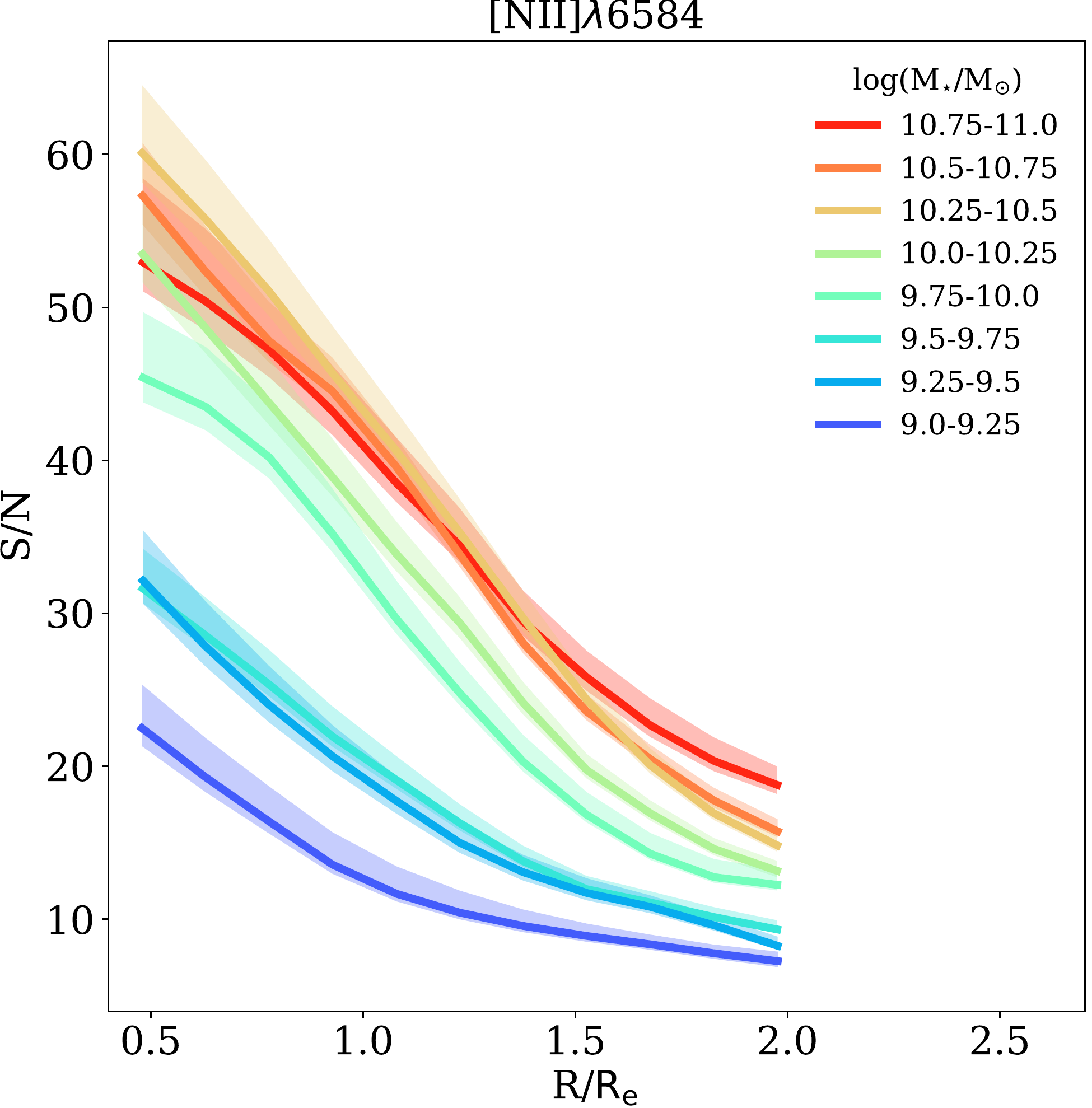}
    \end{minipage}
    
    \begin{minipage}{0.66\columnwidth}
                \includegraphics[width=1\columnwidth]{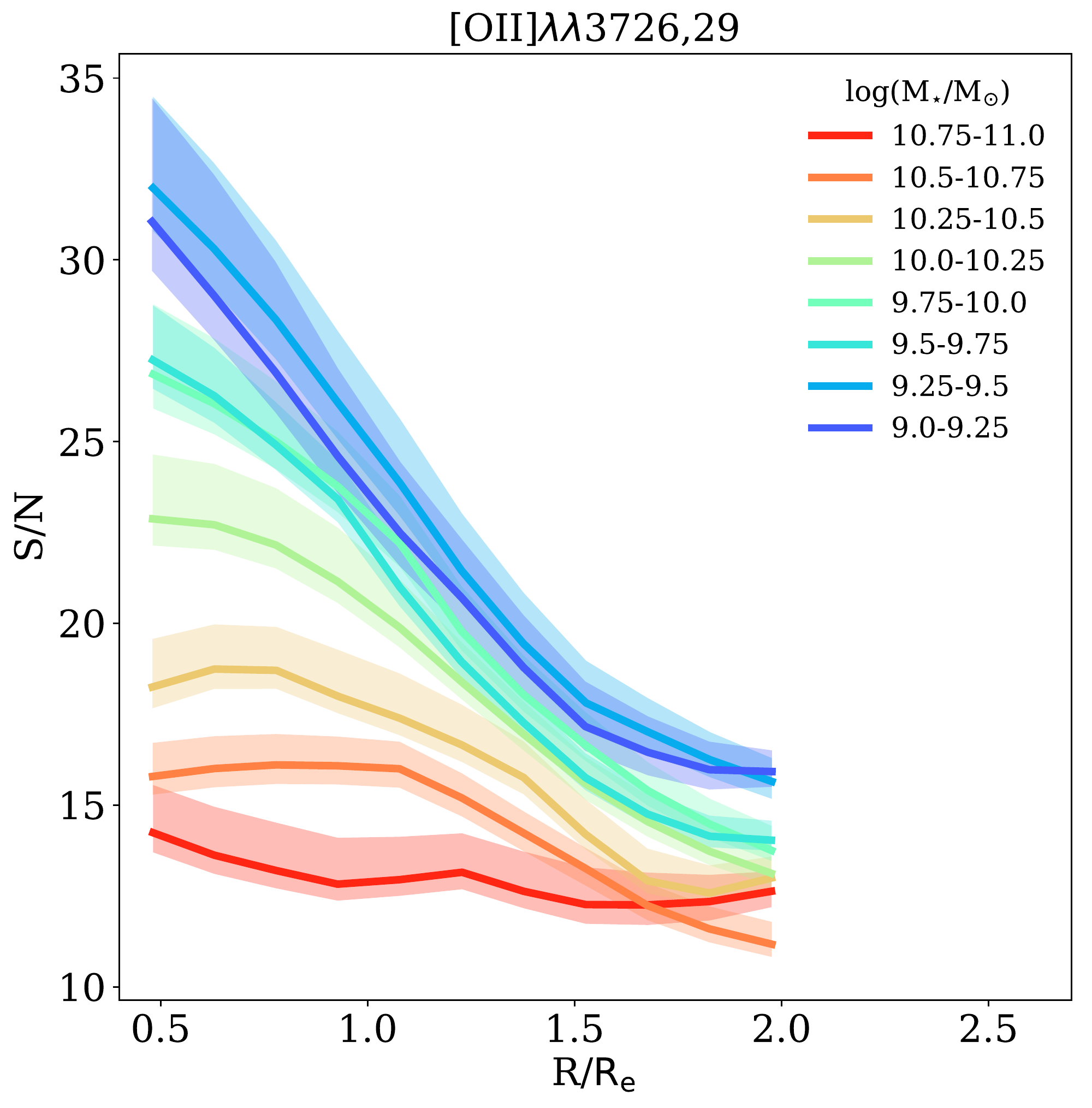}
    \end{minipage} 
    \begin{minipage}{0.66\columnwidth}
                \includegraphics[width=1\columnwidth]{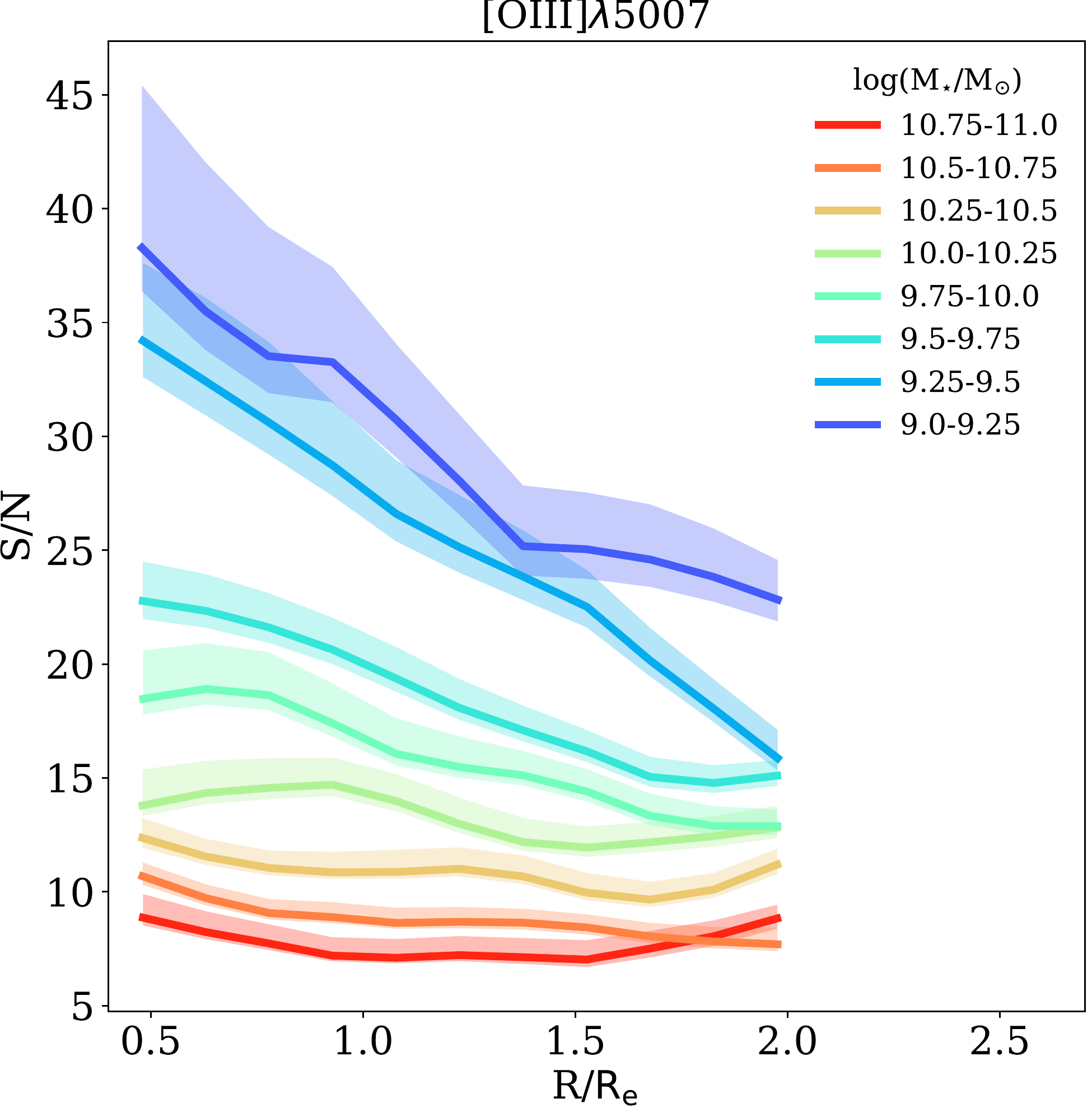}
    \end{minipage}    
    \begin{minipage}{0.66\columnwidth}
                \includegraphics[width=1\columnwidth]{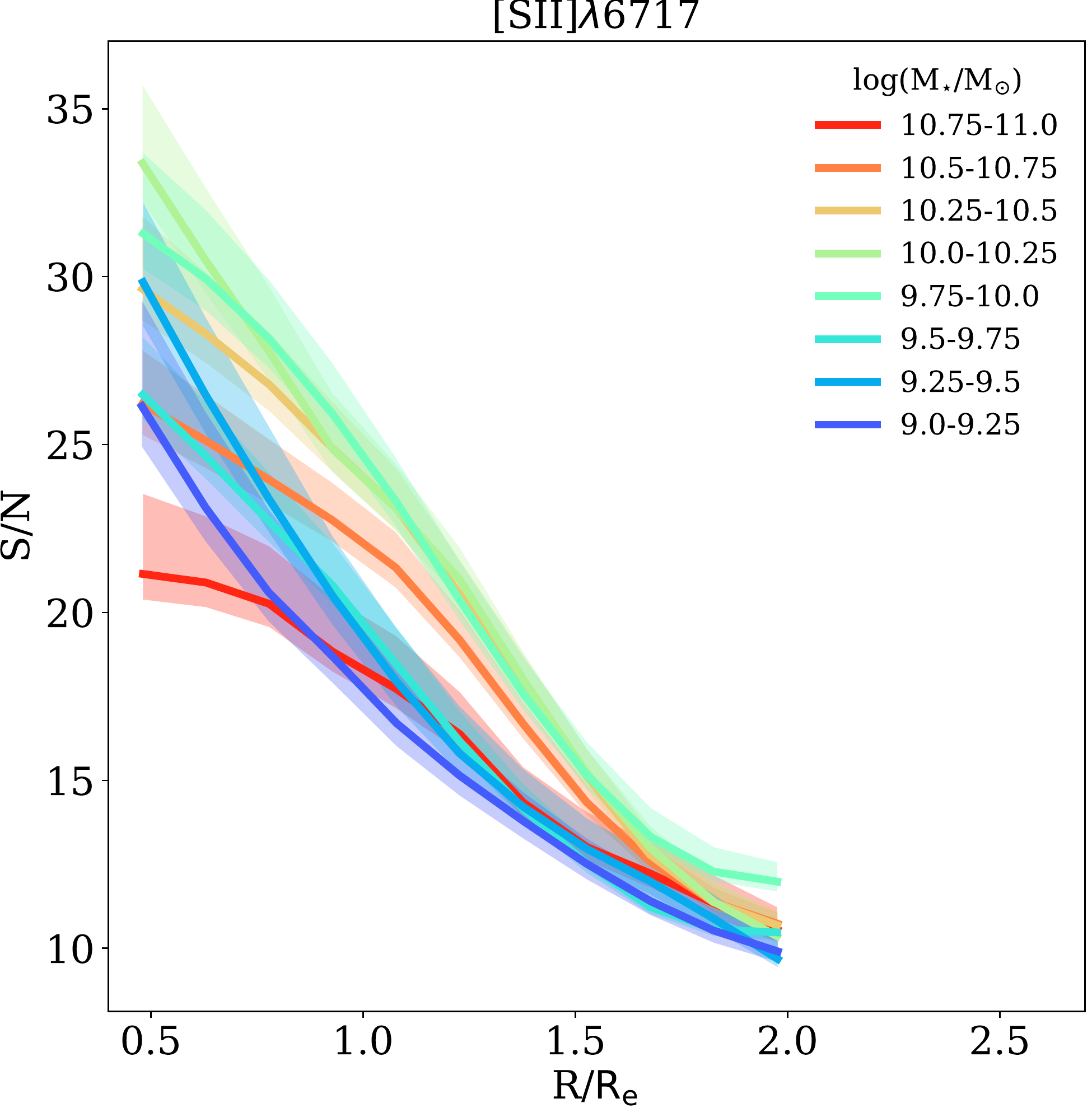}
    \end{minipage} 
    
    \begin{center}
    \begin{minipage}{0.66\columnwidth}
                \includegraphics[width=1\columnwidth]{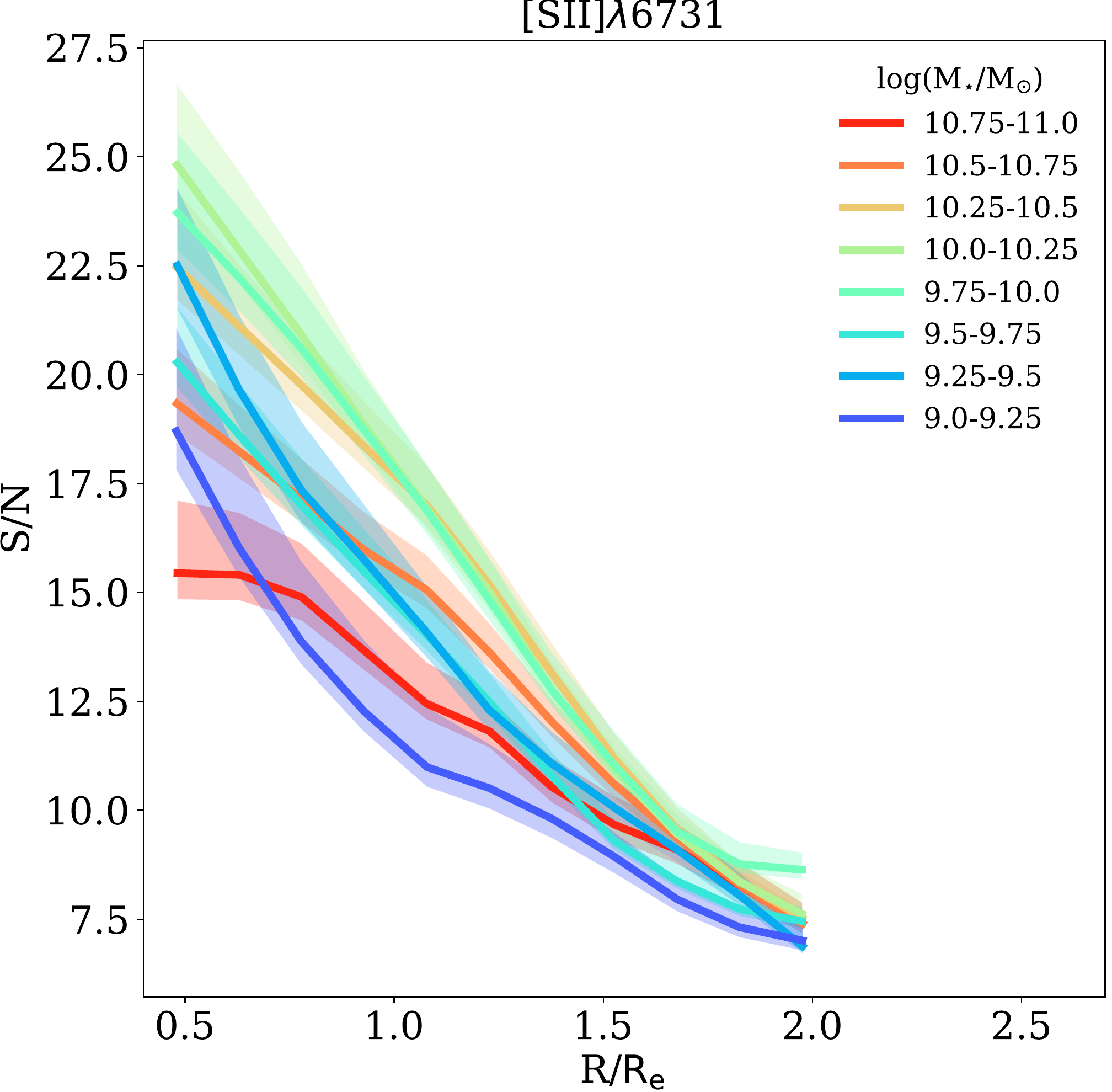}
    \end{minipage}   
    \begin{minipage}{0.66\columnwidth}
                \includegraphics[width=1\columnwidth]{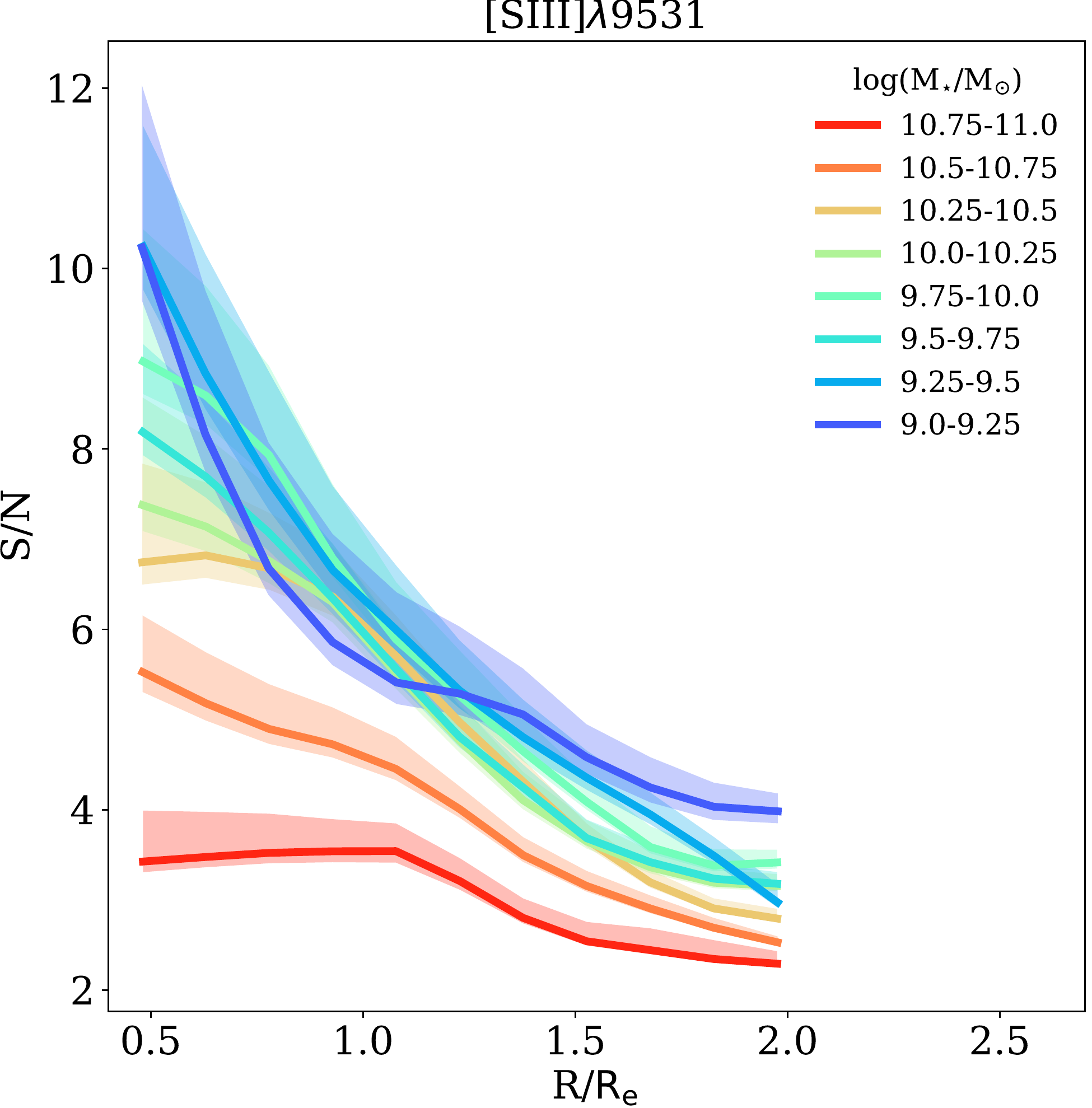}
    \end{minipage} 
    \end{center}
        \caption{Radial distribution of the signal to noise of \ha, \hb, \nii$\lambda$6584, \oii, \oiii$\lambda$5007, \sii$\lambda$6717, \sii$\lambda$6731 and \siii$\lambda$9532, in eight stellar mass $M_\star$ bins, as reported in the legend.}
        \label{fig:signaltonoise_lines}
   \end{figure*}

Fig.~\ref{fig:signaltonoise_lines} shows the S/N radial gradients of \ha, \hb, \nii$\lambda$6584, \oii, \oiii$\lambda$5007, \sii$\lambda$6717, \sii$\lambda$6731 and \siii$\lambda$9532. To compute these gradient we considered all the spaxels in our galaxy sample, subdivided in bins of stellar mass, as reported in the legend. The radial distance is normalised to the elliptical Petrosian effective radius (\re), like for all other gradients presented in this work.
For each stellar mass bin, the radial profile is computed as the median of the galaxies contributing to the bin at that radius. The upper and lower error bars are obtained by calculating the $16^{\mathrm{th}}$ and $84^{\mathrm{th}}$ percentiles of the distribution for the sample and dividing by $\sqrt{N}$, where N is the number of profiles at each radius.
The figure highlights that the MaNGA galaxy sample used in this work shows sufficient S/N in all the strong lines considered, except for \siii$\lambda$9532 at large radii and for high-mass galaxies.

\section{Quality test}
\label{sec:quality_test}



\begin{figure*}
   \centering
    \begin{minipage}{1\columnwidth}
                \includegraphics[width=0.8\columnwidth]{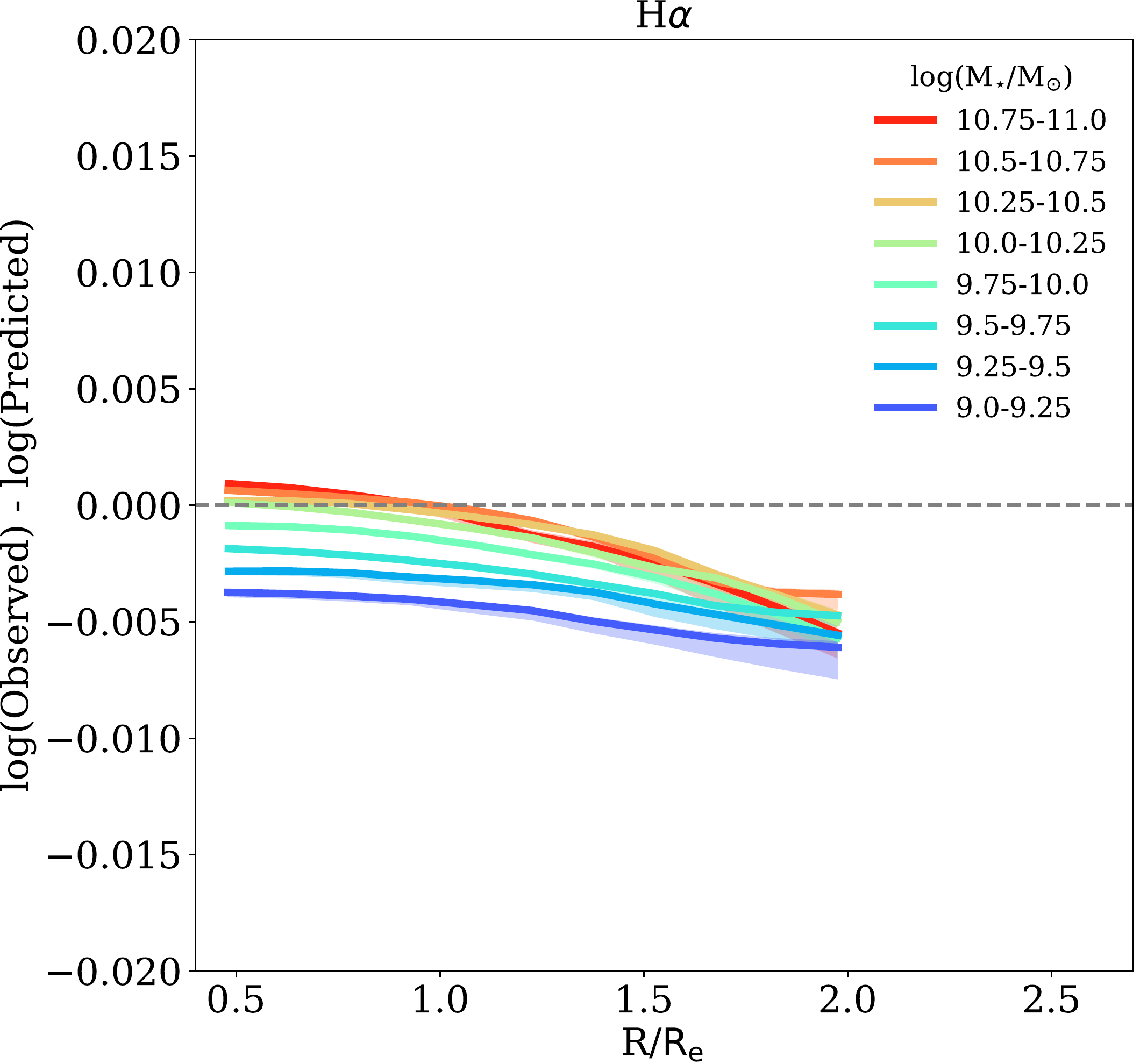}
    \end{minipage}
    \begin{minipage}{1\columnwidth}
                \includegraphics[width=0.8\columnwidth]{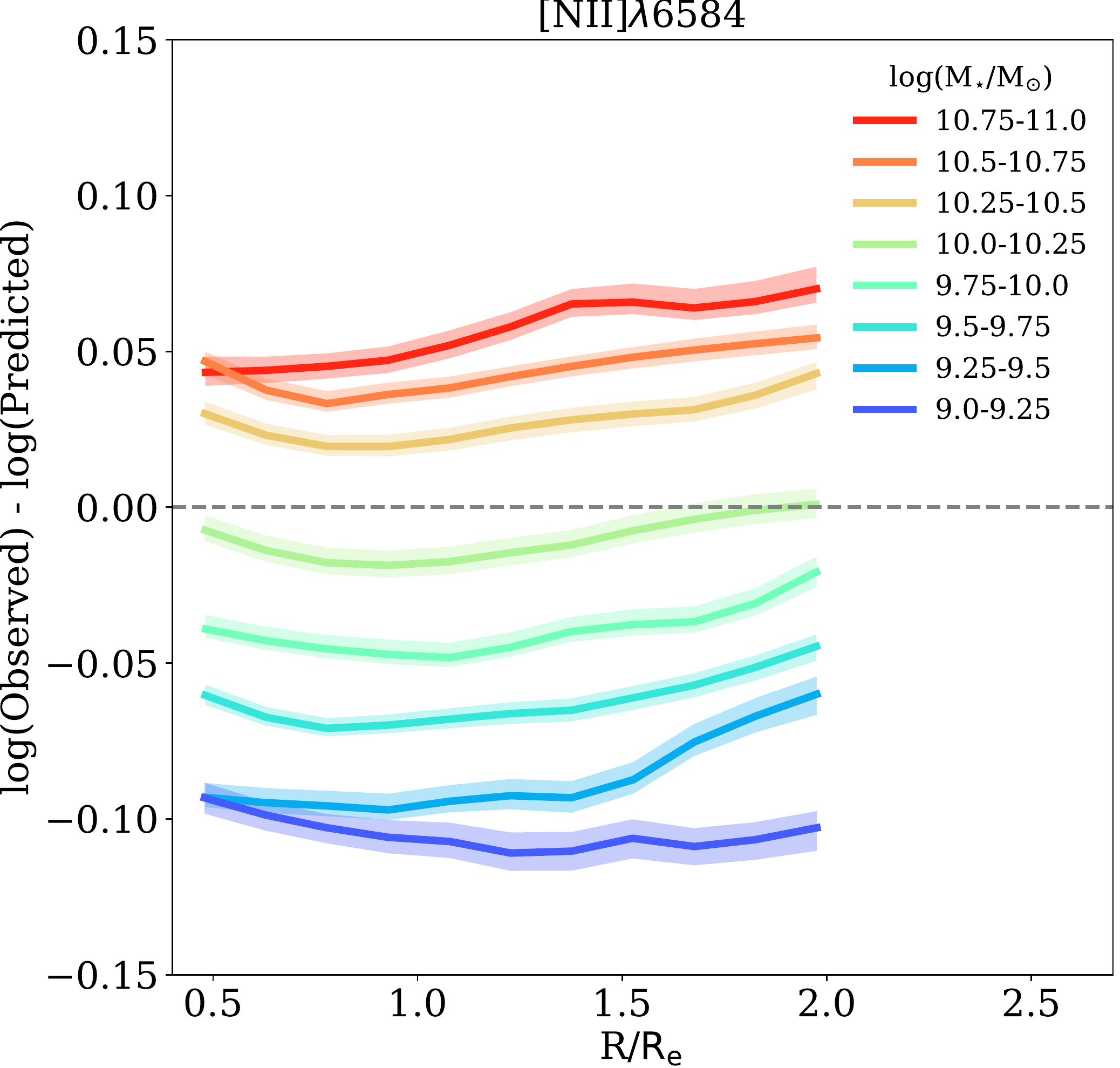}
    \end{minipage}
    
    \begin{minipage}{1\columnwidth}
                \includegraphics[width=0.8\columnwidth]{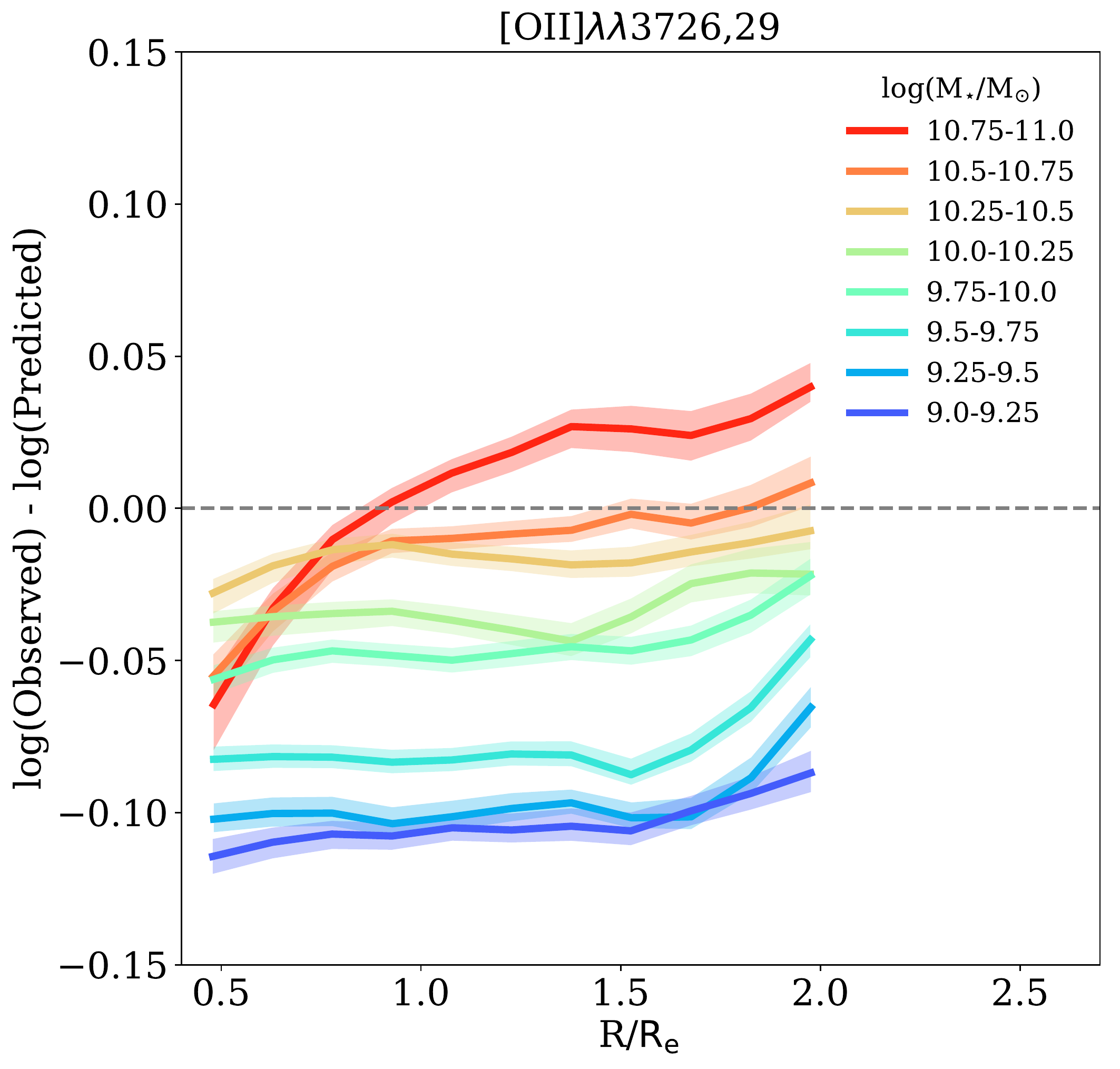}
    \end{minipage} 
    \begin{minipage}{1\columnwidth}
                \includegraphics[width=0.8\columnwidth]{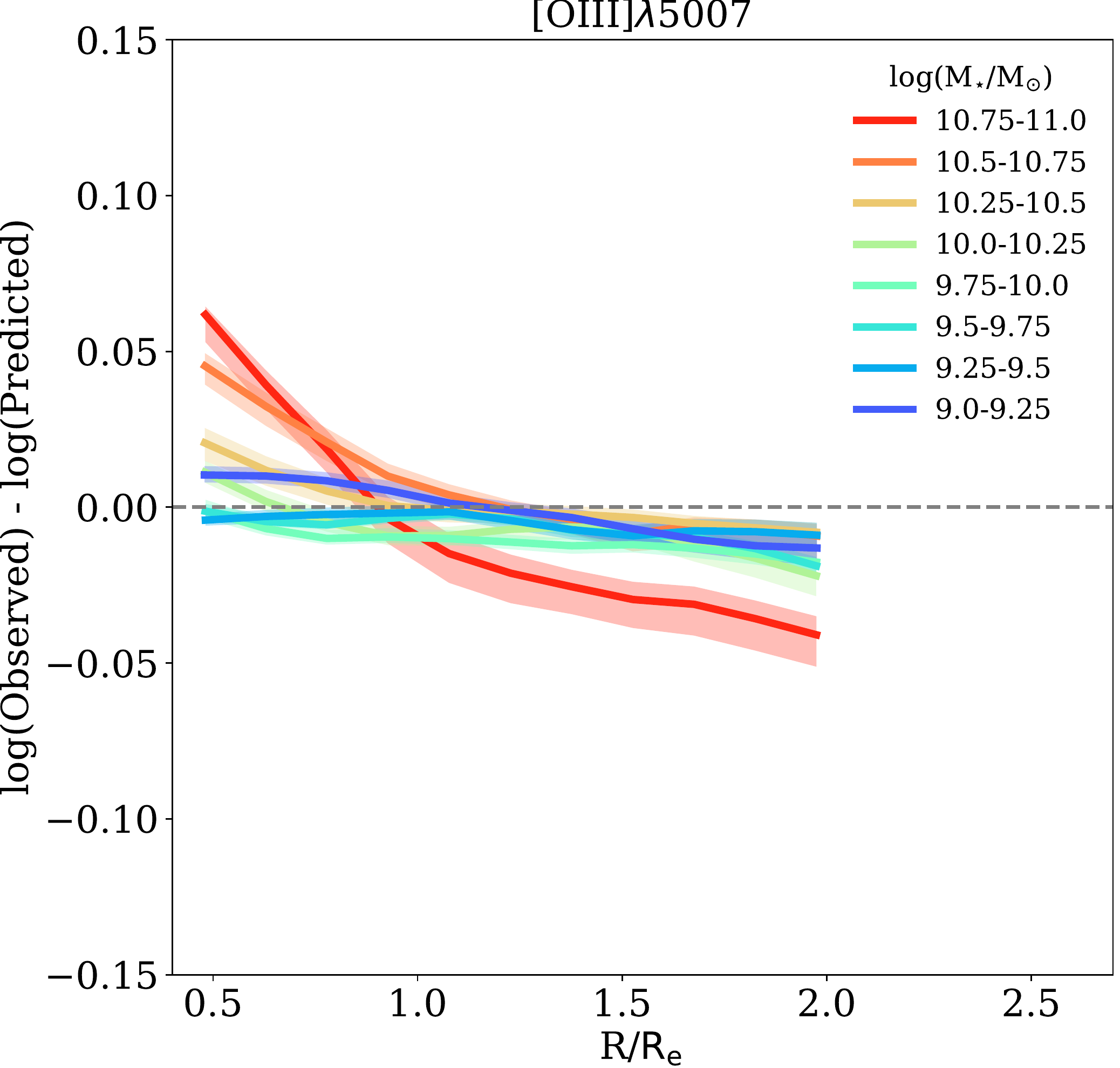}
    \end{minipage}
    
    \begin{minipage}{1\columnwidth}
                \includegraphics[width=0.8\columnwidth]{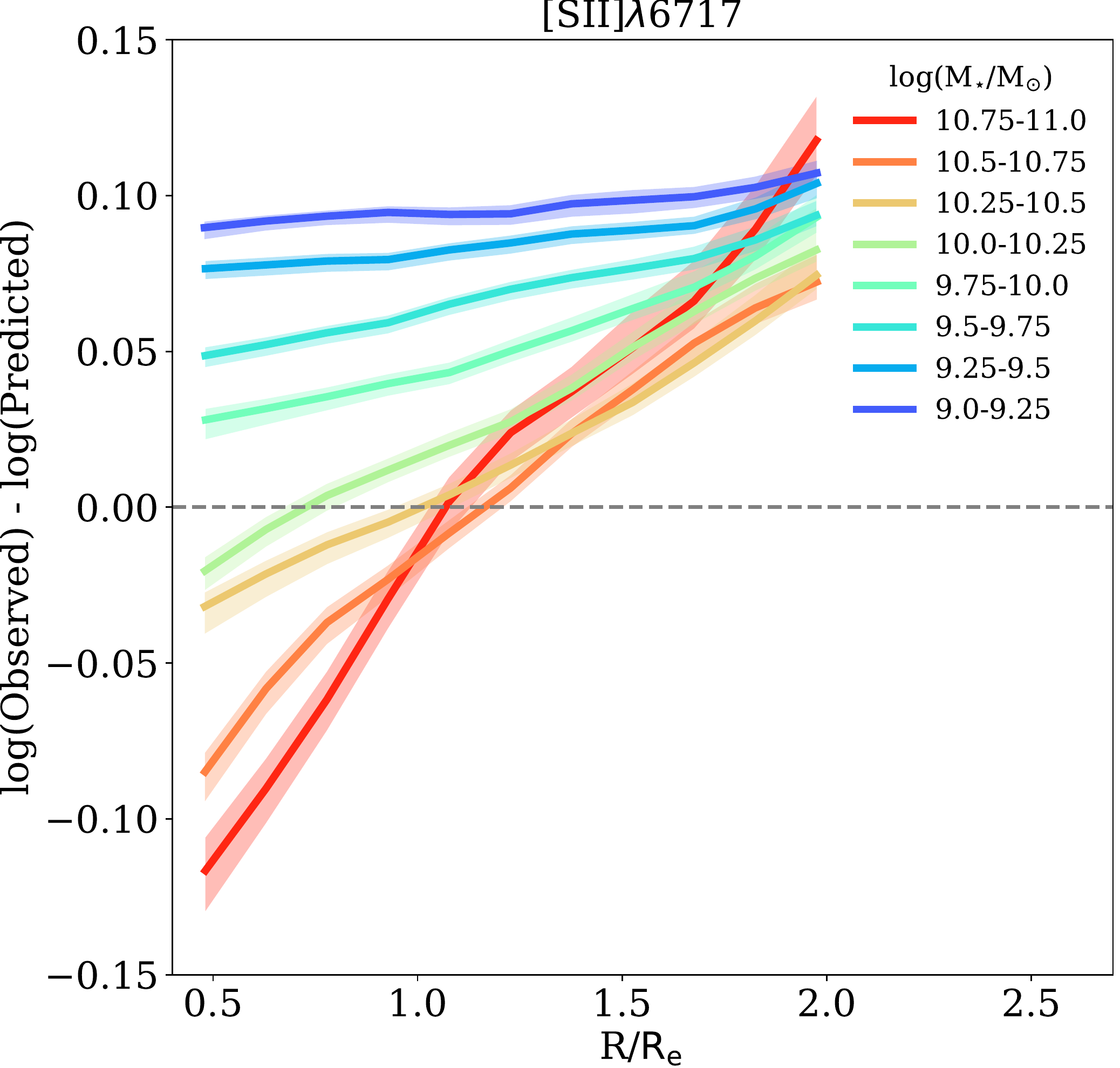}
    \end{minipage} 
    \begin{minipage}{1\columnwidth}
                \includegraphics[width=0.8\columnwidth]{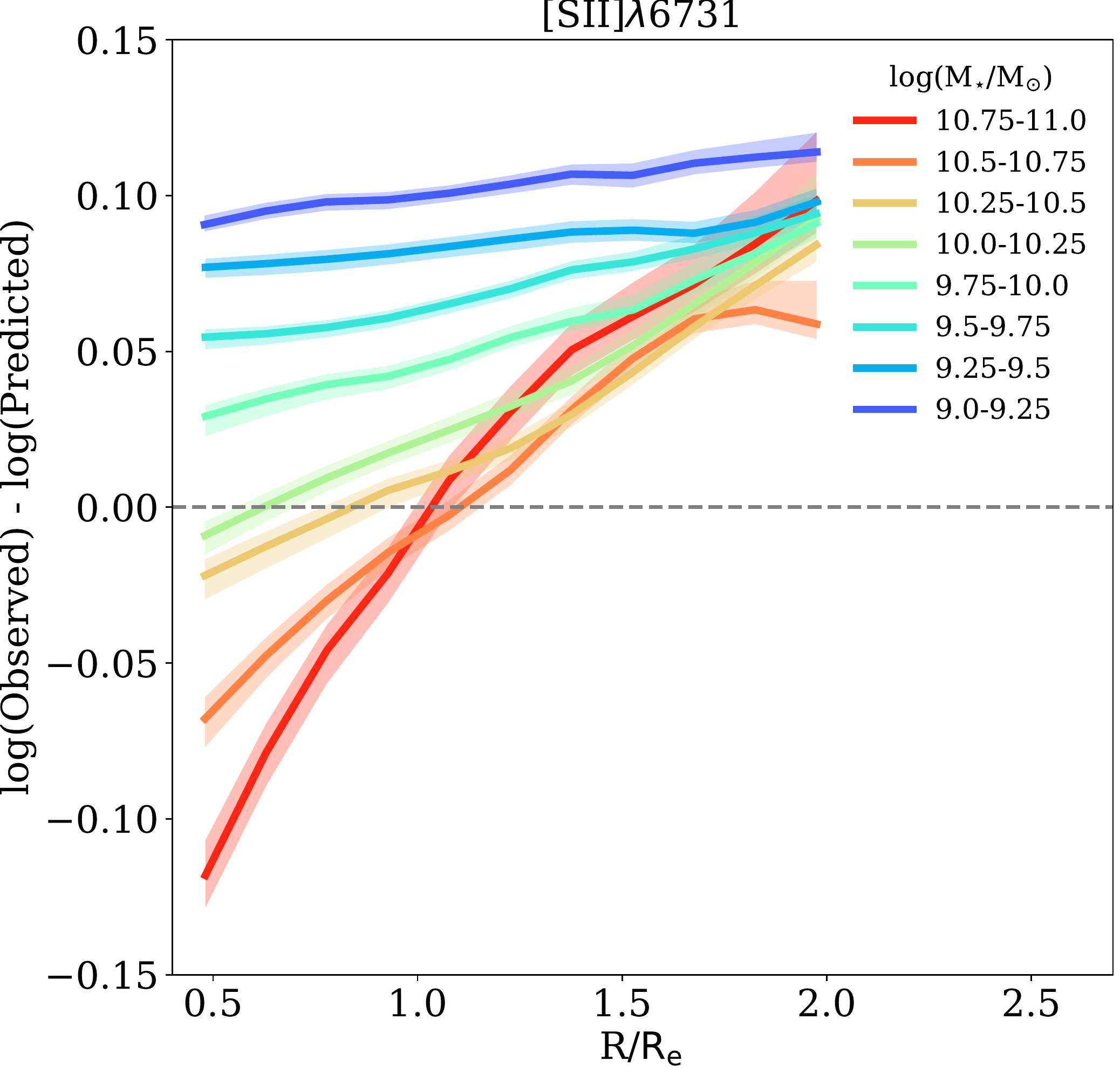}
    \end{minipage}
        \caption{Radial distribution of the difference between the logarithm of the observed flux and that of the best-fit model (both normalised to \hb) obtained with IZI and a Gaussian prior on the ionisation parameter, for all the emission lines taken into account, in eight stellar mass $M_\star$ bins, as reported in the legend. Since these radial distributions lie in the range [-0.1,0.1]~dex, it means that they are all consistent within the intrinsic error taken into account for the models.}
        \label{fig:a_lines}
   \end{figure*}

In order to show the reliability of our method in Fig.~\ref{fig:a_lines} we show the distribution of the difference between the logarithm of the observed value of the emission line fluxes (\ha, \nii$\lambda6584$, \oii, \oiii$\lambda5007$, \sii$\lambda6717$ and \sii$\lambda6731$) and that of the best-fitting model (both normalised to \hb) obtained with IZI, taking into account D13 models and a Gaussian prior on the ionisation parameter (See Sec.~\ref{sec:priorvsnoprior}), for all the spaxels used in this work, as a function of radius and stellar mass. The radial distributions are divided in eight bins of stellar mass, in the range $\rm log(M_{\star}/M_\odot)=9-11$ in bins of 0.25 dex, as reported in the legend. The shaded areas represent the $16^{\mathrm{th}}$ and $84^{\mathrm{th}}$ percentiles of the distribution in each stellar mass bin divided by $\sqrt{N}$, where N is the number of galaxies lying in each bin. For each mass bin, a profile is computed only if more than 100 galaxies have a valid measured radial profile.
Specifically, the \ha\, and \oiii$\lambda5007$ line fluxes are faithfully reproduced at all radii and stellar masses.
However, IZI tends to overpredict the \nii$\lambda6584$ and \oii\, line fluxes at low stellar masses and to underestimate the \nii$\lambda6584$ at high stellar masses, and the \sii$\lambda6717$ and \sii$\lambda6731$ line fluxes at low stellar masses. 
Interestingly, concerning the \sii\, lines there is a dependence also on radius, since in the most massive galaxies the observed \sii\, fluxes are slightly overestimated at radii $R<1$~\re, while they are slightly underestimated at radii $R>1.5$~\re.
Overall, we claim that these stellar-mass- and radius-dependent discrepancies are not affecting the results shown in this paper, because all the radial distributions shown in Fig.~\ref{fig:a_lines} lie in the range [-0.1,0.1]~dex, meaning that they are all consistent within the assumed uncertainty of the photoionisation models of 0.1 dex (0.01 dex for \ha).

\begin{figure}
   \centering
        \includegraphics[width=1\columnwidth]{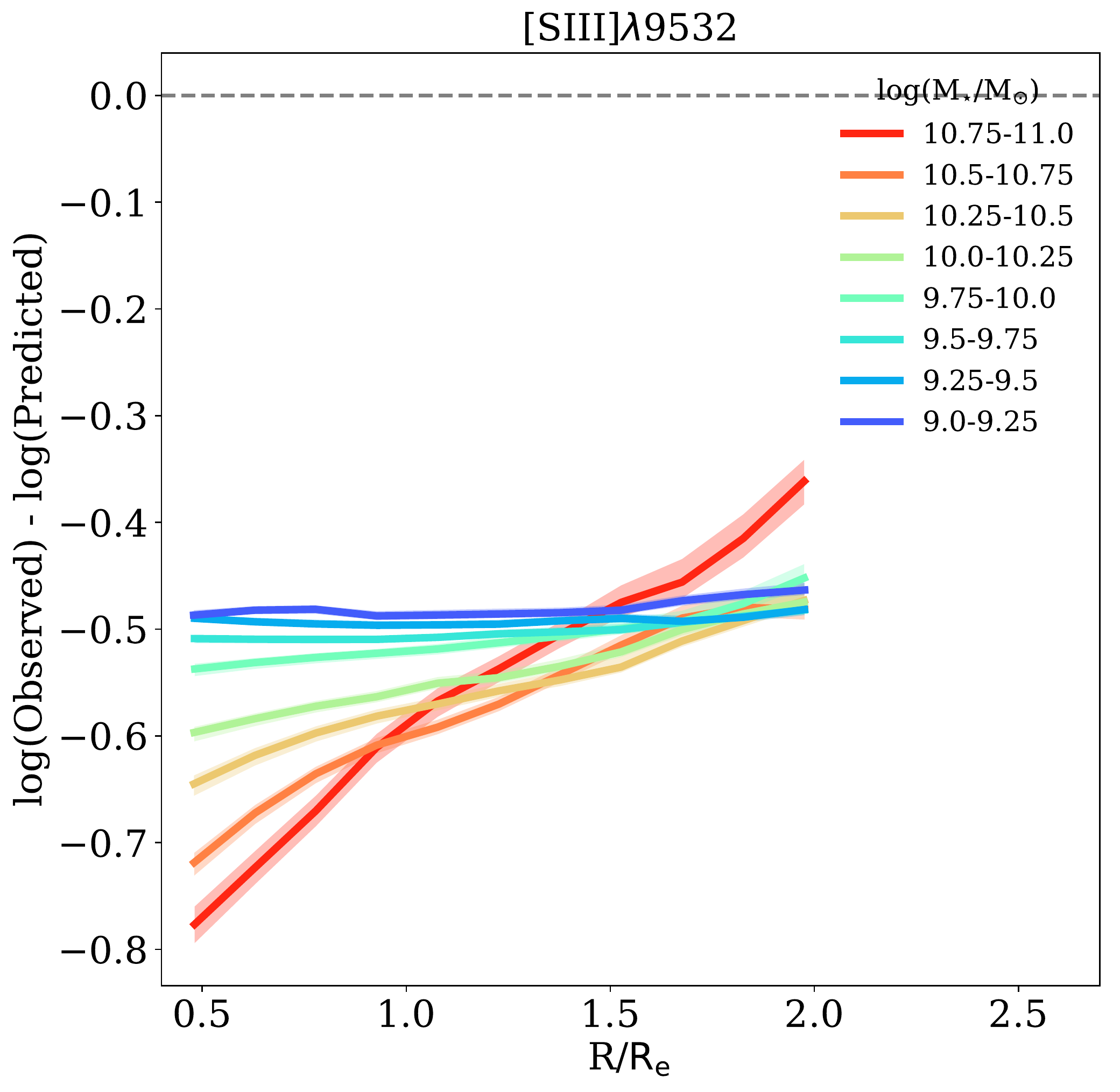}
        \caption{Radial distribution of the difference between the logarithm of the observed \siii$\lambda$9532 flux and that of the best-fit model (both normalised to \hb) predicted by IZI, in eight stellar mass $M_\star$ bins, as reported in the legend. This clearly shows that \siii$\lambda$9532 is largely overestimated, with differences between observed and predicted fluxes up to 0.8~dex.}
        \label{fig:a_lines_siii}
   \end{figure}

Fig.~\ref{fig:a_lines_siii} instead shows the distribution of the difference between the logarithm of the observed flux of the \siii$\lambda$9532 and that of the best-fit model (both normalised to \hb) predicted by IZI, taking into account D13 models and a Gaussian prior on the ionisation parameter, for all the spaxels used in this work, as a function of radius and stellar mass, analogously to Fig.~\ref{fig:a_lines}. Unlike the other emission lines, \siii$\lambda$9532 is largely overestimated, with differences between observed and predicted fluxes up to 0.8~dex. Similarly to \sii\, lines, there is a dependence also on radius, since in the inner regions ($R<1$~\re) of the most massive galaxies the observed \siii\, fluxes are overestimated of 0.3 dex more with respect to the values in the outer regions ($R>1$~\re).

\end{document}